\documentclass[aps,preprintnumbers,amsmath,amssymb,nofootinbib,superscriptaddress,notitlepage,11pt]{revtex4-1}

\usepackage{epsfig}
\usepackage[utf8]{inputenc}
\usepackage{subfigure}
\usepackage{multirow}
\usepackage[figuresright]{rotating}
\usepackage{graphicx}
\usepackage{color}
\usepackage{xcolor}
\usepackage[normalem]{ulem}
\usepackage{hyperref}
\usepackage{pifont}
\usepackage{booktabs}
\usepackage{multirow}
\usepackage{longtable,booktabs}
\usepackage{diagbox}
\usepackage{afterpage}

\newcommand{\itp}{\affiliation{CAS Key Laboratory of Theoretical Physics, Institute of Theoretical Physics,\\ Chinese Academy of Sciences, Beijing 100190, China}}

\newcommand{\ucas}{\affiliation{School of Physical Sciences, University of Chinese Academy of Sciences, Beijing 100049, China}}

\newcommand{\qfnu}{\affiliation{College of Physics and Engineering, Qufu Normal University, Qufu 273165, China}}

\newcommand{\peng}{\affiliation{Peng Huanwu Collaborative Center for Research and Education, Beihang University, Beijing 100191, China}}

\begin{document}

\title{Hadronic decays of the heavy-quark-spin molecular partner of $T_{cc}^+$}

\author{Zhao-Sai Jia}\qfnu \itp
\author{Mao-Jun Yan}\itp
\author{Zhen-Hua Zhang}\itp \ucas
\author{Pan-Pan Shi}\itp \ucas
\author{Gang Li}\email{gli@qfnu.edu.cn} \qfnu 
\author{Feng-Kun Guo} \email{fkguo@itp.ac.cn}
\itp \ucas \peng

\date{\today}

\begin{abstract} 
\rule{0ex}{3ex}
Starting from the hypothesis that the $T_{cc}^+$ discovered at LHCb is a $D^{\ast+} D^0/D^{\ast 0}D^+$ hadronic molecule, we consider the partial width of its heavy quark spin partner, the $T_{cc}^{\ast +}$ as a $D^{\ast +} D^{\ast 0}$ shallow bound state, decaying into the $D^{\ast}D\pi$ final states including the contributions of the $D^{\ast} D$ and $D^{\ast} \pi$ final state interaction by using a nonrelativistic effective field theory. Because of the existence of the $T_{cc}^+$ pole, the $I=0$ $D^{\ast} D$ rescattering can give a sizeable correction up to about 40\% to the decay widths considering only the tree diagrams, and the $D^{\ast} \pi$ rescattering correction is about $10\%$.
The four-body partial widths of the $T_{cc}^{*+}$ into $D D\pi\pi$ are also explicitly calculated, and we find that the interference effect between different intermediate $D^*D\pi$ states is small. The total width of the $T_{cc}^{*+}$ is predicted to be about 41~keV.
\end{abstract}

\maketitle

\section{Introduction}
Recently, a double-charm exotic candidate, the $T_{cc}^+$ with probable quantum numbers $I(J^P)=0(1^+)$, was discovered by the LHCb Collaboration in the $D^0D^0\pi^+$ invariant mass distribution~\cite{LHCb:2021vvq,LHCb:2021auc}. The difference between its mass and the $D^0D^{*+}$ threshold, $\delta m$, and its decay width, $\Gamma$, were obtained in two different models. A fit using a relativistic $P$-wave two-body Breit-Wigner function with a Blatt-Weisskopf form factor gave \cite{LHCb:2021vvq,LHCb:2021auc}
\begin{align}
    \delta m_{\rm BW}&=-273\pm61 \pm 5_{-14}^{+11}\, \rm{keV},\nonumber\\
    \Gamma_{\rm BW}&=410\pm 165\pm43_{-38}^{+18}\, \rm{keV};
    \label{Eq:BW1}
\end{align}
while a unitarized Breit-Wigner profile showed \cite{LHCb:2021auc}\footnote{An analysis of the LHCb data with the full $DD\pi$ three-body effects taken into account gives $\delta m_{\rm pole}=-356_{-38}^{+39}$~keV and $\Gamma_{\rm pole} = (56 \pm 2)$~keV~\cite{Du:2021zzh}.}
\begin{align}
    \delta m_{\rm pole}&=-360\pm 40_{-0}^{+4}\, \rm{keV},\nonumber\\
    \Gamma_{\rm pole}&=48\pm 2_{-14}^{+0}\, \rm{keV}.
    \label{Eq:UBW2}
\end{align}
Both results demonstrate the closeness of the 
$T_{cc}^+$ mass to the $DD^{\ast}$ threshold, and therefore the  $T_{cc}^{+}$ is an excellent candidate of a hadronic molecule, as analyzed in Refs.~\cite{Dong:2021bvy,Chen:2022asf,Feijoo:2021ppq,Meng:2021jnw,Ling:2021bir,Du:2021zzh}.
Based on the assumption that the $T_{cc}^{+}$ is a $DD^{\ast}$ molecular state with respect to the heavy quark spin symmetry (HQSS) \cite{Isgur:1989vq,Neubert:1993mb, Manohar:2000dt}, the $T_{cc}^{\ast+}$, as a cousin of the $T_{cc}^+$, is predicted as a $D^{\ast +}D^{\ast 0}$ hadronic molecule with the quantum numbers $I(J^P)=0(1^+)$ in Refs.~\cite{Du:2021zzh, Albaladejo:2021vln}. In particular, the mass of the $T_{cc}^{\ast+}$ relative to the $D^{\ast}D^{\ast}$ threshold is predicted to be $\mathcal{B} = 2m_{D^*} - m_{T_{cc}^*}=(503 \pm 40) \, \rm{keV}$ in Ref.~\cite{Du:2021zzh}, which is called the binding energy of the $T_{cc}^{\ast+}$ in the following. As a heavy-quark-spin partner of the $T_{cc}^+$, the $T_{cc}^{\ast+}$ is plausible to be observed in the strong decay process $T_{cc}^{\ast } \to D^{\ast}D\pi$, whose partial width can be calculated in a nonrelativistic effective field theory called the XEFT~\cite{Fleming:2007rp,Fleming:2008yn, Fleming:2011xa, Mehen:2011ds, Margaryan:2013tta, Braaten:2010mg, Canham:2009zq, Jansen:2013cba, Mehen:2015efa, Jansen:2015lha, Alhakami:2015uea, Braaten:2015tga, Dai:2019hrf, Braaten:2020nmc, Braaten:2020iye, Braaten:2020iqw}.

The XEFT was first constructed in Ref.~\cite{Fleming:2007rp} to systematically study the properties of the exotic $X(3872)$ \cite{Belle:2003nnu,ParticleDataGroup:2022pth}, also known as $\chi_{c1}(3872)$, including the effects of dynamic pions. With a mass coinciding with the $D^0\bar{D}^{\ast 0}$ threshold,\footnote{The recently updated difference between the $D^0\bar{D}^{\ast 0}$ threshold and the mass of the $X(3872)$ is $\delta m_{X(3872)}=m_{D^0}+m_{D^{\ast 0}}-m_{X(3872)}=(0.01\pm 0.14)~\mathrm{MeV}$ in Ref.~\cite{LHCb:2020xds}, and $\delta m_{X(3872)}=(0.12\pm 0.13)~\mathrm{MeV}$ in Ref.~\cite{LHCb:2020fvo}; the values of the $X(3872)$ mass in both measurements were determined from a Breit-Wigner fit.}
the $X(3872)$ is assumed to be a hadronic molecule composed of $D^0 \bar{D}^{\ast 0}+c.c$ with an extremely small binding energy, and thus the elementary degrees of freedom, the $D$, $D^{\ast}$, $\bar{D}$, $\bar{D}^{\ast }$, and $\pi$, are all treated nonrelativistically in the XEFT. The decay width of $X(3872) \to D^0\bar{D}^0\pi^0$ was calculated to the next-to-leading order (NLO) in Ref.~\cite{Fleming:2007rp} by using the XEFT and the leading order (LO) results are consistent with those in Ref.~\cite{Voloshin:2003nt}, which exploits the universal behavior of the long-range $D^0\bar D^{\ast0}+c.c.$ part of the $X(3872)$ wave function with a small binding energy. The $\pi^0D^0$, $\pi^0\Bar{D}^0$ and $D^0\Bar{D}^0$ rescattering effects, which were neglected in Ref.~\cite{Fleming:2007rp}, were shown to be significant at NLO \cite{Dai:2019hrf} in the XEFT calculation of the $X(3872) \to D^0\bar{D}^0\pi^0$ partial width as it doubles the uncertainty of the partial width as a function of the $X(3872)$ binding energy predicted in Ref.~\cite{Fleming:2007rp}. Since the mass of the $T_{cc}^{(\ast)+}$ is very close to the thresholds of $D^{(\ast)}D^{\ast}$ and $D^{(\ast)}D\pi$, the XEFT is also valid for the study of the  $T_{cc}^{(\ast)+}$ properties. The partial widths for the decays $T_{cc}^+ \to D^0D^0\pi^+$, $D^+D^0\pi^0$, and $D^+D^0\gamma$ were calculated in Refs.~\cite{Fleming:2021wmk,Dai:2023mxm} by using the XEFT, and the total width obtained therein is close to the value given in Eq.~\eqref{Eq:UBW2} extracted from fitting the experimental data with a unitarized Breit-Wigner model~\cite{LHCb:2021auc}. The calculation of the $T_{cc}^{\ast}\to D^{\ast}D\pi$ decay widths in the XEFT can be helpful for searching the $T_{cc}^{*+}$ in the $D^{\ast}D\pi$ invariant mass distributions.

In this paper, we assume that the $T_{cc}^{\ast +}$ is a $D^{\ast +} D^{\ast 0}$ shallow bound state with a binding energy $\mathcal{B}=(503 \pm 40) \, \rm{keV}$ predicted in Ref.~\cite{Du:2021zzh}, and use the XEFT to calculate the partial decay widths of $T_{cc}^{\ast +}\rightarrow D^{*+}D^0\pi^0, D^{*0}D^+\pi^0$, and $D^{*0}D^0\pi^+$, including the corrections from the $D^{\ast}\pi$ and $D^{\ast} D$ final state interactions (FSIs). 
Due to the existence of the $T_{cc}^{+}$, the $S$-wave isoscalar $D^{\ast} D$ rescattering can give a sizeable contribution (about $20\% \sim 40\%)$ to the $T_{cc}^{\ast+}$ decay width at LO, despite that the $T_{cc}^*\to T_{cc}\pi$ is isospin breaking. Since the $D^*$ in the final state is unstable, it is reconstructed experimentally in the $D\pi$ or $D\gamma$ final state. In the former case, the  $D^*D\pi$ decay modes become the $DD\pi\pi$ ones and different $D^*D\pi$ intermediate states can interfere. Thus, we also calculate the 4-body decay widths for $T_{cc}^{*+}\to D^+D^0\pi^0\pi^0$ and $D^0D^0\pi^+\pi^0$, and show that they can be well approximated by the 3-body decay widths multiplying the $D^*\to D\pi$ branching fractions.

This paper is organized as follows. In Sec.~\ref{sec:LAGRANGIAN}, we introduce the XEFT effective Lagrangian for the charmed mesons and pions, and the power countings of the Feynman diagrams in the $T_{cc}^{\ast}\to D^{\ast}D\pi$ processes. The amplitudes and partial decay rates of the $T_{cc}^{\ast} \to D^{\ast} D \pi$ including the corrections from the $D^{\ast}\pi$ and $D^{\ast} D$ FSIs in the XEFT are derived in Sec.~\ref{sec:3BodyAmplitudes}, and the numerical results for the partial decay widths of the  $T_{cc}^{*+} \to D^*D\pi$ are shown in Sec.~\ref{sec:3BodyResults}. The 4-body decay $T_{cc}^{*} \to DD\pi\pi$ including the corrections from the $D^*\pi$ and $D^*D$ FSIs are given in Sec.~\ref{sec:4BodyResults}. Finally, all the results are summarized in Sec.~\ref{sec:SUMMARY}. Some derivations and expressions are relegated to appendices.

\section{Effective Lagrangian and power counting}\label{sec:LAGRANGIAN}

In this section, we introduce the effective Lagrangian for the decays of the $T_{cc}^{\ast +}$ and the power counting rules of the diagrams in the decay processes. As a heavy-quark-spin partner of the $T_{cc}^+$, the $T_{cc}^{\ast +}$ is predicted as an $S$-wave isoscalar $D^{\ast +}D^{\ast 0}$ shallow bound state with $J^P=1^+$ and a binding energy  $\mathcal{B}=(503 \pm 40) ~ \rm{keV}$~\cite{Du:2021zzh}. With such a binding energy, the upper bounds of the typical momentum and velocity of the $D^*$ mesons in the $T_{cc}^{*+}$ bound state are  
$p_{D^*}\sim\gamma\equiv \sqrt{2\mu_{D^*}\mathcal{B}}\lesssim 33~\rm{MeV}$ and $v_{D^*}\simeq\sqrt{{\mathcal{B}/}({2\mu_{D^*})}}\lesssim 0.02$, respectively, where $\mu_{D^*}$ is the reduced mass of $D^{*+}$ and $D^{*0}$, and therefore the nonrelativistic approximation is valid for the $D^*$ and $D$ mesons. The maximum kinetic energy of the emitted pion in the $T_{cc}^*$ decays
is 
\begin{align}
    E_{\mathbf{p}\pi}=\frac{m_{T_{cc}^*}^2-(m_{D}+m_{D^*})^2+m_{\pi}^2}{2m_{T_{cc}^{*}}}-m_\pi\simeq 3.4~\rm{MeV},
\end{align}
which leads to the upper bounds of the typical momentum and velocity of the emitted pion to be $p_{\pi}\simeq \sqrt{2m_{\pi}E_{\mathbf{p}\pi}}\lesssim 31~\rm{MeV}$ and $v_{\pi}\simeq{p_{\pi}}/{m_{\pi}}\lesssim 0.22$. Here $m_{T_{cc}^*}, m_{D}, m_{D^*} $, and $m_{\pi}$ are the masses of $T_{cc}^*$, $D$, $D^*$, and $\pi$, respectively. Clearly, the pions can also
be treated nonrelativistically in the $T_{cc}^{\ast}\to D^{\ast}D \pi$ and $T_{cc}^{*} \to DD\pi\pi$ decays.

The elementary degrees of freedom in the effective Lagrangian are the nonrelativistic $D^*$ and $D$ mesons, which are written as isodoublets of the pseudoscalar and vector fields \cite{Fleming:2007rp}
\begin{align}
H=\left(\begin{array}{c}
D^{0} \\
D^{+}
\end{array}\right), \quad H^{\ast }=\left(\begin{array}{c}
D^{\ast 0 } \\
D^{\ast+}
\end{array}\right),
\end{align}
and the pions, which are in the isospin adjoint representation,
\begin{align}
\pi=\left(\begin{array}{cc}
\pi^{0} & \sqrt{2}\pi^{+} \\
\sqrt{2}\pi^{-} & -\pi^{0}
\end{array}\right) .
\end{align} 

\begin{figure}
    \centering
    \includegraphics[scale=0.34]{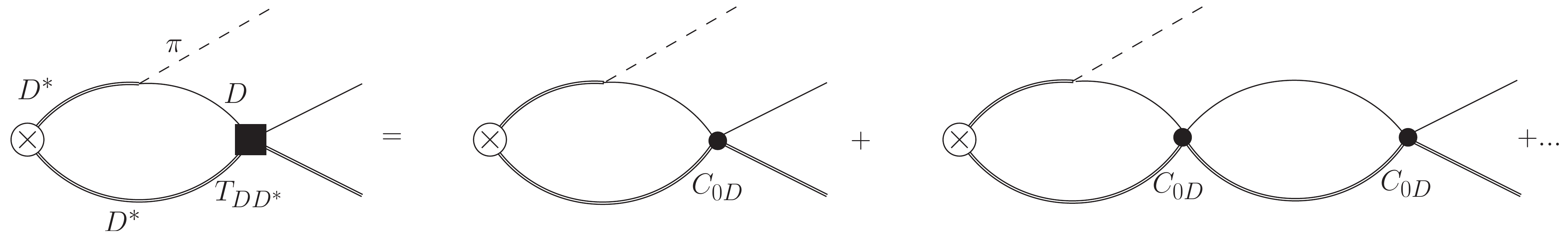}
    \caption{Resumming the $D^*D$ rescattering diagrams. The single thin lines represent the $D^+(D^0)$, the double lines represent the $D^{*0}(D^{*+})$, and the dashed lines represent the $\pi^{0}(\pi^{+})$.}
    \label{Fig.Resummation}
\end{figure}

The LO XEFT effective Lagrangian for the $T_{cc}^+$ has been given in Ref.~\cite{Fleming:2021wmk}. As an analogy, the XEFT Lagrangian we use for the $T_{cc}^{*+}$ as a heavy-quark-spin partner of the $T_{cc}^{+}$ reads~\cite{Fleming:2007rp,Guo:2017jvc}
\begin{align}
\mathcal{L}=&\, H_i^{\ast  \dagger}\left(i \partial^{0}+\frac{\nabla^{2}}{2 m_{H^{\ast}}}\right) H^{\ast i}+H^{\dagger}\left(i \partial^{0}+\frac{\nabla^{2}}{2 m_{H}}\right) H +\frac{1}{2}\left\langle\pi^{\dagger}\left(i \partial^{0}+\frac{\nabla^{2}}{2 m_{\pi}}+\delta\right) \pi \right\rangle\nonumber\\
&-C_{0}\left(H_i^{\ast T} \tau_{2} H^{\ast i}\right)^{\dagger}\left(H_i^{\ast T} \tau_{2} H^{\ast i}\right)-C_{1}\left(H_i^{\ast T} \tau_{2} \tau_{a} H^{\ast i}\right)^{\dagger}\left(H_i^{\ast T} \tau_{2} \tau_{a} H^{\ast i}\right)\nonumber\\
&+\frac{\bar{g}}{F_{\pi} \sqrt{2m_{\pi}}} \left(H^{\dagger} \partial_{i} \pi H^{\ast i}+\mathrm{H.c.} \right)+\frac{C_{0D}}{2}(H^T\tau_2H_i^{\ast})^ {\dagger}(H^T\tau_2 H^{\ast i})+\frac{C_{\frac{1}{2}\pi}}{6m_{\pi}}\left(\pi\tau_1H_i^*\right)^\dagger\left(\pi\tau_1H^{*i}\right)\nonumber\\
&+\frac{C_{\frac{3}{2}\pi}}{12m_{\pi}}\left\{\left[\left(\pi\tau_3+\frac{1}{2}\langle\pi\tau_3\rangle\right)\tau_1 H_i^{*}\right]^{\dagger}\left[\left(\pi\tau_3+\frac{1}{2}\langle\pi\tau_3\rangle\right)\tau_1 H^{*i}\right]\right.\nonumber\\
&\left.+3\left[\left(\pi\tau_3-\frac{1}{2}\langle\pi\tau_3\rangle\right) H_i^{*}\right]^{\dagger}\left[\left(\pi\tau_3-\frac{1}{2}\langle\pi\tau_3\rangle\right) H^{*i}\right]\right\},
\label{eq:XEFTlagrangian}
\end{align}
where $m_{H}$, $m_{H^{\ast}}$, and $m_{\pi}$ are the masses of the $H$, $H^{\ast}$, and $\pi$ particles, respectively;\footnote{Here we use the physical masses for $D^{*+}$ and $D^{*0}$, $D^+$ and $D^0$, $\pi^+$ and $\pi^0$. In this way, the isospin violating effects due to the mass splitting for mesons within the same isospin multiplet are included.} $\delta=\Delta-m_\pi\simeq 7~\rm{MeV}$ with $\Delta=m_{D^{*0}}-m_{D^0}$ comes from the shift of the residual mass from the $D^*$ 
kinetic term~\cite{Fleming:2007rp} and is related to a small scale $\mu=\sqrt{\Delta^2-m_{\pi}^2}\simeq\sqrt{2m_{\pi}\delta}\approx 45~\rm{MeV}$ appearing in the pion propagator~\cite{Fleming:2007rp,Dai:2019hrf}; the pion decay constant is taken as $F_{\pi}=92.2~\rm{MeV}$, and $\tau_a$ with $a=1,2,3$ are the Pauli matrices in the isospin space, in which the traces $\left(\langle\, \rangle\right)$ act. 
Notice that in Eq.~\eqref{eq:XEFTlagrangian}, both pions and $D^{(*)}$ mesons are nonrelativistic particles, which means the $\pi$ operator annihilates and the $\pi^\dagger$ operator creates the $\pi$ quanta, so as the $D^{(*)}$ and $D^{(*)\dagger}$~\cite{Fleming:2007rp}, and one has 
\begin{align}
    \pi^{\dagger}=\begin{pmatrix}
    (\pi^0)^{\dagger} & \sqrt{2}(\pi^-)^{\dagger}\\
    \sqrt{2}(\pi^+)^{\dagger}&-(\pi^0)^{\dagger} 
    \end{pmatrix},\quad
    H^\dagger=\left(\begin{array}{c}
(D^{0})^{\dagger} \\
(D^{+})^{\dagger}
\end{array}\right)^T, \quad H^{\ast\dagger }=\left(\begin{array}{c}
(D^{\ast 0 })^{\dagger} \\
(D^{\ast+})^{\dagger}
\end{array}\right)^T.
\end{align}
The first line of Eq.~\eqref{eq:XEFTlagrangian} includes the kinetic terms for the charmed mesons and pions. The second line contains the contact interactions of the $D^{*+}$ and $D^{*0}$, where the term with $C_{0}$ mediates the $D^{\ast}D^{\ast}$ scattering in the $I=0$ channel, and the term with $C_1$ mediates the scattering in the $I=1$ channel. The first term in the third line is the same term in Ref.~\cite{Fleming:2021wmk} which couples the charmed mesons to pions derived from the heavy hadron chiral perturbation theory (HH$\chi$PT), and the coupling constant $\bar{g}\simeq 0.27$,\footnote{Notice that $\bar{g}$ is related to the $g$ in Ref.~\cite{Fleming:2021wmk} by $\bar g=g/2$.} is determined from the updated $D^{*+}$ decay width~\cite{ParticleDataGroup:2022pth}. 
The second term in the third line is the contact interaction for $D^{*}D\to D^{*}D$ with $I=0$, and the resummation effect of the coupling $C_{0D}$ shown in Fig.~\ref{Fig.Resummation} needs to be considered \cite{Dai:2019hrf} due to the existence of the $T_{cc}^+$ shallow bound state. The resummation effect is equivalent to replacing $C_{0D}$ with the near-threshold $T$-matrix~\cite{Kaplan:1998we}
\begin{align}
	C_{0D}\to T_{DD^*}= -\frac{2\pi}{\mu_{D}} \frac{1}{-1/a+ i p}, 
	\label{Eq:unitary TDDstar}
\end{align}
where $\mu_{D}$ is the reduced mass of $D^{\ast+}(D^{\ast0})$ and $D^0(D^+)$, $p=\vert \vec{p}_{D^{\ast}}-\vec{p}_D \vert/2$ is the relative momentum between $D^{\ast+}(D^{\ast0})$ and $D^0(D^+)$ in the $D^*D$ center-of-mass (c.m.) frame, and the $D^{*+}D^0(D^{*0}D^+)$ scattering length $a$ is set to be $a=\left[-\left(6.72^{+0.36}_{-0.45}\right)-i\left(0.10^{+0.03}_{-0.03}\right)\right]\, \rm{fm}$~\cite{Du:2021zzh}. Here we neglect the isospin breaking effect in the $I=0$ $D^*D\to D^*D$ rescattering, which is a higher order effect~\cite{Guo:2019fdo}. There is no isovector state like the $T_{cc}^+$ found near the $D^*D$ threshold, so there should be no near-threshold pole singularity in the $I=1$ scattering
amplitude for $D^*D\to D^*D$; thus, the isovector $DD^*$ FSI should be much weaker than the isoscalar one and is neglected in our calculation.
The last two terms with $C_{\frac{1}{2}\pi}$ and $C_{\frac{3}{2}\pi}$ are the $D^*\pi \to D^*\pi$ contact interactions for $I=\frac{1}{2}$ and $I=\frac{3}{2}$, respectively, where $ C_{\frac{1}{2}\pi}=25.2~\mathrm{GeV}^{-1}, C_{\frac{3}{2}\pi}=-6.8~\mathrm{GeV}^{-1}$ are derived by matching to the $D^*\pi$ scattering lengths given in Ref.~\cite{Liu:2012zya} (for detailed derivations, see Appendix~\ref{sec:Contact interactions}).

The square of effective coupling between the $T_{cc}^{*+}$ hadronic molecule and the
$D^{*+}D^{*0}$ components can be derived from the residue of the $D^{*+}D^{*0}$ scattering amplitude at the $T_{cc}^{*+}$ pole as~\cite{Weinberg:1965zz,Baru:2003qq,Guo:2017jvc}
\begin{align}
{g_0^2} = \frac{2\pi \gamma}{\mu^2_{D^*}},
\end{align} 
and thus the effective Lagrangian for the $T_{cc}^{\ast +}$ coupling to $D^{\ast+} D^{\ast0}$ can be written as 
\begin{align}
	\mathcal{L}_0=& \frac{g_0}{\sqrt{2}} \varepsilon_{ijk} T_{cc}^{\ast +,i}  D^{\ast+,j} D^{\ast0,k},
	\label{Eq:L0}
\end{align}
where $\varepsilon_{ijk}$ is the 3-dimensional antisymmetric Levi-Civita tensor. 

\begin{figure}[tb]
    \subfigure[] {
        \includegraphics[scale=0.5]{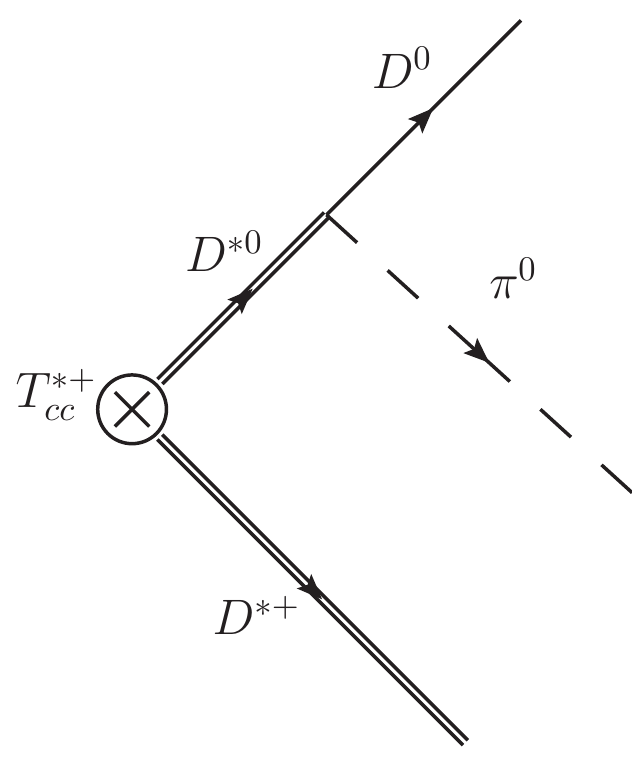}\label{Fig.Tccstar+_Dstar+D0pi0_LO}
    }
    \subfigure[] {
        \includegraphics[scale=0.5]{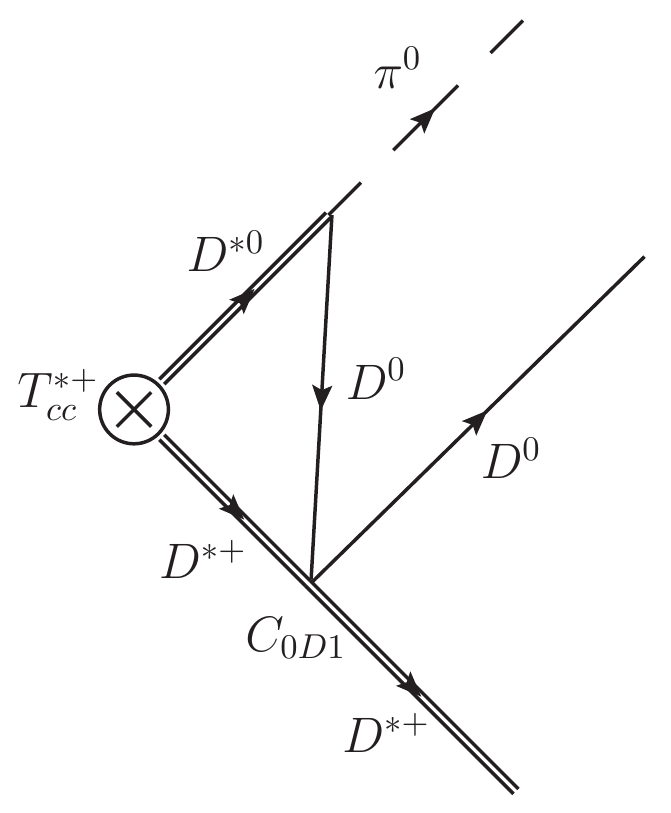}\label{Fig.Tccstar+_Dstar+D0pi0_C0D1}
    }
    \subfigure[] {
        \includegraphics[scale=0.5]{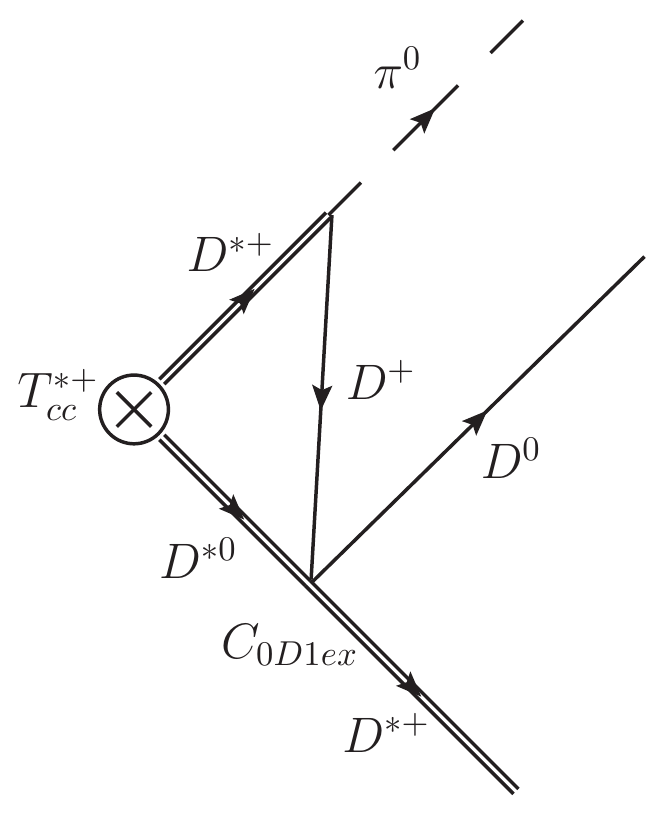}\label{Fig.Tccstar+_Dstar+D0pi0_C0D2}
    }\\
    \subfigure[] {
        \includegraphics[scale=0.5]{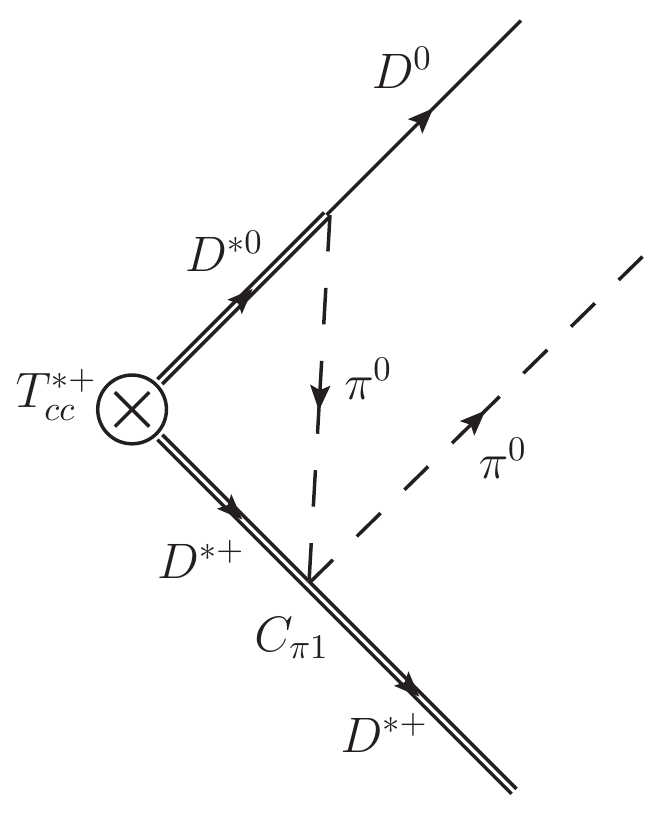}\label{Fig.Tccstar+_Dstar+D0pi0_NLO1}
    }
    \subfigure[] {
        \includegraphics[scale=0.5]{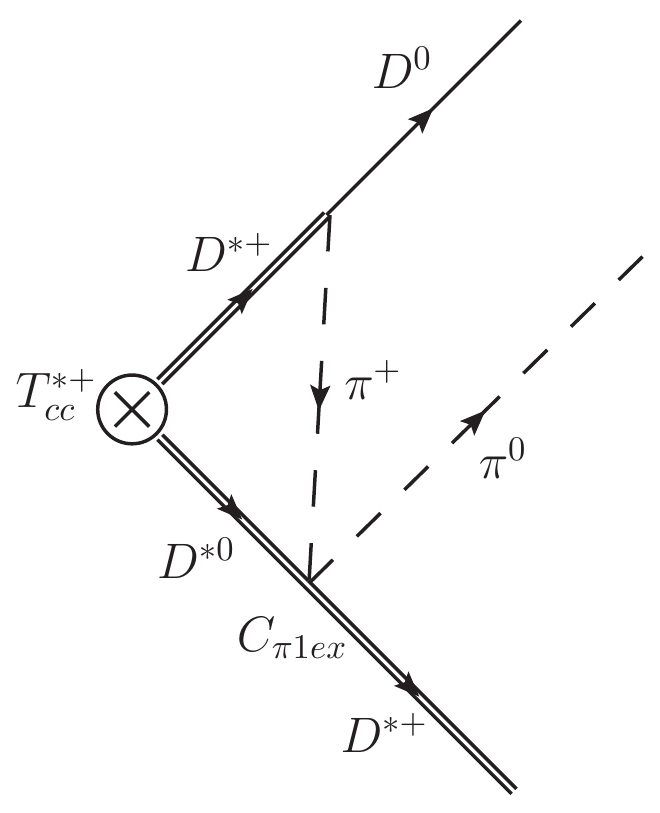}\label{Fig.Tccstar+_Dstar+D0pi0_NLO2}
    }
    \caption{Feynman diagrams for calculating the partial decay width of $T_{cc}^{*+}\rightarrow D^{*+}D^0\pi^0$. The circled cross is the $T_{cc}^{*+}$ state, the single thin lines represent the $D^+(D^0)$, the double lines represent the $D^{*0}(D^{*+})$, and the dashed lines represent the $\pi^{0}(\pi^{+})$.}
    \label{Fig.Tccstar+_Dstar+D0pi0}
\end{figure}

With the above Lagrangians in Eqs.~\eqref{eq:XEFTlagrangian} and \eqref{Eq:L0}, the LO amplitude for the $T_{cc}^{*}\to D^*D\pi$ including the effects of the $D^*D$ and $D^*\pi$ FSIs are shown in Fig.~\ref{Fig.Tccstar+_Dstar+D0pi0}. Here we only show the diagrams for the decay $T_{cc}^{*+}\to D^{*+}D^0\pi^0$, and there are also similar diagrams for the $T_{cc}^{*+}\to D^{*0}D^+\pi^0$ and the $T_{cc}^{*+}\to D^{*0}D^0\pi^+$, except that no $D^{*0}D^{0}$ FSI
diagram is included in the $T_{cc}^{*+}\to D^{*0}D^0\pi^+$ as it only contains $D^*D$ pair with $I=1$. 

In the following, we will give a brief power counting to the contributions of all these diagrams. The power counting for the decays of the $X(3872)$ has been given in detail in Refs.~\cite{Fleming:2007rp,Dai:2019hrf,Yan:2021wdl}, and the discussions here for the $T_{cc}^{*}\to D^*D\pi$ are similar. 
The relevant small momenta involved in the decays of the $T_{cc}^{\ast +}$ are $\{p_D,p_{D^{\ast}},p_{\pi},\gamma, \mu\}$, which are at the same order and denoted by $Q$ to be the power counting scale. In the decay diagrams, each pion vertex contributes at $\mathcal{O}(Q)$, and each nonrelativistic propagator contributes at $\mathcal{O}(Q^{-2})$. As the nonrelativistic energy counts as $\mathcal{O}(Q^{2})$, each loop integral is of $\mathcal{O}(Q^{5})$. The $C_{\pi}$ contact term is related to the $D\pi$ contact term in Ref.~\cite{Dai:2019hrf} via the HQSS and therefore $C_\pi$ is of the same order as the $D\pi$ vertex in Ref.~\cite{Dai:2019hrf}, i.e., $\mathcal{O}(Q^{0})$. 
As the $I=0$ contact interaction between the $D^{*}$ and $D$, 
the $C_{0D}$ should be replaced with $T_{DD^*}$ in Eq.~\eqref{Eq:unitary TDDstar} due to the near threshold $T_{cc}^+$ pole, and contribute at $\mathcal{O}(Q^{-1})$~\cite{Fleming:2007rp,Guo:2017jvc}. 
For the diagrams in Fig.~\ref{Fig.Tccstar+_Dstar+D0pi0}, the amplitude from the diagram in Fig.~\ref{Fig.Tccstar+_Dstar+D0pi0_LO} scales as $\mathcal{O}(Q / Q^{2})=\mathcal{O}(Q^{-1})$ since there are one nonrelativistic propagator and one $P$-wave pion vertex which gives a factor of $p_{\pi} \sim \mathcal{O}(Q)$. The isospin breaking diagrams in Figs.~\ref{Fig.Tccstar+_Dstar+D0pi0_C0D1} and \ref{Fig.Tccstar+_Dstar+D0pi0_C0D2} also scale as $\mathcal{O}(Q^{-1})$ for the decays $T_{cc}^{\ast+ }\to D^{*+}D^0\pi^0$, $T_{cc}^{\ast+ }\to D^{*0}D^+\pi^0$ considering the resummation effect (with $C_{0D}$ replaced by $T_{DD^*}$ which has a near-threshold $T_{cc}$ pole) and can contribute at LO; if one uses isospin averaged masses for all the involved mesons, the contributions of these diagrams vanish. Isospin breaking effects are enhanced due to the presence of the $T_{cc}^+$ pole which is located much closer to the $D^{*+}D^0$ threshold than to the $D^{*0}D^+$ one. The amplitudes from diagrams in Figs.~\ref{Fig.Tccstar+_Dstar+D0pi0_NLO1} and \ref{Fig.Tccstar+_Dstar+D0pi0_NLO2} scale as $\mathcal{O}\left(Q^{0}\right)$ and can only contribute at NLO. 

\section{Differential decay rate of  $T_{cc}^{\ast}\to D^{\ast}D\pi$}\label{sec:3BodyAmplitudes}
In this section, we give all the decay amplitudes of $T_{cc}^{\ast }\to D^*D\pi$ in Fig.~\ref{Fig.Tccstar+_Dstar+D0pi0}, including the processes $T_{cc}^{\ast+ }\to D^{*+}D^0\pi^0$, $T_{cc}^{\ast+ }\to D^{*0}D^+\pi^0$ and $T_{cc}^{\ast+ }\to D^{*0}D^0\pi^+$, and give the partial differential decay rates including the effects of $D^*D$ and $D^*\pi$ rescattering. The Breit-Wigner form of the $D^*$ propagator, $G_{D^*}(p)$, is used to include the contribution of the $D^{*}$ self-energy, i.e.,
\begin{align}
  G_{D^*}(p)=  \frac{i}{p^0_{D^*}-m_{D^*}-\frac{\vec{p}_{D^*}^2}{2m_{D^*}}+i\frac{\Gamma_{D^*}}{2}},
  \label{Eq.Dstar_self_energy}
\end{align}
where $D^*$ denotes $D^{*+}$ or $D^{*0}$, $p=(p^0_{D^*},\vec{p}_{D^*})$ is the 4-momentum of the $D^*$, $\Gamma_{D^{*+}}=83.4~\rm{keV}$, and $\Gamma_{D^{*0}}=55.3~\rm{keV}$ is settled by isospin symmetry \cite{Guo:2019qcn}.

\subsection{Partial decay rate of $T_{cc}^{\ast +}\to D^{\ast +}D^0\pi^0$}\label{sec:Tccstar_Dstar+D0pi0Amp}
First, we consider the partial decay rate of $T_{cc}^{\ast +}\to D^{\ast +}D^0\pi^0$.
The LO amplitude from the tree diagram in Fig.~\ref{Fig.Tccstar+_Dstar+D0pi0_LO} reads
\begin{align}
 i\mathcal{A}_{a}=&\frac{-g_0 \bar{g}\mu_{D^*}}{\sqrt{m_{\pi^0}}F_{\pi}} \frac{1}{\vec{p}_{D^{\ast+}}^2+ \gamma^2-i\mu_{D^*}\Gamma_{D^{*0}}} \varepsilon_{ijk} \epsilon^i\left(T_{cc}^{\ast+}\right) \epsilon^{j*}\left(D^{\ast +}\right) p^k_{\pi^0},
\end{align}
where $\vec{p}_{D^{\ast +}}$ is the three-momentum of the external $D^{\ast+}$, $\vec{p}_{\pi^0}$ is the three-momentum of the final state $\pi^0$, and the  $\epsilon^i(T_{cc}^{*+})$ and
$\epsilon^j(D^{*+})$ are the polarization vectors of the $T_{cc}^{*+}$ and $D^{*+}$, respectively.

The LO amplitude from the $D^{*+}D^0/D^{*0}D^+$ rescattering diagrams in Figs.~\ref{Fig.Tccstar+_Dstar+D0pi0_C0D1} and \ref{Fig.Tccstar+_Dstar+D0pi0_C0D2} reads
\begin{align}
	i \mathcal{A}_{bc}=&\frac{g_0 \bar{g}}{2 \sqrt{m_{\pi^0}} F_{\pi}} \left \{-C_{0D1} I_{b}(p_{\pi^0})-C_{0D1\text{ex}}I_{c} (p_{\pi^0}) \right \} \varepsilon_{ijk} \epsilon^i \left(T_{cc}^{\ast +}\right)\epsilon^{j*}\left(D^{\ast +}\right)  p^k_{\pi^0},
\end{align}
where $C_{0D1}=+\frac{1}{2} C_{0D}$ and $C_{0D1\text{ex}}=-\frac{1}{2} C_{0D}$ are the contact interactions for the $D^{*+}D^0\to D^{*+}D^0$ and the $D^{*0}D^+\to D^{*+}D^0$, respectively, and  $C_{0D}$ needs to be replaced by $T_{DD^*}$ considering the resummation effect due to the existence of the nearby $T_{cc}$ pole, and the exact form of the 3-point scalar loop integral $I(p)$ is given in Appendix~\ref{sec:Triangal loop}~\cite{Guo:2010ak, Dai:2019hrf}, with the masses of the three particles in the loop represented by $m_1$, $m_2$ and $m_3$. Here $m_1$, $m_2$, and $m_3$ are taken to be the masses of $D^{\ast 0}$, $D^{\ast +}$, and $D^0$ in the integral $I_{b}(p_{\pi^0})$ appearing in Fig.~\ref{Fig.Tccstar+_Dstar+D0pi0_C0D1}, and the masses of $D^{\ast +}$, $D^{\ast 0}$, and $D^+$ in the integral $I_{c}(p_{\pi^0})$ appearing in Fig.~\ref{Fig.Tccstar+_Dstar+D0pi0_C0D2}, respectively.

The NLO amplitude from the $D^{\ast}\pi$ rescattering diagrams in Figs.~\ref{Fig.Tccstar+_Dstar+D0pi0_NLO1} and \ref{Fig.Tccstar+_Dstar+D0pi0_NLO2} is
\begin{align}
	i\mathcal{A}_{de}=& \frac{g_0 \bar{g}}{4 m_{\pi}^{3/2} F_{\pi}} \left \{ -C_{\pi 1}\left[I_{d}^{(1)}(p_D)-I_{d}(p_D)\right]-\sqrt{2}C_{\pi 1\text{ex}} \left[I_{e}^{(1)}(p_D)+I_{e}(p_D)\right] \right \}\nonumber\\
	&\times\varepsilon_{ijk} \epsilon^i\left(T_{cc}^{\ast +}\right)\epsilon^{j*}\left(D^{\ast +}\right)  p_{D^0}^k,
\end{align}
where the couplings $C_{\pi 1}=\frac{2}{3}C_{\frac{3}{2}\pi}+\frac{1}{3}C_{\frac{1}{2}\pi}=4.1 \, \rm{GeV}^{-1}$ and $C_{\pi 1\text{ex}}=-\frac{\sqrt{2}}{3}C_{\frac{3}{2}\pi}+\frac{\sqrt{2}}{3}C_{\frac{1}{2}\pi}=15.1 \, \rm{GeV}^{-1}$ are the contact interactions for the $D^{*+}\pi^0 \to D^{*+}\pi^0$ and the $D^{*0}\pi^+ \to D^{*+}\pi^0$, respectively,
and the 3-point vector loop integral $I^{(1)}\left(p\right)$ is given in Appendix~\ref{sec:Triangal loop}~\cite{Guo:2010ak, Dai:2019hrf}, with the masses of the three particles in the loop represented by $m_1$, $m_2$, and $m_3$; see Eq.~\eqref{Eq:loop_integral}. Here $m_1$, $m_2$, and $m_3$ are taken to be the masses of $D^{\ast 0}$, $D^{\ast +}$, and $\pi^0$ in the integrals $I_{d}(p_{D})$ and $I_{d}^{(1)}\left(p_{D}\right)$ appearing in Fig.~\ref{Fig.Tccstar+_Dstar+D0pi0_NLO1}, and the masses of $D^{\ast +}$, $D^{\ast 0}$, and $\pi^+$ in the integrals $I_{e}(p_{D})$ and $I_{e}^{(1)}\left(p_{D}\right)$ appearing in Fig.~\ref{Fig.Tccstar+_Dstar+D0pi0_NLO2}, respectively.

The decay rate is given by
\begin{align}
d\Gamma=&2M 2E_1 2E_2 2E_3 \frac{1}{2M}\frac{1}{2j+1} \sum_{\mathrm{spins}} \vert \mathcal{A}\vert^2 d\Phi_{3},
\label{dGamma1}
\end{align}
where the overall factor comes from the normalization of nonrelativistic particles, with $M$ being the mass of the initial particle, $E_1$, $E_2$, and $E_3$ being the energies of three finial-state particles in the $T_{cc}^{*}$ rest frame, respectively, $j$ is the total spin of the initial particle, and there is a sum over all the polarizations of the final state particles. Here the three-body phase space 
\begin{align}
\int d\Phi_{3}=&\int\frac{1}{32 \pi^3} \frac{1}{4 E_1 E_2} d\vert \vec{p}_1 \vert^2 d\vert \vec{p}_2 \vert^2,
\label{Eq.three_body_phase_space}
\end{align}
is derived in Appendix~\ref{sec:Three-body phase space}, where $\vec{p}_1$ and $\vec{p}_2$ are the three-momenta for two of the final state particles.

The NLO partial differential rate for the $T_{cc}^{\ast +} \to D^{\ast +}D^0\pi^0$ including the corrections from the $D^*D$, $D^*\pi$ rescattering reads
\begin{align}
	\frac{d \Gamma_{T_{cc}^{\ast +} \to D^{\ast +}D^0\pi^0}}{dp^2_{D^0}dp^2_{D^{\ast +}}}=&\frac{1}{3}\frac{m_{\pi^0}}{16 \pi^3}\sum_{\mathrm{spins}} \vert \mathcal{A}_{a}+\mathcal{A}_{bc}\vert^2+\frac{1}{3}\frac{m_{\pi^0}}{16 \pi^3} 2 \mathrm{Re}\left[ \sum_{\mathrm{spins}} \left(\mathcal{A}_{a}+\mathcal{A}_{bc}\right) \times \mathcal{A}_{de}^*\right],
\end{align}
where the second term includes the correction of the $D^{\ast}\pi$ rescattering, which is the interference term between the amplitudes at LO and NLO.

\subsection{Partial decay rate of $T_{cc}^{\ast +}\to D^{\ast 0}D^+\pi^0$}\label{sec:Tccstar_Dstar0D+pi0Amp}

For the decay $T_{cc}^{\ast +}\to D^{\ast 0}D^+\pi^0$, the LO amplitude from the tree diagram in Fig.~\ref{Fig.Tccstar+_Dstar+D0pi0_LO} reads
\begin{align}
 i\mathcal{A}_{a2}=&-\frac{g_0 \bar{g}\mu_{D^*}}{\sqrt{m_{\pi^0}}F_{\pi}} \frac{1}{\vec{p}_{D^{\ast 0}}^2+ \gamma^2 -i\mu_{D^*}\Gamma_{D^{*+}}} \varepsilon_{ijk} \epsilon^i\left(T_{cc}^{\ast+}\right) \epsilon^{j*}\left(D^{\ast 0}\right) p^k_{\pi^0},
\end{align}
where $\vec{p}_{D^{\ast 0}}$ is the three-momentum of the external $D^{\ast 0}$, and the $\epsilon^j(D^*)$ is the polarization vector of the $D^*$. The LO amplitude from the $D^{*+}D^0/D^{*0}D^+$ rescattering diagrams in Figs.~\ref{Fig.Tccstar+_Dstar+D0pi0_C0D1} and \ref{Fig.Tccstar+_Dstar+D0pi0_C0D2} is
\begin{align}
	i \mathcal{A}_{bc2}=&\frac{g_0 \bar{g}}{2\sqrt{m_{\pi^0}} F_{\pi}} \left[ -C_{0D2} I_{2b}(p_{\pi^0})-C_{0D2\text{ex}} I_{2c}(p_{\pi^0}) \right] \varepsilon_{ijk} \epsilon^i \left(T_{cc}^{\ast +}\right)\epsilon^{j*}\left(D^{\ast 0}\right)  p^k_{\pi^0},
\end{align}
where $C_{0D2}=\frac{1}{2}C_{0D}$ and $C_{0D2\text{ex}}=-\frac{1}{2}C_{0D}$ are the contact interactions for the $D^{*0}D^+\to D^{*0}D^+$ and the $D^{*+}D^0\to D^{*0}D^+$, respectively, and $C_{0D}$ needs to be replaced by $T_{DD^*}$ considering the resummation effect, and the masses $m_1$, $m_2$, and $m_3$ in the loop integrals $I_{2b}(p_{\pi^0})$ appearing in Fig.~\ref{Fig.Tccstar+_Dstar+D0pi0_C0D1} and $I_{2c}(p_{\pi^0})$ appearing in Fig.~\ref{Fig.Tccstar+_Dstar+D0pi0_C0D2} are taken to be the masses of $D^{\ast +}$, $D^{\ast 0}$, and $D^+$ and the masses of $D^{\ast 0}$, $D^{\ast +}$, and $D^0$, respectively. 

The NLO amplitude from the $D^{\ast}\pi$ rescattering diagrams in Figs.~\ref{Fig.Tccstar+_Dstar+D0pi0_NLO1} and \ref{Fig.Tccstar+_Dstar+D0pi0_NLO2} is
\begin{align}
	i\mathcal{A}_{de2}=& \frac{g_0 \bar{g}}{4 m_{\pi}^{3/2} F_{\pi}} \left \{- C_{\pi 2}\left[I_{2d}^{(1)}(p_{D^+})-I_{2d}(p_{D^+})\right]+\sqrt{2}C_{\pi 2\text{ex}}\left[I_{2e}^{(1)}(p_{D^+})+I_{2e}(p_{D^+})\right] \right \}\nonumber \\ &\times\varepsilon_{ijk} \epsilon^i\left(T_{cc}^{\ast +}\right)\epsilon^{j*}\left(D^{\ast 0}\right)  p_{D^+}^k,
\end{align}
where $C_{\pi2}=\frac{2}{3}C_{\frac{3}{2}\pi}+\frac{1}{3}C_{\frac{1}{2}\pi}=4.1\, \mathrm{GeV}^{-1}$ and $C_{\pi 2\text{ex}}=\frac{\sqrt{2}}{3}C_{\frac{3}{2}\pi}-\frac{\sqrt{2}}{3}C_{\frac{1}{2}\pi}=-15.1 \, \rm{GeV}^{-1}$ are the contact interactions for $D^{*0}\pi^0 \to D^{*0}\pi^0$ and $D^{*+}\pi^- \to D^{*0}\pi^0$, respectively; $m_1$, $m_2$, and $m_3$ are the masses of $D^{\ast +}$, $D^{\ast 0}$, and $\pi^0$ in the loop integrals $I_{2d}(p_{D^+})$ and $I_{2d}^{(1)}(p_{D^+})$ appearing in Fig.~\ref{Fig.Tccstar+_Dstar+D0pi0_NLO1}, and are the masses of $D^{\ast 0}$, $D^{\ast +}$, and $\pi^-$ in the loop integrals $I_{2e}(p_{D^+})$ and $I_{2e}^{(1)}(p_{D^+})$ appearing in Fig.~\ref{Fig.Tccstar+_Dstar+D0pi0_NLO2}.

The NLO partial differential rate for the $T_{cc}^{\ast +}\to D^{\ast 0} D^+ \pi^0$ including the corrections from the $D^*D$ and $D^*\pi$ rescattering is 
\begin{align}
	\frac{d \Gamma_{T_{cc}^{\ast +}\to D^{\ast 0}D^+\pi^0}}{dp^2_{D^+}dp^2_{D^{\ast 0}}}=\frac{1}{3}\frac{m_{\pi}}{16 \pi^3}\sum_{\mathrm{spins}} \vert \mathcal{A}_{a2}+\mathcal{A}_{bc2}\vert^2+\frac{1}{3}\frac{m_{\pi}}{16 \pi^3} 2 \mathrm{Re}\left[\sum_{\mathrm{spins}} \left(\mathcal{A}_{a2}+\mathcal{A}_{bc2}\right) \times \mathcal{A}_{de2}^*\right],
\end{align}
where the second term includes the correction of the $D^{\ast}\pi$ rescattering, which is the interference term between the amplitudes at LO and NLO.

\subsection{Partial decay rate of $T_{cc}^{\ast +}\to D^{\ast 0}D^0\pi^+$}\label{sec:Tccstar_Dstar0D0picAmp}

For the decay $T_{cc}^{\ast +}\to D^{\ast 0}D^0\pi^+$, the LO amplitude from the tree diagram in Fig.~\ref{Fig.Tccstar+_Dstar+D0pi0_LO} reads
\begin{align}
 i\mathcal{A}_{a3}=&\frac{2 g_0 \bar{g}\mu_{D^*}}{\sqrt{2m_{\pi^+}}F_{\pi}} \frac{1}{\vec{p}_{D^{\ast 0}}^2+ \gamma^2-i\mu_{D^*}\Gamma_{D^{*+}}} \varepsilon_{ijk} \epsilon^i\left(T_{cc}^{\ast +}\right) \epsilon^{j*}\left(D^{\ast 0}\right) p^k_{\pi^+},
\end{align}
where $\vec{p}_{\pi^+}$ is the three-momentum of the final state $\pi^+$. 
There are no terms with $I=0$ in the $D^{\ast 0}D^0$ rescattering that can be related to the  $T_{cc}^+$ and thus there is no LO contribution from the $D^{\ast 0}D^0$ rescattering.

The NLO amplitude from the $D^{\ast}\pi$ rescattering diagrams in Figs.~\ref{Fig.Tccstar+_Dstar+D0pi0_NLO1} and \ref{Fig.Tccstar+_Dstar+D0pi0_NLO2} is
\begin{align}
	i\mathcal{A}_{de3}=&\frac{g_0 \bar{g}}{4 m_{\pi}^{3/2} F_{\pi}} \left \{ 
	\sqrt{2}C_{\pi 3} \left[I_{3d}^{(1)}(p_{D^0})-I_{3d}(p_{D^0})\right]+C_{\pi 3\text{ex}} \left[I_{3e}^{(1)}(p_{D^0})+I_{3e}(p_{D^0})\right] \right \}\nonumber \\
	&\times\varepsilon_{ijk} \epsilon^i\left(T_{cc}^{\ast +}\right)\epsilon^{j*}\left(D^{\ast 0}\right)  p_{D^0}^k,
\end{align}
where $C_{\pi3}=\frac{1}{3}C_{\frac{3}{2}\pi}+\frac{2}{3}C_{\frac{1}{2}\pi}=14.4 \, \rm{GeV}^{-1}$ and $C_{\pi 3\text{ex}}=-\frac{\sqrt{2}}{3}C_{\frac{3}{2}\pi}+\frac{\sqrt{2}}{3}C_{\frac{1}{2}\pi}=15.1 \, \rm{GeV}^{-1}$ are the contact interactions for the $D^{*0}\pi^+ \to D^{*0}\pi^+$ and the $D^{*+}\pi^0 \to D^{*0}\pi^+$, respectively, and the masses $m_1$, $m_2$, and $m_3$ are the masses of $D^{\ast +}$, $D^{\ast 0}$, and $\pi^0$ in the loop integrals $I_{3d}(p_{D^0})$ and $I_{3d}^{(1)}(p_{D^0})$ appearing in Fig.~\ref{Fig.Tccstar+_Dstar+D0pi0_NLO1}, and are the masses of $D^{\ast 0}$, $D^{\ast +}$ and $\pi^0$ in the loop integrals $I_{3e}(p_{D^+})$ and $I_{3e}^{(1)}(p_{D^+})$ appearing in Fig.~\ref{Fig.Tccstar+_Dstar+D0pi0_NLO2}.

The NLO partial differential rate for the $T_{cc}^{\ast +} \to D^{\ast 0}D^0\pi^+$ including the corrections from the $D^*D$ and $D^*\pi$ rescattering is 
\begin{align}
	\frac{d \Gamma_{T_{cc}^{\ast +}\to D^{\ast 0}D^0\pi^+}}{dp^2_{D^0}dp^2_{D^{\ast 0}}}=&\frac{1}{3}\frac{m_{\pi}}{16 \pi^3}\sum_{\mathrm{spins}} \vert \mathcal{A}_{a3}\vert^2+\frac{1}{3}\frac{m_{\pi}}{16 \pi^3} 2\mathrm{Re}\left[ \sum_{\mathrm{spins}}\mathcal{A}_{a3} \times \mathcal{A}_{de3}^*\right],
\end{align}
where again the second term includes the correction of the $D^{\ast}\pi$ rescattering, which is the interference term between the amplitudes at LO and NLO.

\section{Partial decay widths for $T_{cc}^{\ast} \to D^{\ast}D\pi$}\label{sec:3BodyResults}

\begin{table}[bt]
\caption{\label{Tab:Tccstar+3Body} Partial decay widths of the $T_{cc}^{\ast +}$ with a binding energy $\mathcal{B}=(503 \pm 40)\, \rm{keV}$. “Tree” contains the contributions from the tree-level diagrams,“LO” is the LO decay width which includes the contributions from the tree-level and $D^{\ast}D$ rescattering diagrams, “NLO” is the decay width which includes the corrections from the $D^{\ast}\pi$ rescattering to the $\Gamma_{\text{LO}}$. 
Since no isovector $D^*D$ rescattering is considered, $\Gamma_\text{LO}=\Gamma_\text{Tree}$ in the last row. The errors come from that of the predicted binding energy $\mathcal{B}$.}
\renewcommand{\arraystretch}{1.2}
\setlength{\tabcolsep}{30pt}{
\begin{tabular*}{\columnwidth}{l|c|c|c}
\hline\hline 
          $\Gamma$ [keV]
        &$\text{Tree}$
        &$\text{LO}$
        &$\text{NLO}$
        \\[3pt]        
\hline
$\Gamma[T_{cc}^{\ast +}\rightarrow D^{\ast +}D^0\pi^0]$ &$12.8 ^{+ 0.6}_{-0.5}$ &$17.4  \pm 0.7$ &$15.3 ^{+0.7}_{-0.6}$
  \\[3pt]
\hline
  $\Gamma[T_{cc}^{\ast +}\rightarrow D^{\ast 0}D^+\pi^0]$ & $8.0 \pm 0.4$ & $9.2 \pm 0.5$ & $8.3 ^{+0.5}_{-0.4}$
  \\[3pt]
  \hline
  $\Gamma[T_{cc}^{\ast +}\rightarrow D^{\ast 0}D^0\pi^+]$ & $18.2 \pm 0.9$ & $18.2 \pm 0.9$ & $17.6 \pm 0.9$ 
  \\[3pt]
\hline\hline
\end{tabular*}}
\end{table}
\begin{figure}[tbh]
    \subfigure[$T_{cc}^{\ast +}\to D^{\ast +}D^0\pi^0$] {\includegraphics[scale=0.295]{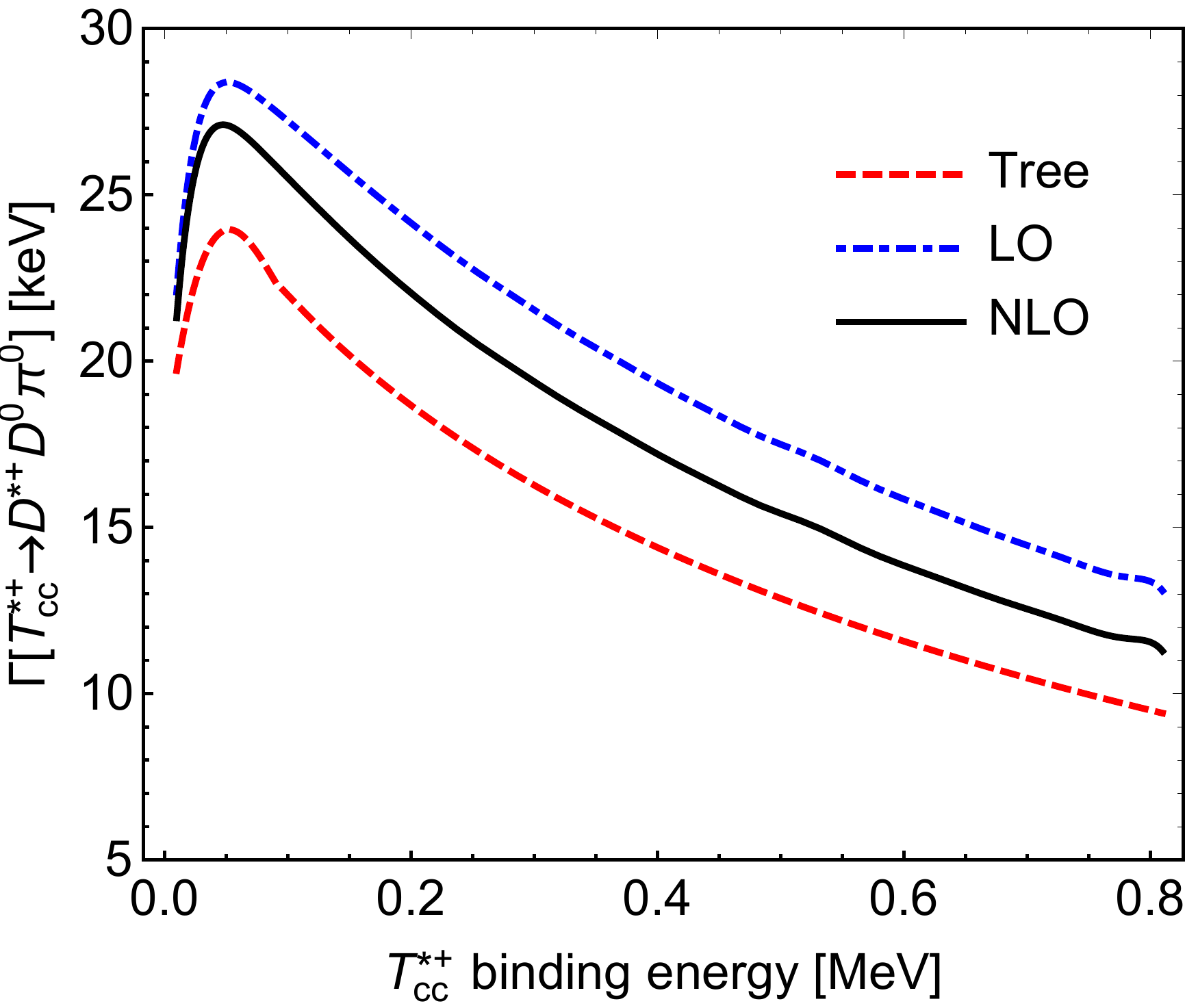} \label{Fig.Tccstar_Dstar+D0pi0_decay_width_LO}
    }
    \subfigure[$ T_{cc}^{\ast +}\to D^{\ast 0}D^+\pi^0$] {\includegraphics[scale=0.295]{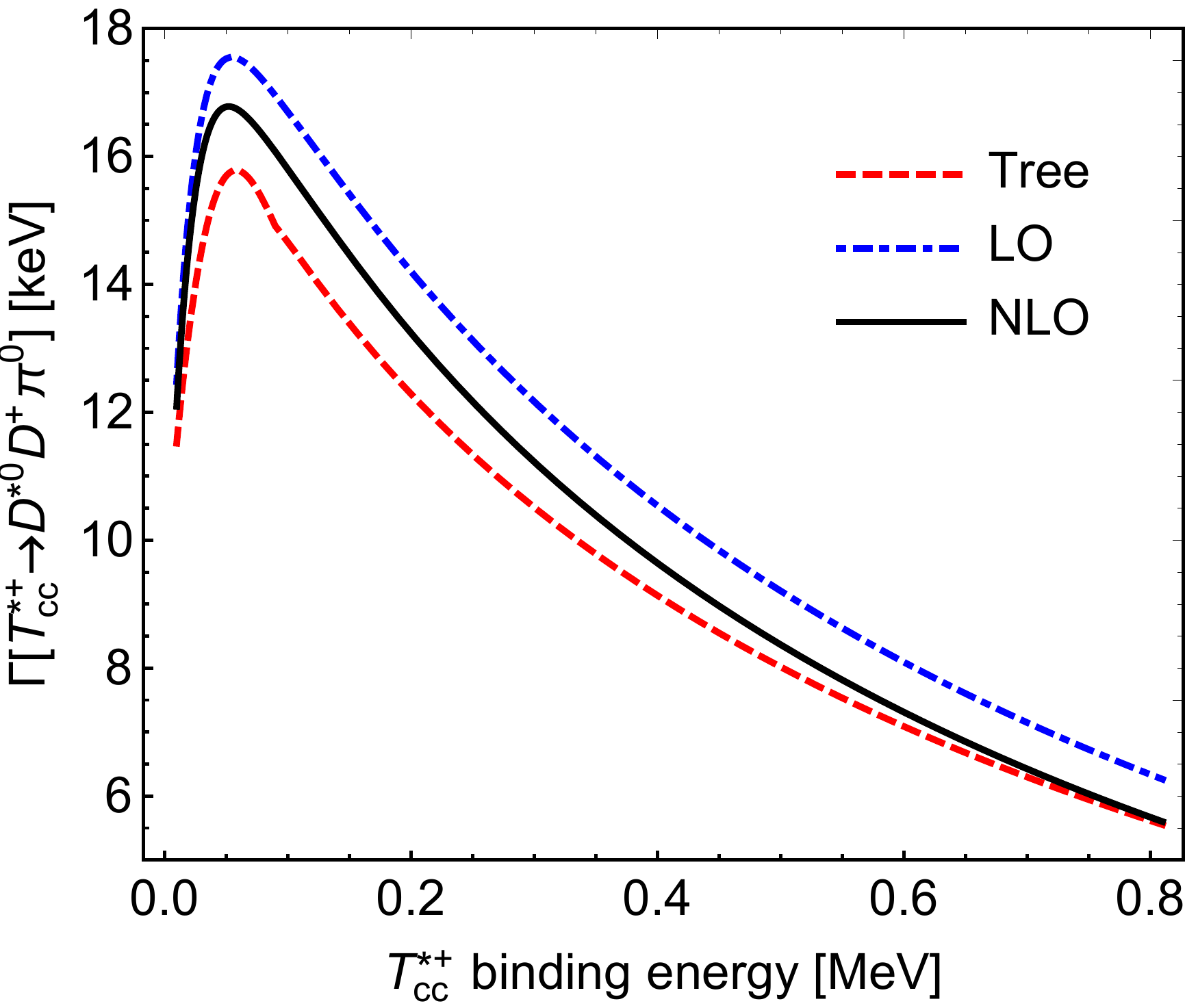} \label{Fig.Tccstar_Dstar0D+pi0_decay_width_LO}
    }
    \subfigure[$T_{cc}^{\ast 0}\to D^{\ast 0}D^0\pi^+$] {
    \includegraphics[scale=0.295]{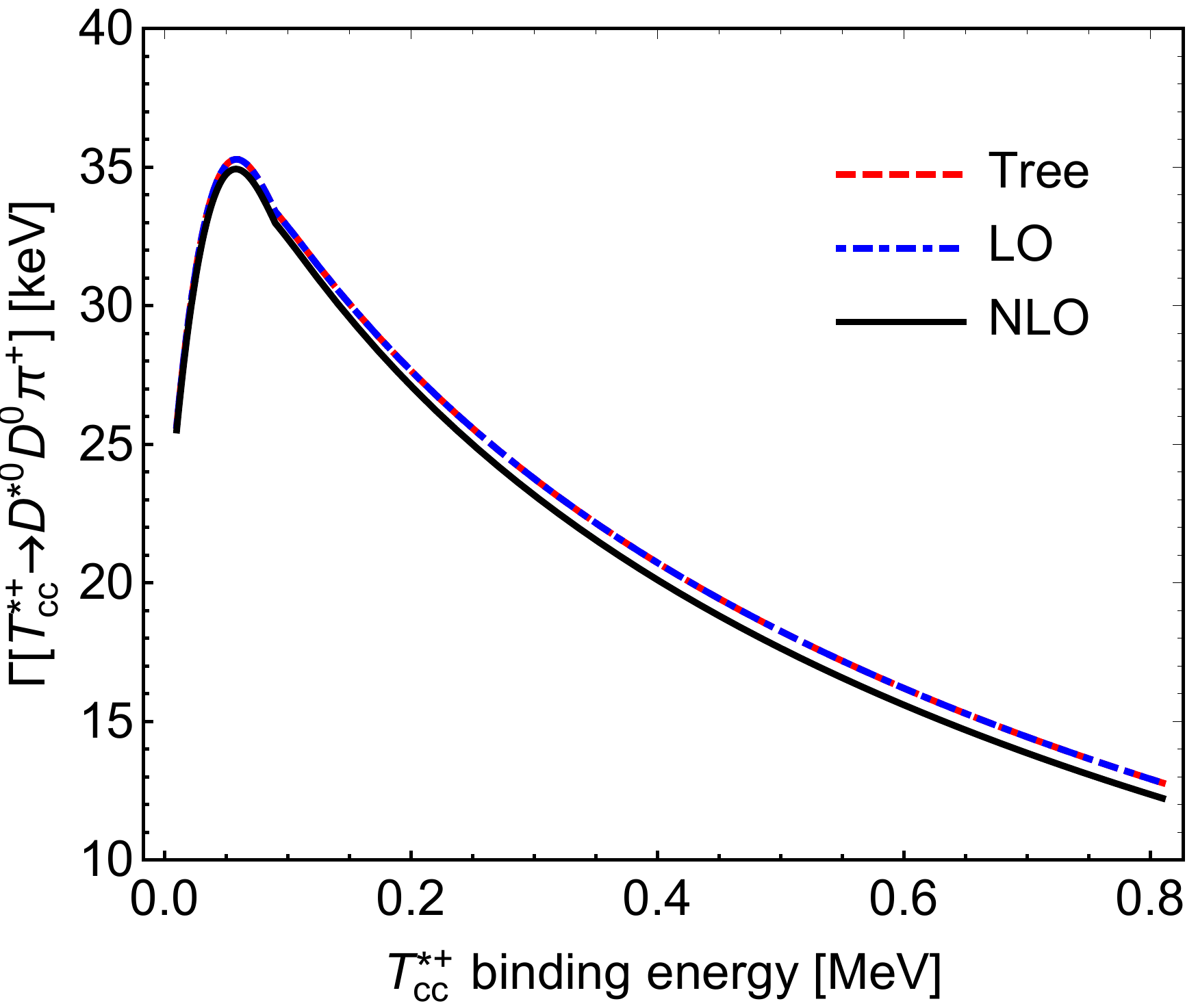} \label{Fig.Tccstar_Dstar0D0pi+_decay_width_LO}
    }
   \caption{Partial decay widths of the $T_{cc}^{\ast }\rightarrow D^{\ast} D \pi$ versus the binding energy of the $T_{cc}^{\ast +}$.}
    \label{Fig.Tccstar_Dstar+D0pi0 decay width}
\end{figure}
In this section, we give the partial decay widths for the decays $T_{cc}^{\ast} \to D^{\ast}D\pi$.
Table~\ref{Tab:Tccstar+3Body} shows the decay widths with the binding energy of the $T_{cc}^{*}$ being $\mathcal{B}=(503 \pm 40) \, \rm{keV}$. The second column of Table~\ref{Tab:Tccstar+3Body} is the decay width only including the contribution from the tree-level diagram denoted by $\Gamma_{\text{Tree}}$. 
The third column is the LO decay width including the tree-level and the $D^{\ast}D$ rescattering contributions marked by $\Gamma_{\rm{LO}}$. One sees that the isoscalar $D^{\ast}D$ rescattering which contains a $T_{cc}$ pole indeed increases the results by about $36\%$ and $15\%$ for $T_{cc}^{*+} \rightarrow D^{*+}D^0\pi^0$ and $D^{*0} D^+ \pi^0$. The $D^{\ast 0}D^0$ is an isovector system; since no near-threshold isovector double-charm tetraquark state has been found, the $D^{\ast 0}D^0$ rescattering for $T_{cc}^{*+} \rightarrow D^{*0}D^0\pi^+$ remains an NLO effect, thus $\Gamma_{\rm{LO}}=\Gamma_{\text{Tree}}$ in the last row of Table~\ref{Tab:Tccstar+3Body}.
The fourth column of Table~\ref{Tab:Tccstar+3Body} is the decay width considering the NLO corrections only from the $D^{\ast} \pi$ rescattering represented by $\Gamma_{\rm{NLO}}$, which should be regarded as the final predictions in this work. The $D^{\ast} \pi$ rescattering reduces the LO decay widths by about $12\%$, $9\%$, and $3\%$ for the three decays $T_{cc}^{\ast +} \to D^{\ast +} D^0 \pi^0$, $D^{\ast 0}D^+\pi^0$, and $D^{\ast 0}D^0\pi^+$, respectively.

\begin{figure}[tb]
    \subfigure[$T_{cc}^{\ast +}\to D^{\ast +}D^0\pi^0$] {\includegraphics[scale=0.295]{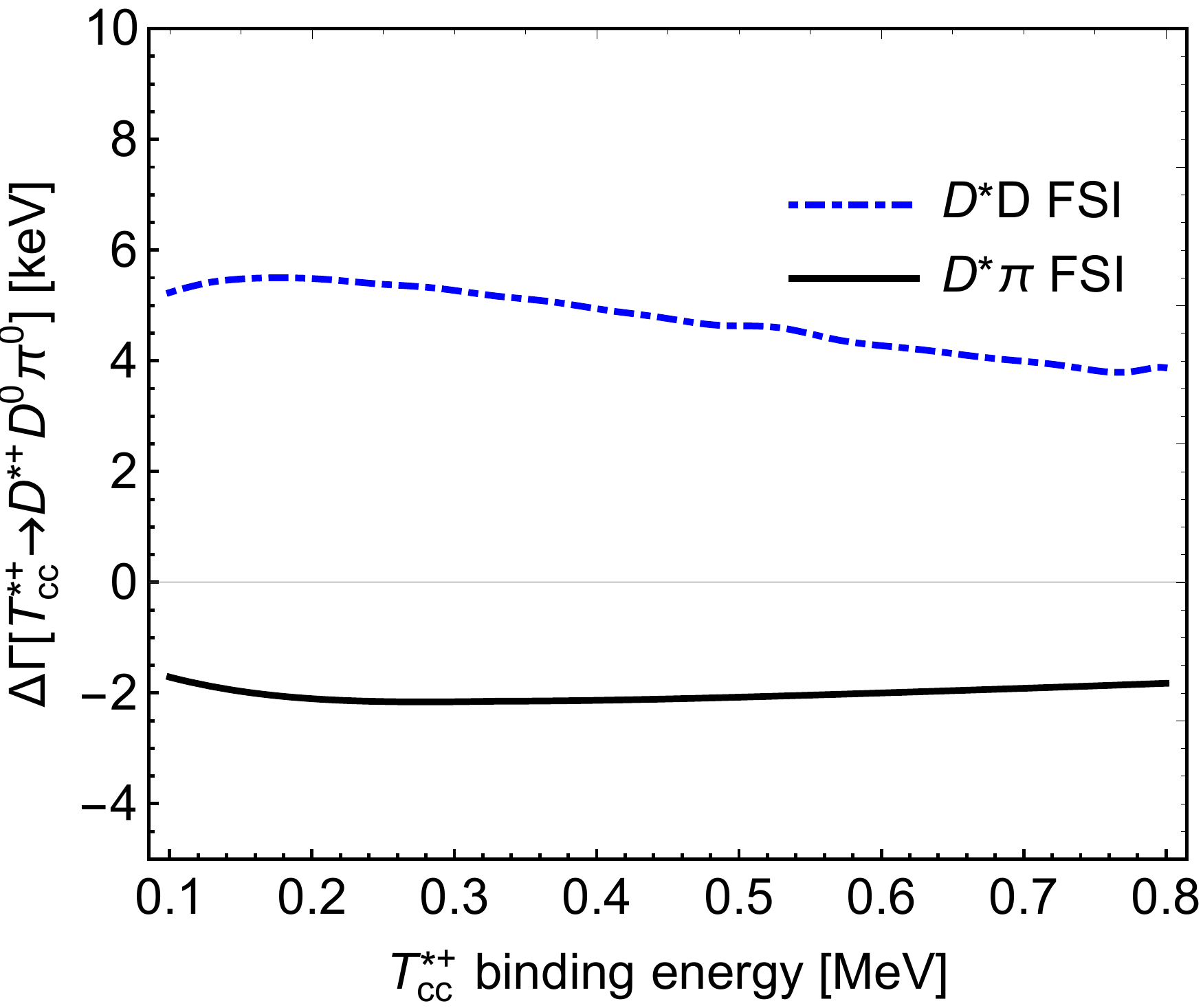} \label{Fig.FSI_Dstar+D0pi0}
    }
    \subfigure[$T_{cc}^{\ast +}\to D^{\ast 0}D^+\pi^0$] {\includegraphics[scale=0.295]{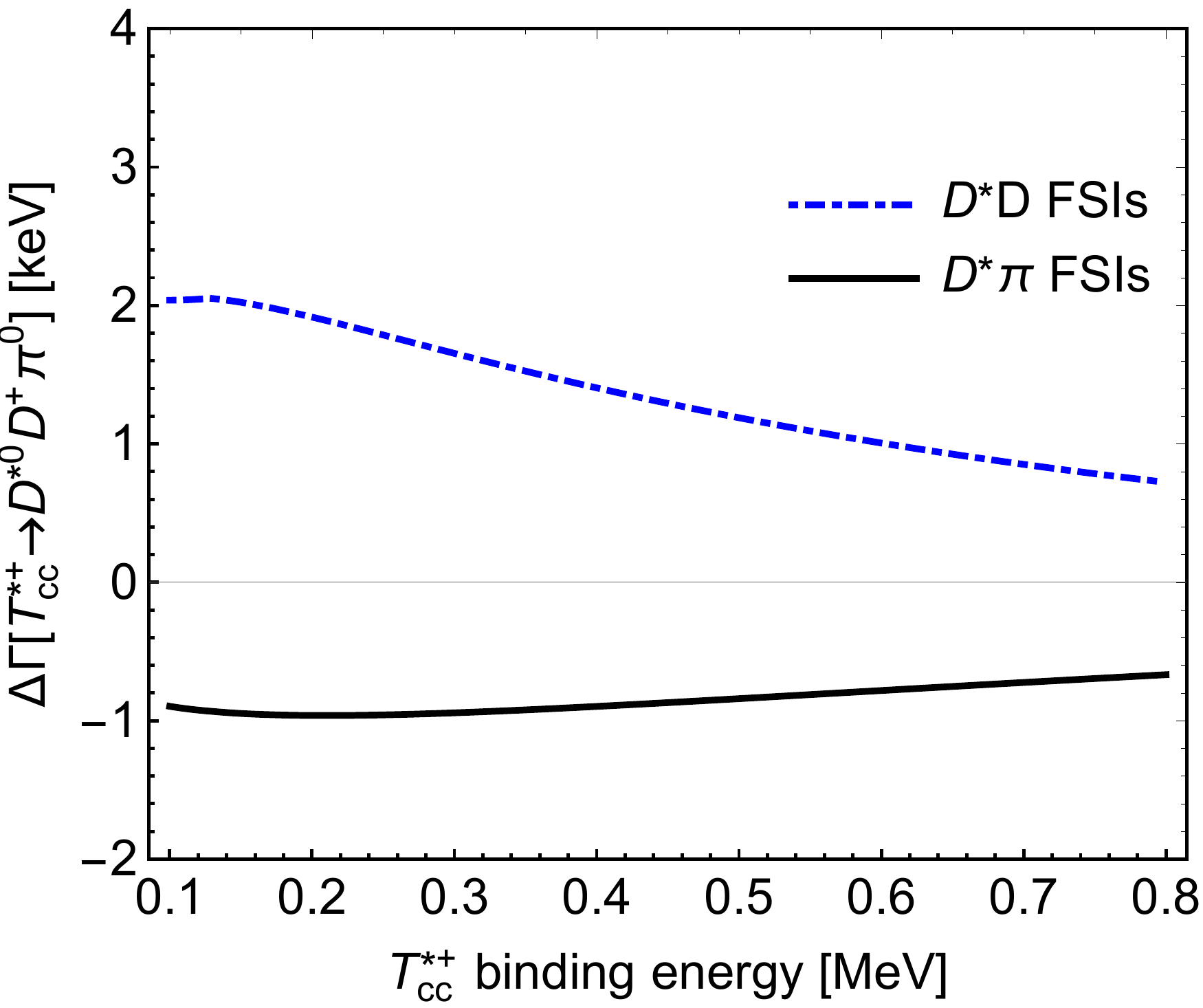} \label{Fig.FSI_Dstar0D+pi0}
    }
    \subfigure[$T_{cc}^{\ast 0}\to D^{\ast 0}D^0\pi^+$] {
    \includegraphics[scale=0.295]{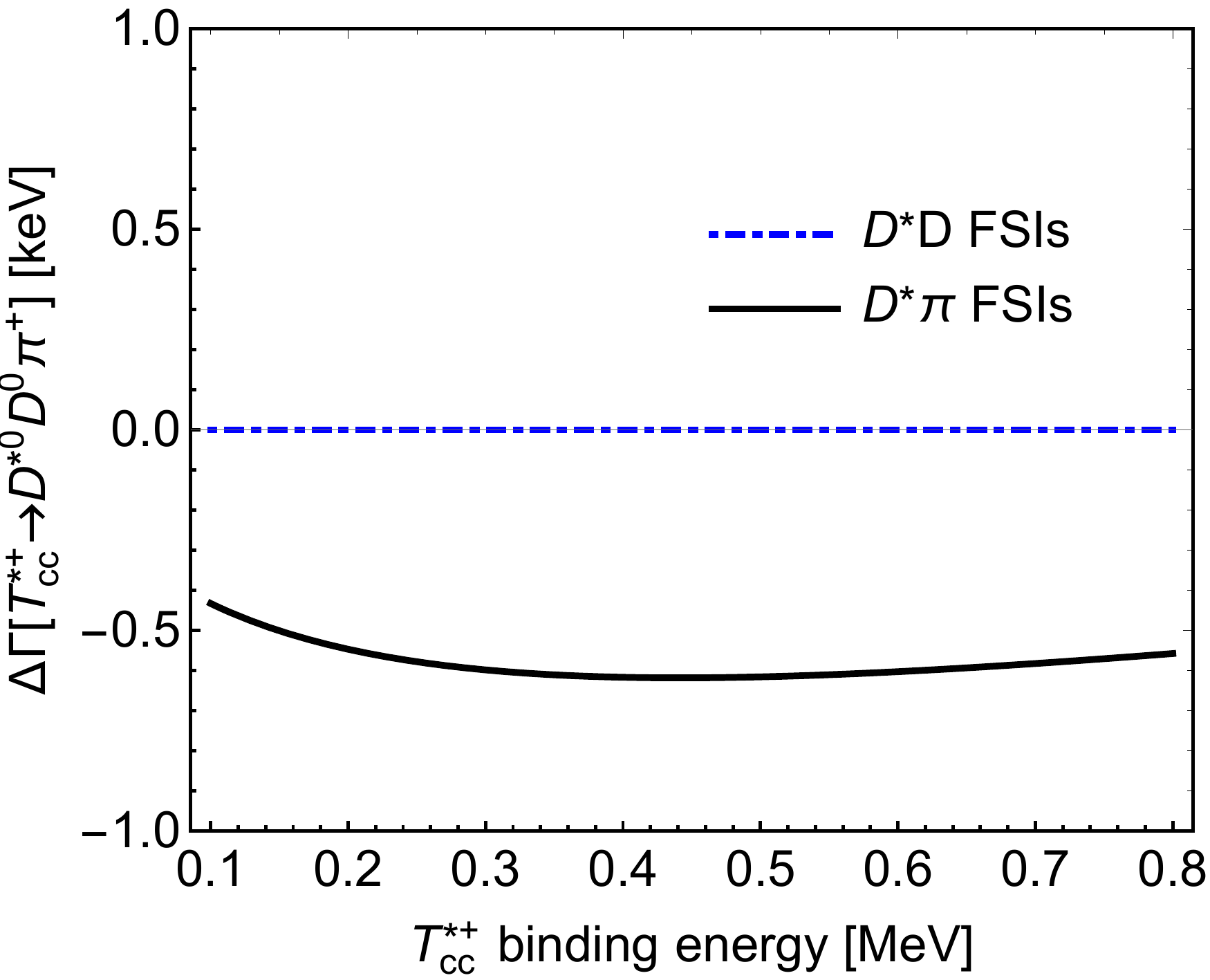} \label{Fig.FSI_Dstar0D0pi+}
    }
   \caption{Corrections from the $D^{\ast}D$ and $D^{\ast}\pi$ FSIs to the LO partial decay widths of $T_{cc}^{\ast }\rightarrow D^{\ast} D \pi$ versus the binding energy of $T_{cc}^{\ast +}$.}
    \label{Fig.FSI_DstarDpi decay width}
\end{figure}

Since the binding energy of the $T_{cc}^{\ast +}$ is uncertain, we further give the partial width of $T_{cc}^{\ast} \to D^{\ast}D\pi$ with the binding energy varying from $0.01~\rm{MeV}$ to $0.80~\rm{MeV}$ in Fig.~\ref{Fig.Tccstar_Dstar+D0pi0 decay width},
where the red dashed lines show the decay widths from the tree-level diagram, the blue dot-dashed lines show the LO decay width including the tree-level and the $D^{\ast}D$ rescattering contributions, and the black solid lines show the decay width including the corrections from the $D^{\ast}\pi$ rescattering to the LO results. To see the contributions of the $D^{\ast}D$ and $D^{\ast}\pi$ rescattering to the decay widths more clearly, the corrections from the $D^{\ast}D$ and $D^{\ast}\pi$ FSIs with the binding energy being $0.01\sim0.80~\rm{MeV}$ are demonstrated by the blue dot-dashed lines and black solid lines in Fig.~\ref{Fig.FSI_DstarDpi decay width}, respectively. 

Summing up these three-body partial decay widths leads to the result for the total width of the $T_{cc}^{*+}$ to be 
\begin{equation}
    \Gamma(T_{cc}^{*+}) = (41 \pm 2)~{\rm keV}. \label{eq:total}
\end{equation}

\section{Partial decay widths for $T_{cc}^{*} \to DD\pi\pi$}\label{sec:4BodyResults}

\begin{figure}[tbh]
    \subfigure[] {
        \includegraphics[scale=0.5]{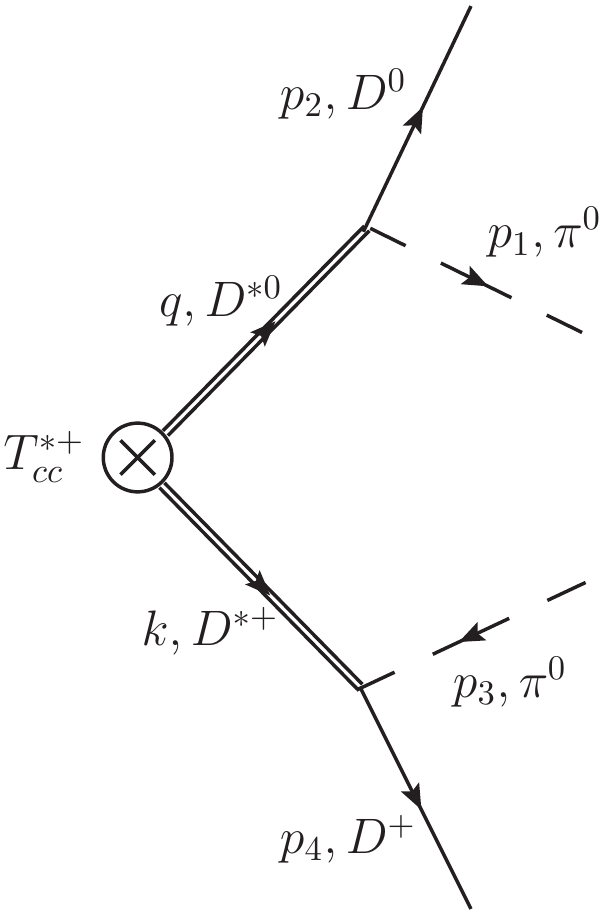}\label{Fig.4bodypi0Tree}
    }
    \subfigure[] {
        \includegraphics[scale=0.5]{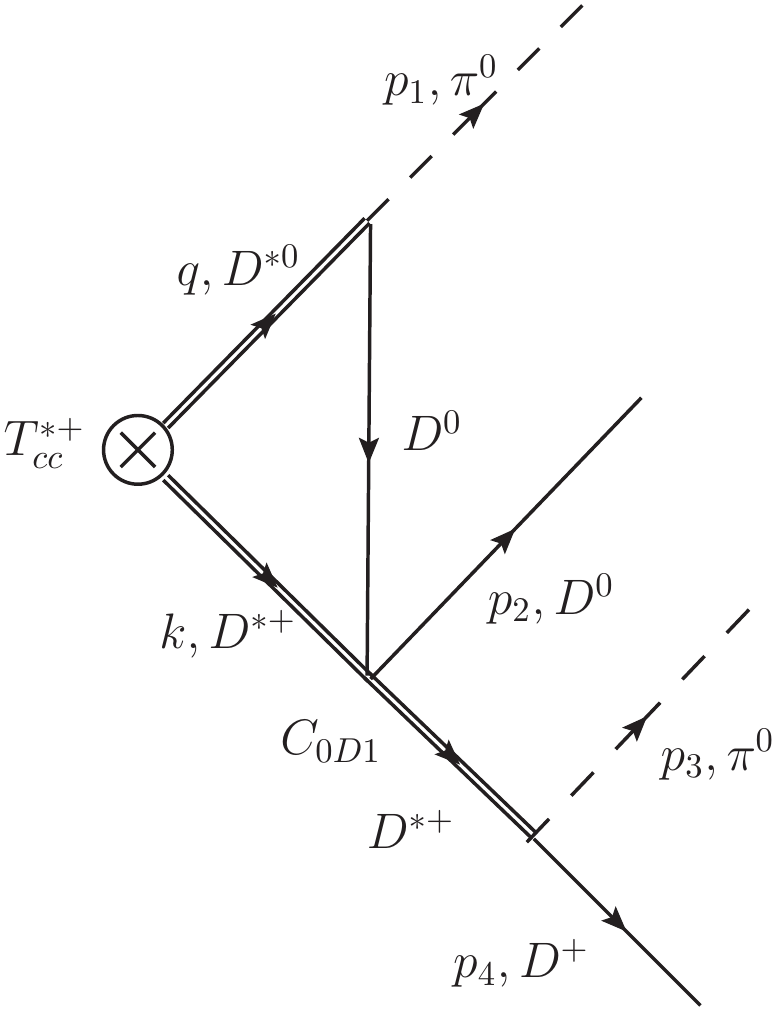}\label{Fig.4bodypi0C0D1}
    }
    \subfigure[] {
        \includegraphics[scale=0.5]{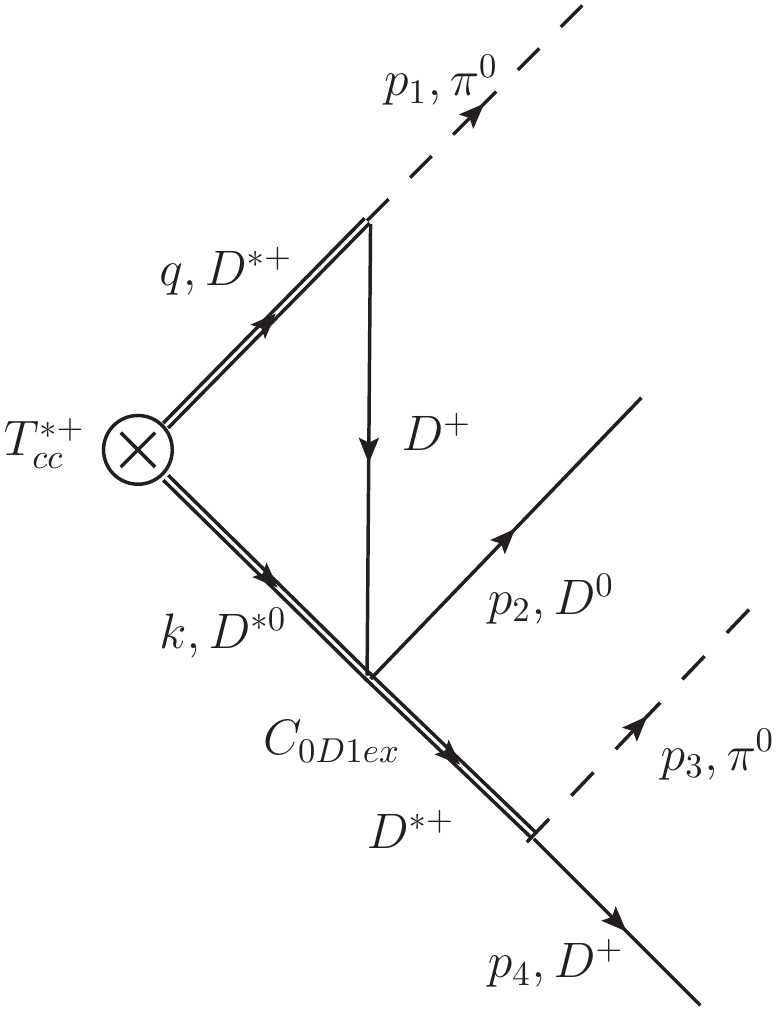}\label{Fig.4bodypi0C0D1ex}
    }\\
    \subfigure[] {
        \includegraphics[scale=0.5]{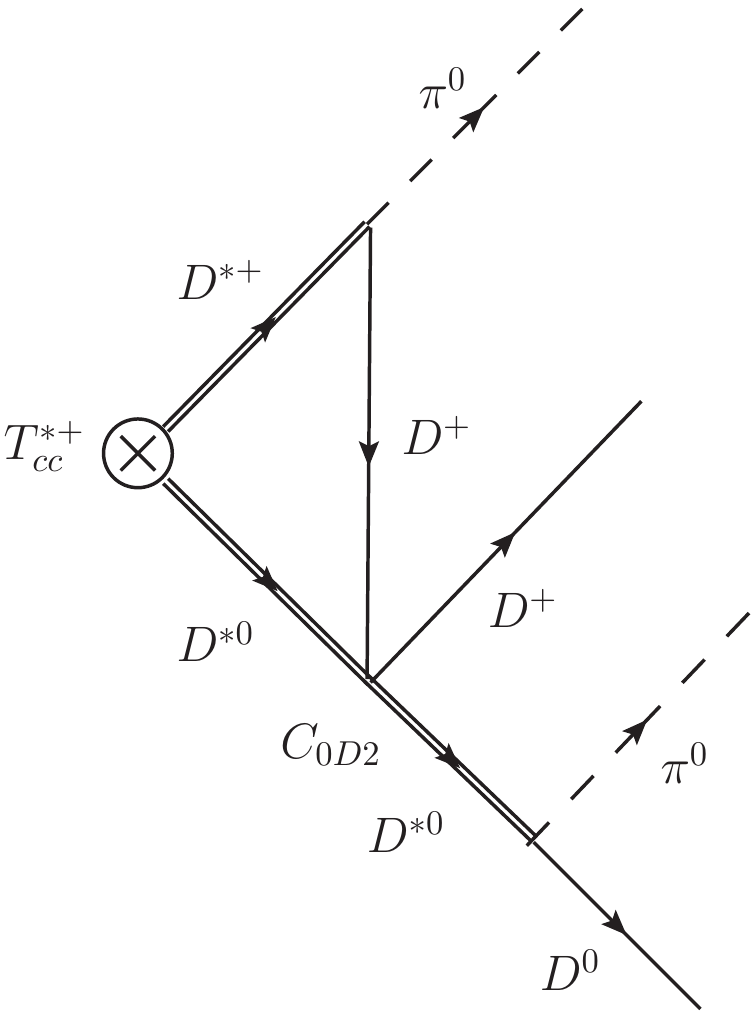}\label{Fig.4bodypi0C0D2}
    }
    \subfigure[] {
        \includegraphics[scale=0.5]{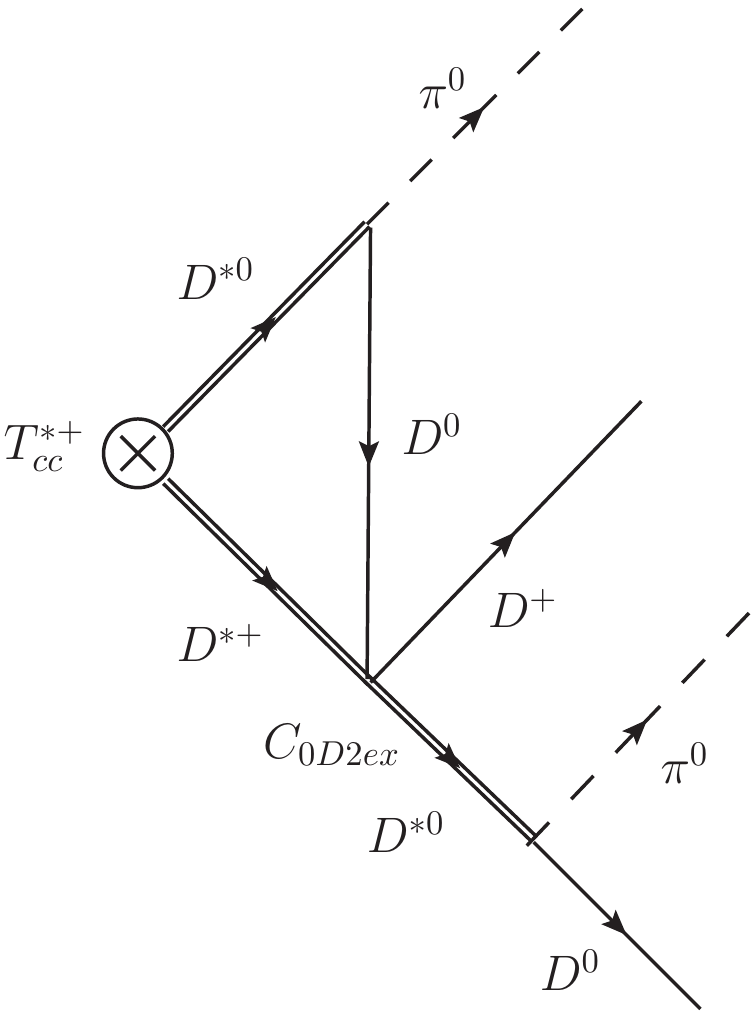}\label{Fig.4bodypi0C0D2ex}
    }
    \subfigure[] {
        \includegraphics[scale=0.5]{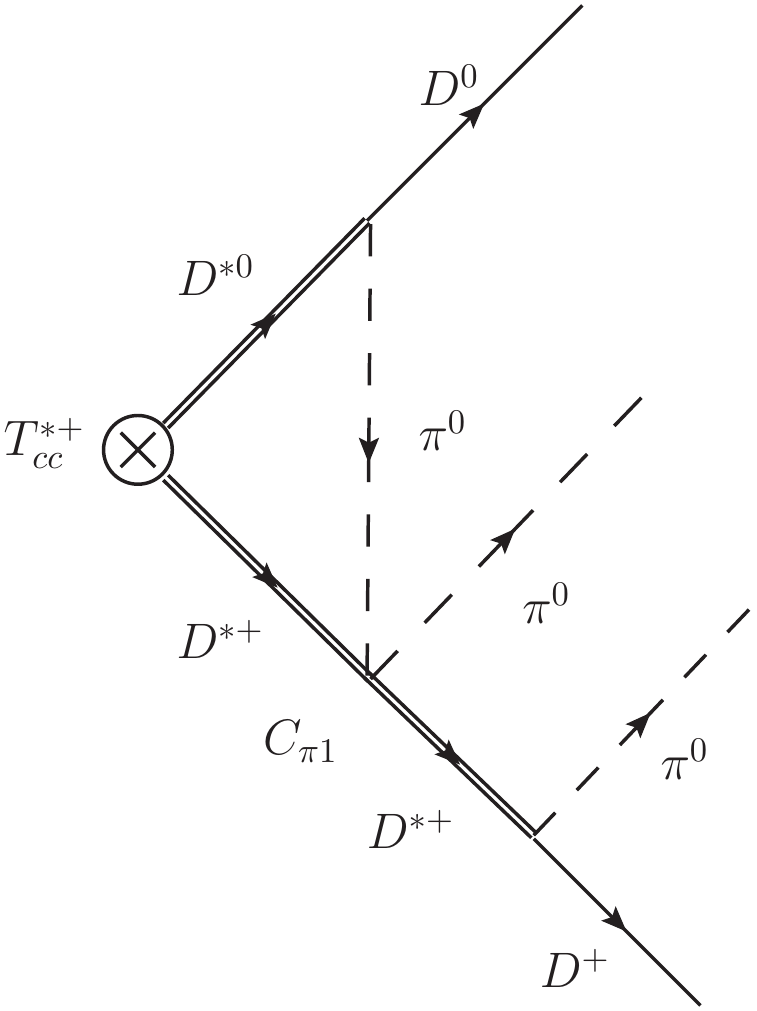}\label{Fig.4bodypi0Cpi1}
    }\\
     \subfigure[] {
        \includegraphics[scale=0.5]{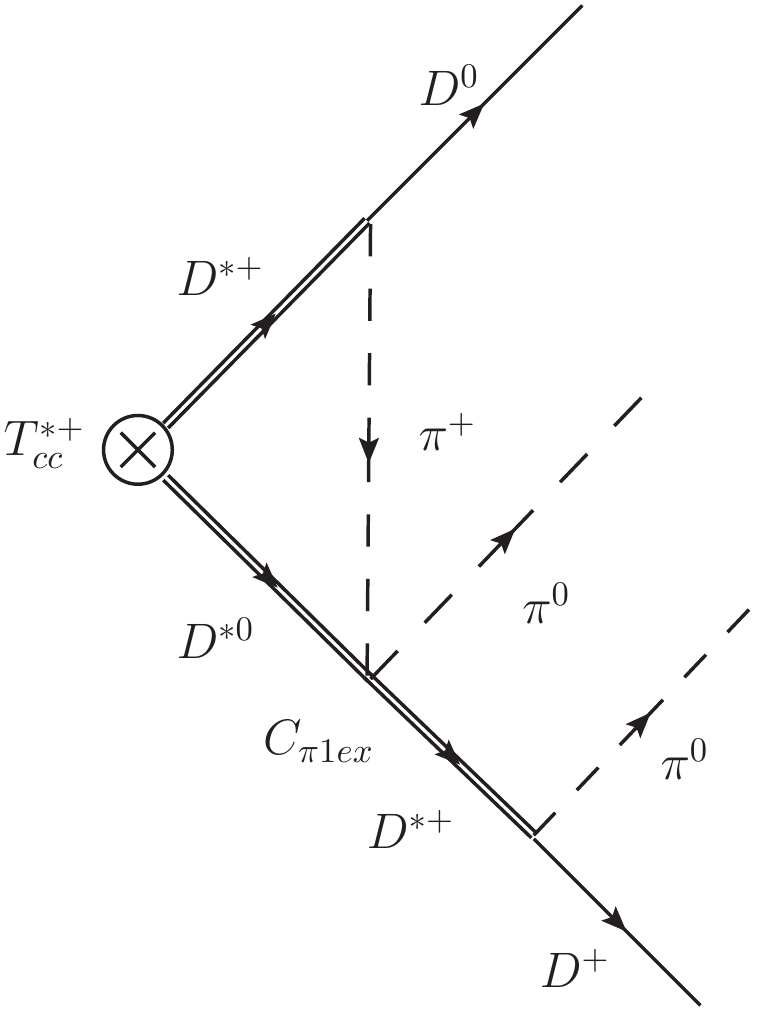}\label{Fig.4bodypi0Cpi1ex}
    }
     \subfigure[] {
        \includegraphics[scale=0.5]{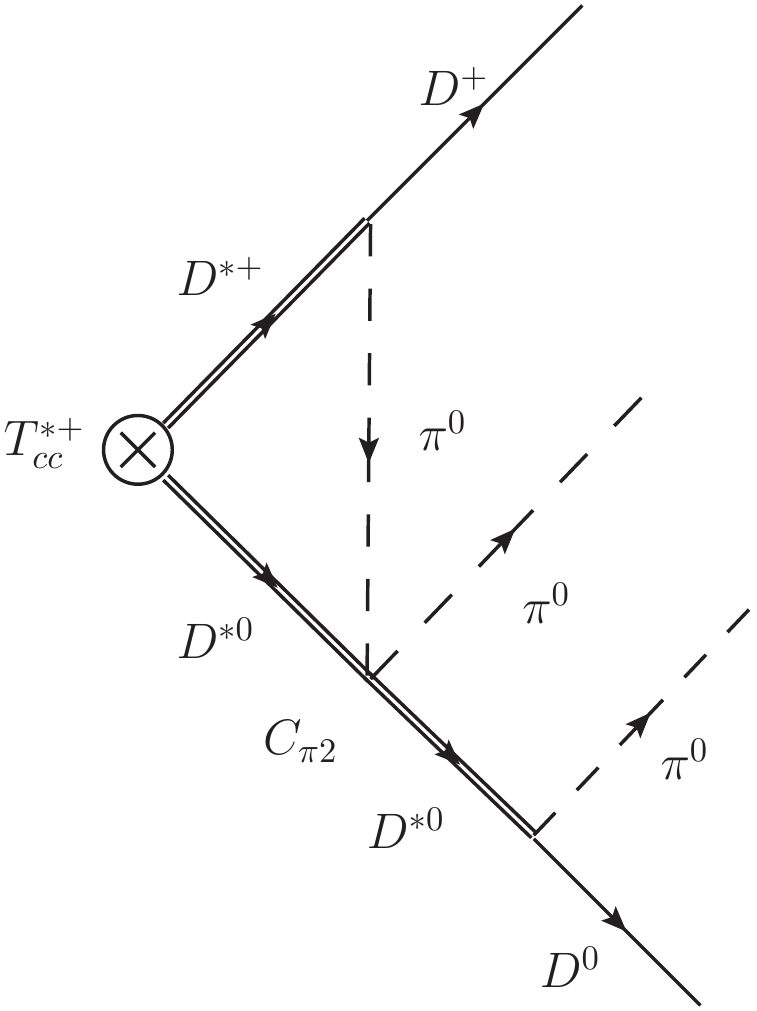}\label{Fig.4bodypi0Cpi2}
    }
     \subfigure[] {
        \includegraphics[scale=0.5]{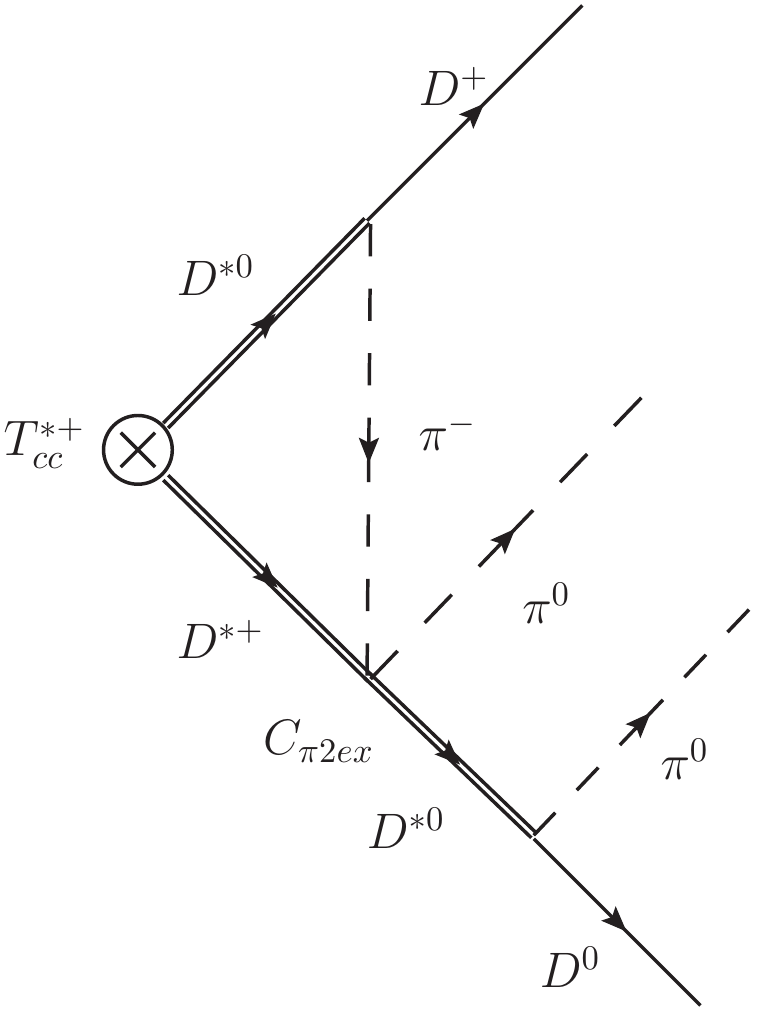}\label{Fig.4bodypi0Cpi2ex}
    }    
    \caption{Feynman diagrams for calculating the partial decay width of $T_{cc}^{*+}\to D^{*}D\pi\rightarrow D^0 D^+\pi^0 \pi^0 $. The circled cross is the $T_{cc}^{*+}$ state, the single thin lines represent the $D^+(D^0)$, the double lines represent the $D^{*0}(D^{*+})$, and the dashed lines represent the $\pi^{0}(\pi^{+})$.}
    \label{Fig.4bodypi0}
\end{figure}

\begin{figure}[tb]
    \subfigure[] {
        \includegraphics[scale=0.5]{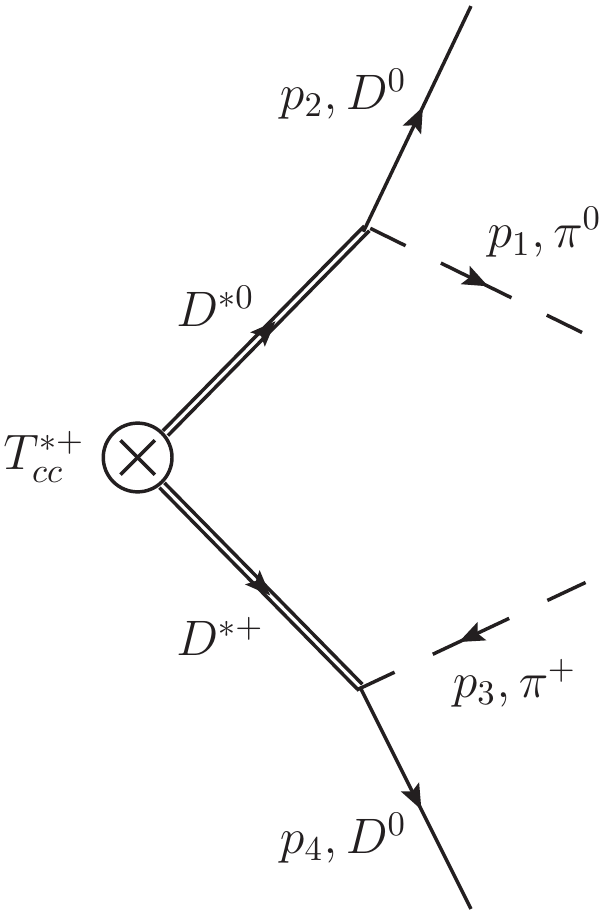}\label{Fig.4bodypi+Tree}
    }
    \subfigure[] {
        \includegraphics[scale=0.5]{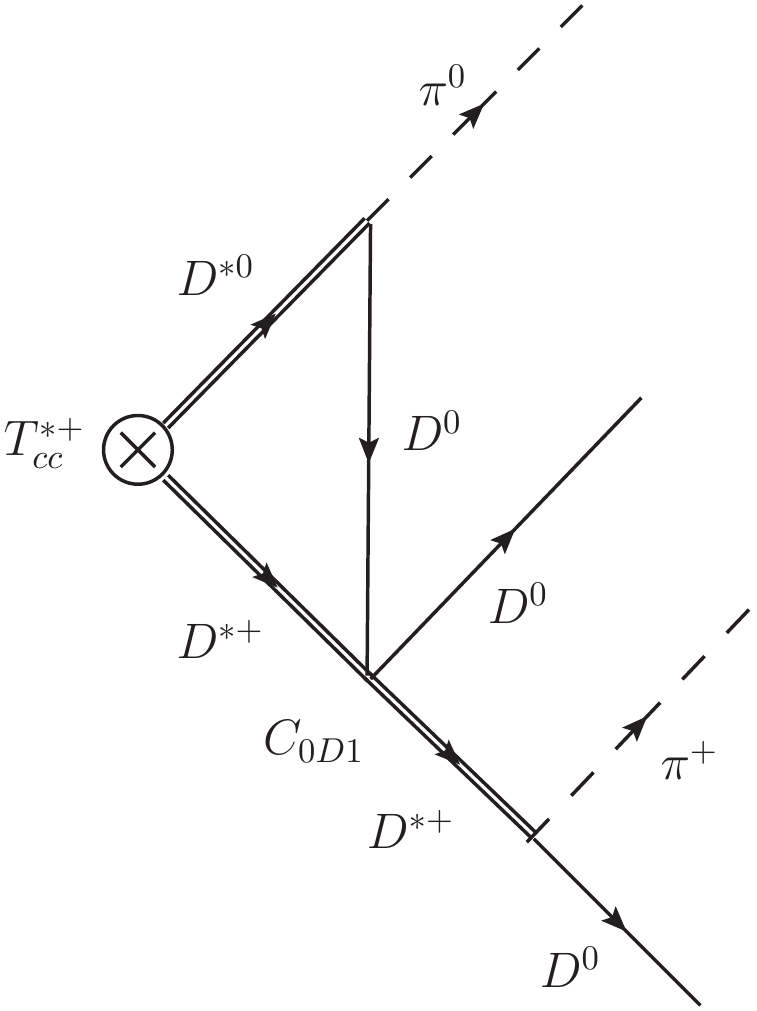}\label{Fig.4bodypi+C0D1}
    }
    \subfigure[] {
        \includegraphics[scale=0.5]{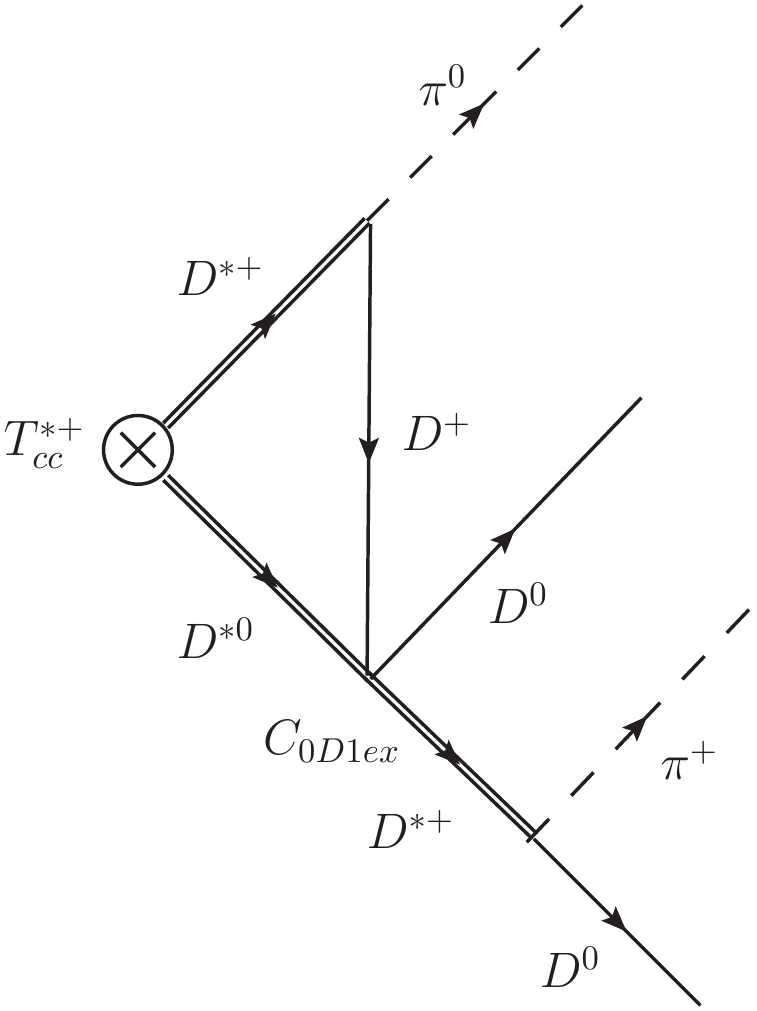}\label{Fig.4bodypi+C0D1ex}
    }\\
    \subfigure[] {
        \includegraphics[scale=0.5]{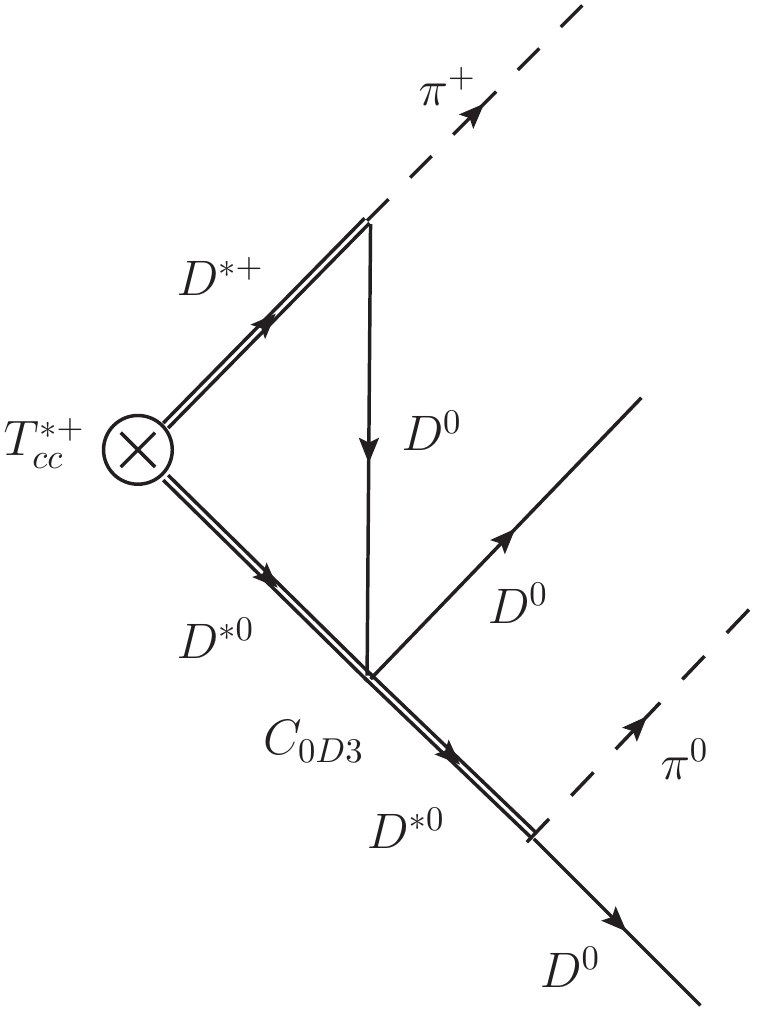}\label{Fig.4bodypi+C0D3}
    }
    \subfigure[] {
        \includegraphics[scale=0.5]{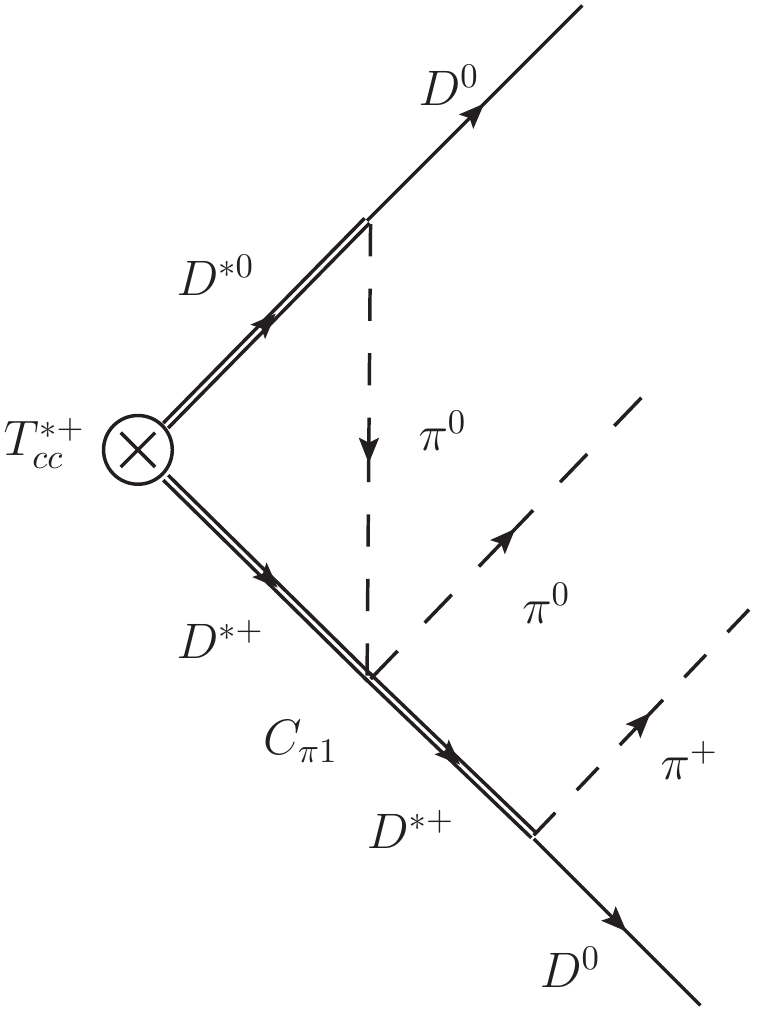}\label{Fig.4bodypi+Cpi1}
    }
    \subfigure[] {
        \includegraphics[scale=0.5]{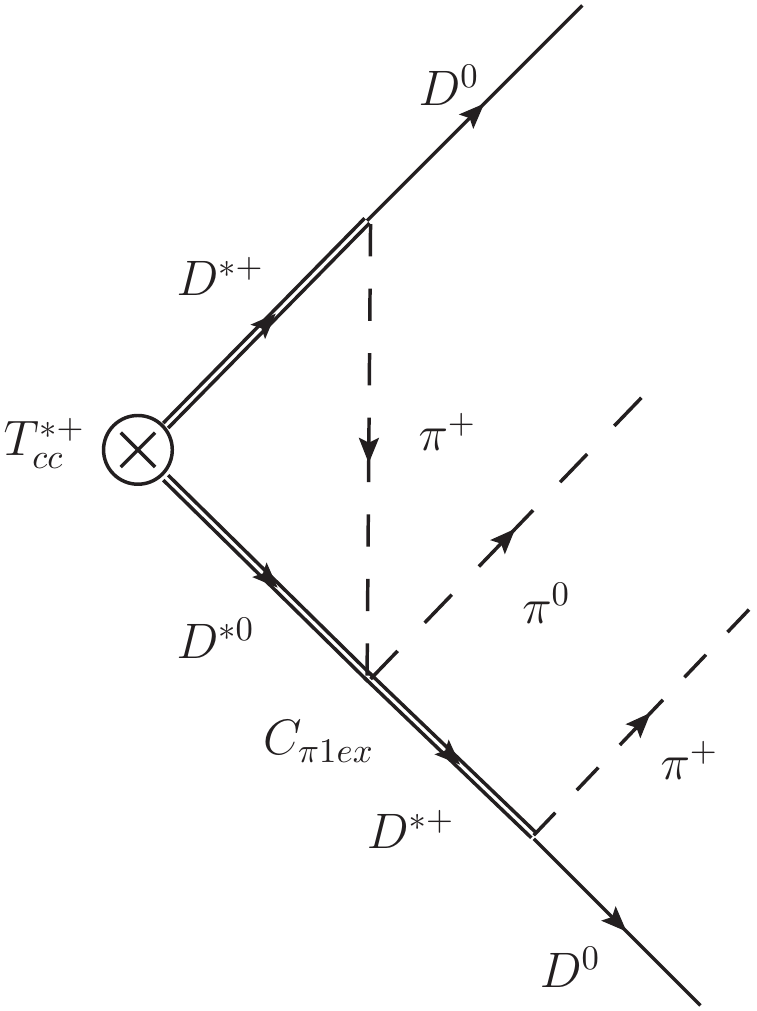}\label{Fig.4bodypi+Cpi1ex}
    }\\
    \subfigure[] {
        \includegraphics[scale=0.5]{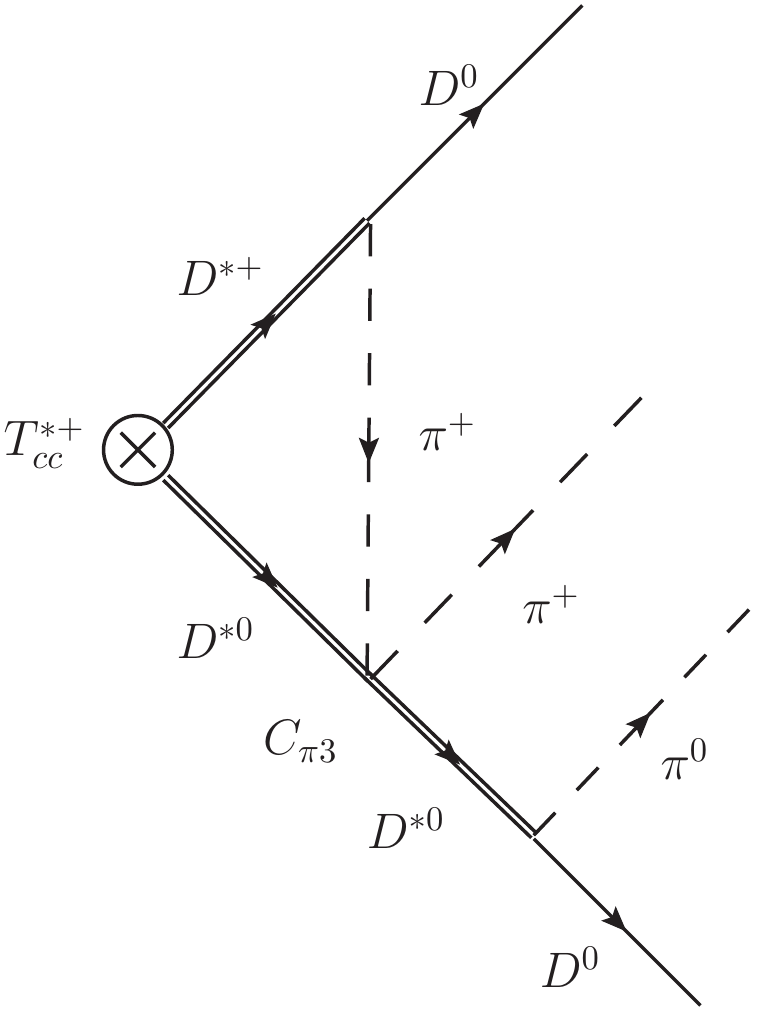}\label{Fig.4bodypi+Cpi3}
    }
    \subfigure[] {
        \includegraphics[scale=0.5]{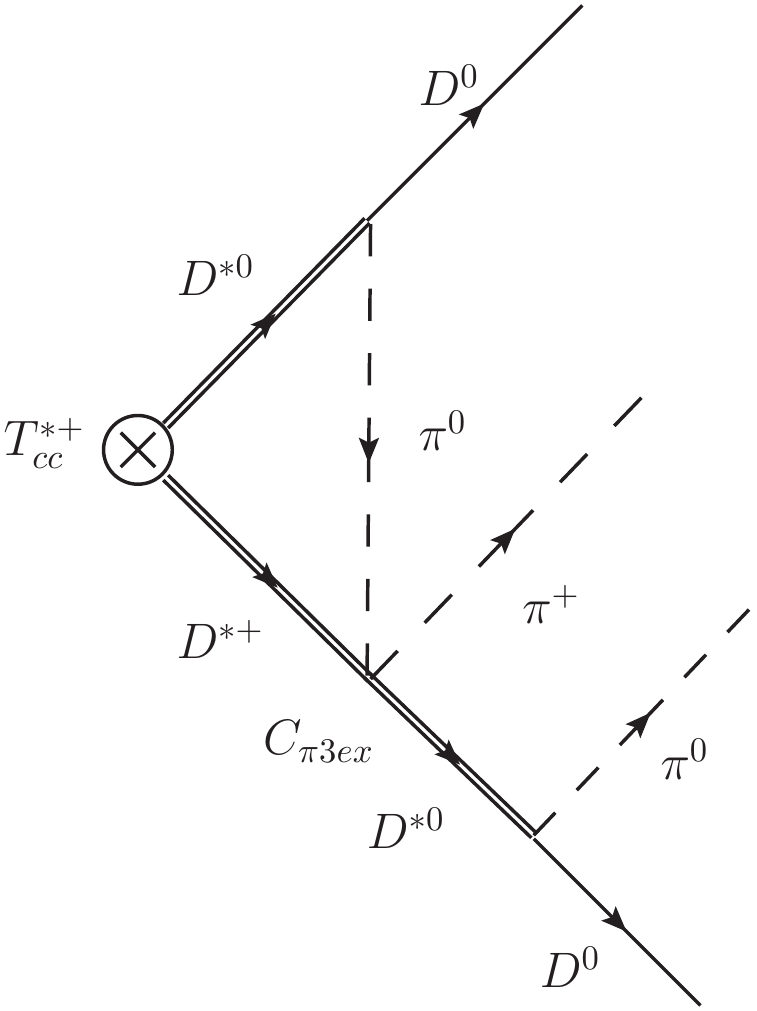}\label{Fig.4bodypi+Cpi3ex}
    }
    \caption{Feynman diagrams for calculating the partial decay width of $T_{cc}^{*+}\rightarrow D^0 D^0\pi^0 \pi^+$. The circled cross is the $T_{cc}^{*+}$ state, the single thin lines represent the $D^+(D^0)$, the double lines represent the $D^{*0}(D^{*+})$, and the dashed lines represent the $\pi^{0}(\pi^{+})$.}
    \label{Fig.4bodypic}
\end{figure}
\afterpage{\clearpage}

Since the $D^*$ mesons are resonances, they need to be reconstructed through the $D\pi$ or $D\gamma$ final state in experimental analysis.
In the former case, the $D^*$ will continue to decay into the $D\pi$, and the stable (against decays through strong and electromagnetic interactions) final states that $T_{cc}^{*}$ decays into are the $D^0 D^+\pi^0 \pi^0 $ and $D^0 D^0\pi^0 \pi^+$. Since the $D^{*+}D^0\pi^0$ can decay into the same four-body final states as $D^{*0}D^+\pi^0$ and $D^{*0}D^0\pi^+$, the three processes $T_{cc}^{*+} \to D^{*+}D^0\pi^0$, $D^{*0}D^+\pi^0$, and $D^{*0}D^0\pi^+$ can interfere.  Therefore, in the following, we will calculate the decay widths of the $T_{cc}^{*+} \to D^0 D^+\pi^0 \pi^0 $ and $D^0 D^0\pi^0 \pi^+$. We will show that the interference between the intermediate three-body states is small, and it is a good approximation that we consider only the 3-body $D^{*}D\pi$ final states to calculate the $T_{cc}^{*}$ decay width.

The diagrams for the four-body decays  $T_{cc}^{*+} \to D^0 D^+\pi^0 \pi^0 $ and $T_{cc}^{*+}\to D^0 D^0\pi^0 \pi^+$ are shown in Fig.~\ref{Fig.4bodypi0} and Fig.~\ref{Fig.4bodypic}, respectively.
The amplitudes for all the diagrams are collected in Appendix~\ref{Appendix:4body_amplitudes}. 

With all the amplitudes, the decay rate for the four-body decay $T_{cc}^{*}\to DD\pi\pi$ is given by
\begin{align}
d\Gamma[T_{cc}^{*} \to DD\pi\pi]=2M 2E_1 2E_2 2E_3 2E_4\frac{1}{2SM}\frac{1}{2j+1} \sum_{\text{spins}} \left\vert \mathcal{A}[T_{cc}^{*} \to DD\pi\pi] \right\vert^2 d\Phi_4,
\label{Eq.4-body_decay_rate}
\end{align}
where the overall factor comes from the normalization of nonrelativistic particles, and $E_i\, (i=1,\ldots,4)$ are the energies of four finial-state particles in the $T_{cc}^{*}$ rest frame, respectively. The symmetry factor $S=2$ comes from the identical  $\pi^0\pi^0$ or $D^0D^0$ particles in the final states.
The four-body phase space in Eq.~\eqref{Eq.4-body_decay_rate} derived in Appendix~\ref{sec:Four-body phase space} reads
\begin{align}
d\Phi_4(P;p_1,...,p_4)= \frac{1}{(8\pi^2)^4M} \int_{m_1+m_2}^{M-m_3-m_4}d\sqrt{s_{12}}\int_{m_3+m_4}^{M-\sqrt{s_{12}}}d\sqrt{s_{34}} \int d\Omega_1^*d\Omega_3^{\prime}d\Omega \vert \vec{p}_1^{\,*} \vert \vert \vec{p}_3^{\, \prime} \vert \vert \vec{q} \, \vert,
\label{Eq.Phi4}
\end{align}
where $\vec{q}$ is the three-momentum for the 1,2 system in the rest frame of the $T_{cc}^{*}$, $\vec{p}_1^{\,*}$ is the three-momentum of particle 1 in the c.m. frame of particles 1 and 2, and $\vec{p}_3^{\,\prime}$ is the three-momentum of particle 3 in the c.m. frame of particles 3 and 4. They are given by
\begin{align}
  \vert \vec{q} \, \vert= \frac{\lambda^{1/2}(M^2,s_{12},s_{34})}{2M}, \quad
  \vert \vec{p}_1^{\,*} \vert= \frac{\lambda^{1/2}(s_{12}, m_1^2, m_2^2)}{2\sqrt{s_{12}}}, \quad
  \vert \vec{p}_3^{\,\prime} \vert=&\, \frac{\lambda^{1/2}(s_{34}, m_3^2, m_4^2)}{2\sqrt{s_{34}}},
\end{align}
with $\lambda(x,y,z)\equiv x^2+y^2+z^2-2(xy+xz+yz)$. $d\Omega_1^*=d\varphi_1^*d\cos\theta_1^*$ is the solid angle of particle 1 in the c.m. frame of particles 1 and 2, $d\Omega_3^{\prime}=d\varphi_3^{\prime}d\cos\theta_3^{\prime}$ is the solid angle of particle 3 in the c.m. frame of particles 3 and 4, and $d\Omega$ is the solid angle of the 1, 2 system in the rest frame of the decay particle $T_{cc}^*$; $\theta_1^*$ is the angle between the directions of $\vec{q}$ and $\vec{p}_1^{\,*}$, and $\theta_3^{\prime}$ is the angle between $\vec{k}=-\vec{q}$ and $\vec{p}_3^{\,\prime}$.

The differential decay rate for the $T_{cc}^{*+} \to DD\pi\pi$ up to NLO including the $D^*D$ and $D^*\pi$ rescattering corrections reads
\begin{align}
\frac{d\Gamma[T_{cc}^{*} \to DD\pi\pi]}{d\sqrt{s_{12}}d\sqrt{s_{34}}}=&\, 2m_{T_{cc}^*}2 p_1^0 2p_2^0 2p_3^0 2p_4^0 \frac{1}{4m_{T_{cc}^*}} \frac{1}{3}  \frac{1}{(8\pi^2)^4 m_{T_{cc}^*}} d\Omega_1^*d\Omega_3^{\prime}d\Omega \vert \vec{p}_{1}^* \vert \vert \vec{p}_{3}^{{\, \prime}} \vert \vert \vec{q} \, \vert \nonumber\\
&\, \times \left\{\sum_{\text{spins}} \vert \mathcal{A}_{\text{LO}} \vert^2+2\text{Re}\left[\sum_{\text{spins}}\mathcal{A}_{\text{LO}} \times \mathcal{A}_{\text{NLO}}\right]\right\},
\end{align}
where $\mathcal{A}_{\text{LO}}$ is the LO amplitude including the contribution from the tree-level and $D^*D$ rescattering diagrams, $\mathcal{A}_{\text{NLO}}$ is the NLO amplitude including only the $D^*\pi$ rescattering diagrams. The second term in the curly brackets includes the correction of the $D^*\pi$ rescattering, which is the interference term between the amplitudes at LO and NLO.

\begin{table}[bth]
\caption{\label{Tab:Tccstar+4Body} Partial decay widths of the $T_{cc}^{\ast } \to DD\pi\pi$ with a binding energy $\mathcal{B}=(503 \pm 40)\, \rm{keV}$. The second column contains the contributions from the tree-level diagrams, the third column is the LO decay width which includes the contributions from the tree-level and $D^{\ast}D$ rescattering diagrams, and the fourth column is the decay width which includes the corrections from the $D^{\ast}\pi$ rescattering to $\Gamma_{\text{LO}}$.}
\renewcommand{\arraystretch}{1.2}
\setlength{\tabcolsep}{20pt}{
\begin{tabular*}{\columnwidth}{l|c|c|c}
\hline\hline                                  
        $\Gamma$[keV]
        &$\text{Tree}$
        &$\text{LO}$ 
        &$\text{NLO}$
        \\[3pt]        
\hline
 $\Gamma[T_{cc}^{\ast +}\rightarrow D^0 D^+\pi^0 \pi^0 ] $ &$8.3^{+0.6}_{-0.3}$ &$10.5^{+0.8}_{-0.4}$ &$9.8^{+0.8}_{-0.4}$ \\[3pt]
 \hline 
 $\Gamma[T_{cc}^{\ast +}\rightarrow D^{*+}\pi D] \times \text{Br}(D^{*+} \rightarrow \pi^0D^+)$ &\multirow{2}{*}{$9.1 \pm 0.4$} &\multirow{2}{*}{$11.3 \pm 0.5$} &\multirow{2}{*}{$10.1^{+0.5}_{-0.4}$}\\[3pt]
 +$\Gamma[T_{cc}^{\ast +}\rightarrow D^{*}\pi D^+] \times \text{Br}(D^{*0} \rightarrow \pi^0 D^0)$ & & \\ [3pt] 
\hline
 $\Gamma[T_{cc}^{\ast +}\rightarrow D^0 D^0\pi^0 \pi^+]$ &$19.5^{+1.3}_{-1.8}$ &$23.9^{+0.1}_{-1.7}$ &$23.2^{+0.1}_{-1.7}$\\[3pt]
 \hline
 $\Gamma[T_{cc}^{\ast +}\rightarrow D^{*+}\pi D] \times \text{Br}(D^{*+} \rightarrow \pi^+D^0)$ &\multirow{2}{*}{$20.4 \pm 0.9$} &\multirow{2}{*}{$23.6 \pm 1.1$} &\multirow{2}{*}{$21.7 \pm 1.0$}\\[3pt]
 +$\Gamma[T_{cc}^{\ast +}\rightarrow D^{*}\pi^+ D] \times \text{Br}(D^{*0}\rightarrow \pi^0D^0)$ & & \\ [3pt] 
\hline\hline
\end{tabular*}}
\end{table}
\begin{figure}[htbp]
    \subfigure[$T_{cc}^{\ast +}\to D^0 D^+\pi^0 \pi^0 $] {\includegraphics[scale=0.295]{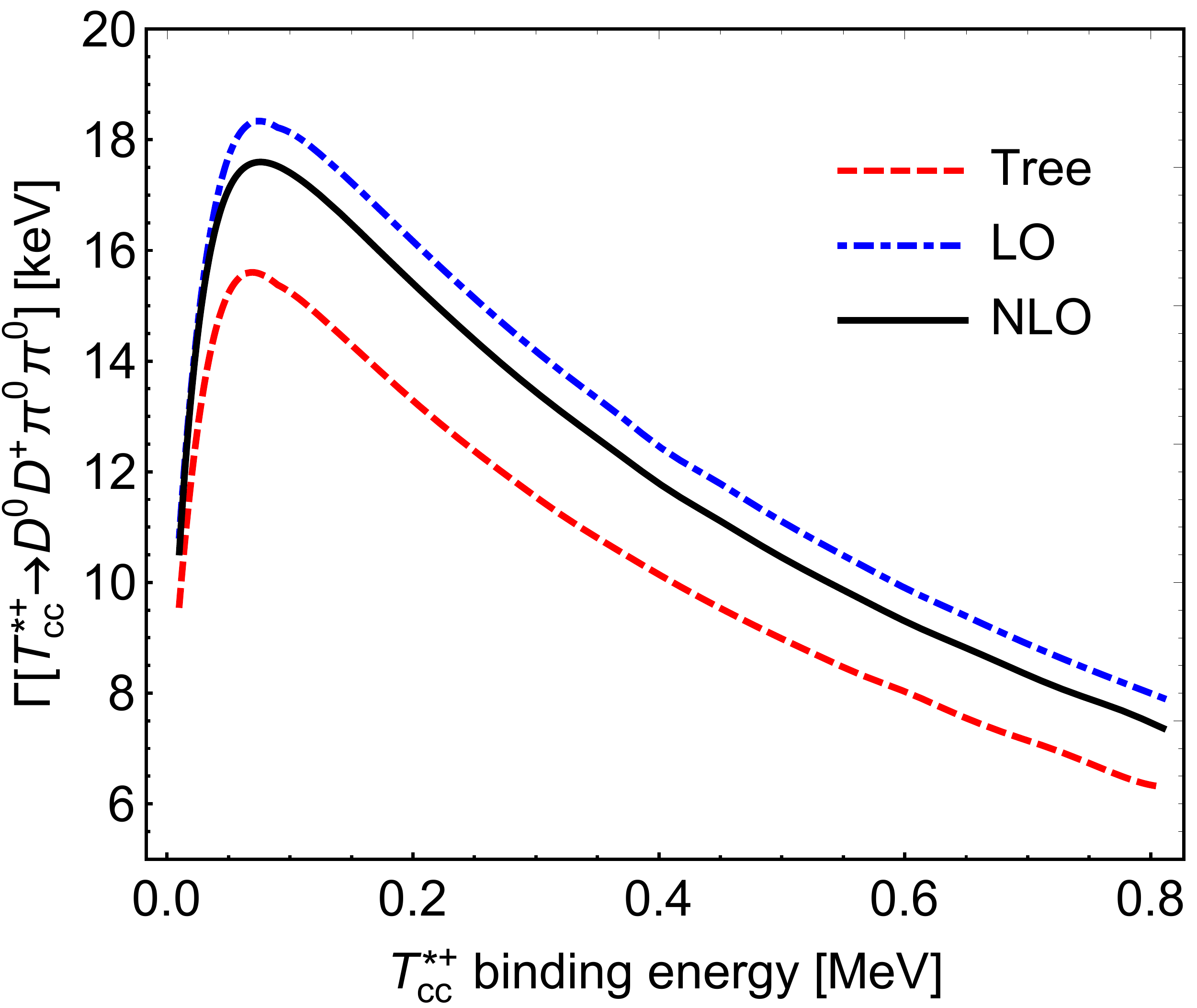} \label{Fig.DcDpipiTreeLONLO}
    }
    \subfigure[$T_{cc}^{\ast +}\to D^0 D^0\pi^0 \pi^+$] {\includegraphics[scale=0.295]{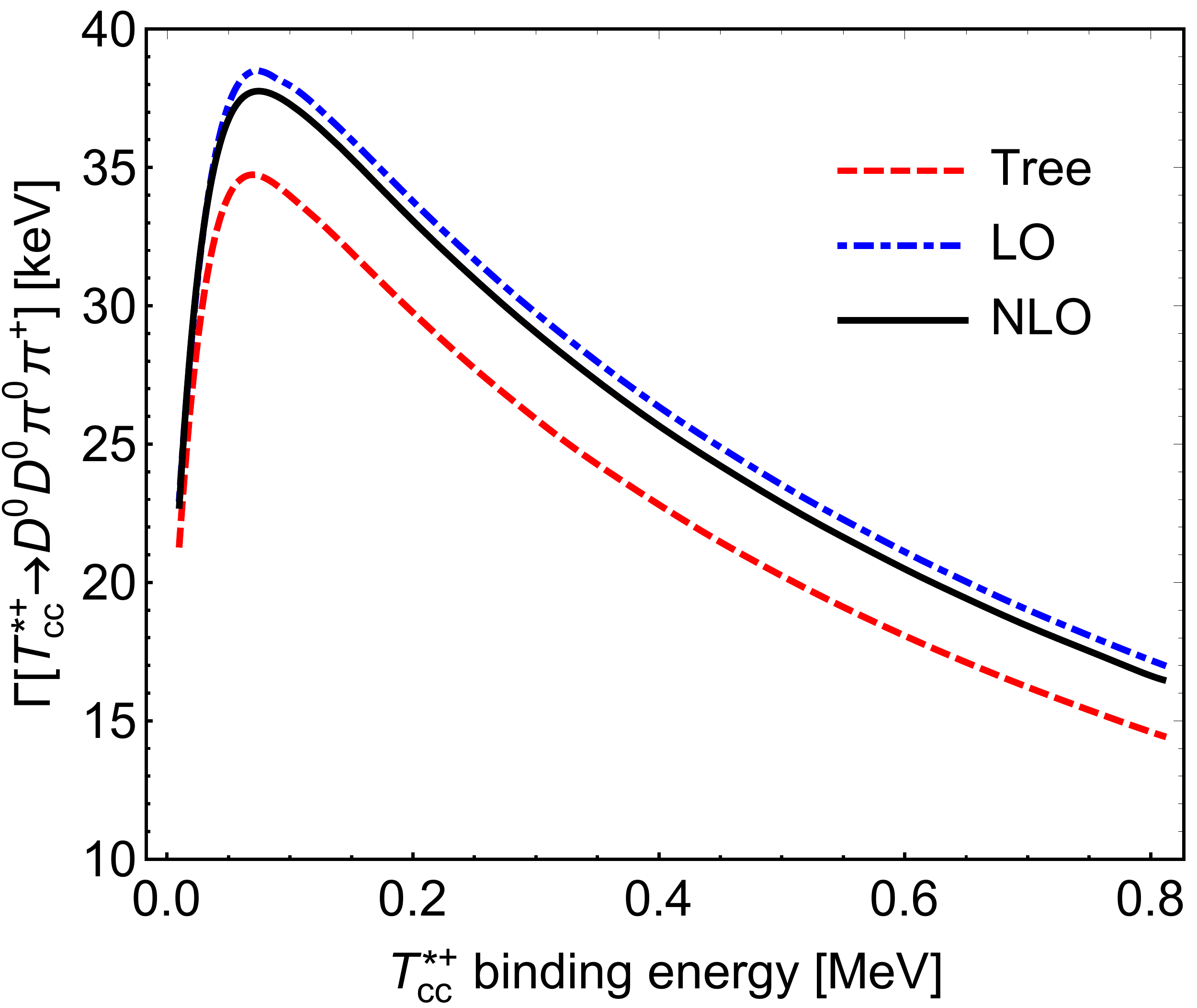} \label{Fig.DDpipicTreeLONLO}
    }
   \caption{Partial decay widths of the $T_{cc}^{\ast }\rightarrow DD\pi\pi$ versus the binding energy of the $T_{cc}^{\ast +}$.}
    \label{Fig.Tccstar_DDpipi decay width}
\end{figure}
\begin{figure}[htbp]
    \subfigure[$T_{cc}^{\ast +}\to D^0 D^+\pi^0 \pi^0 $] {\includegraphics[scale=0.295]{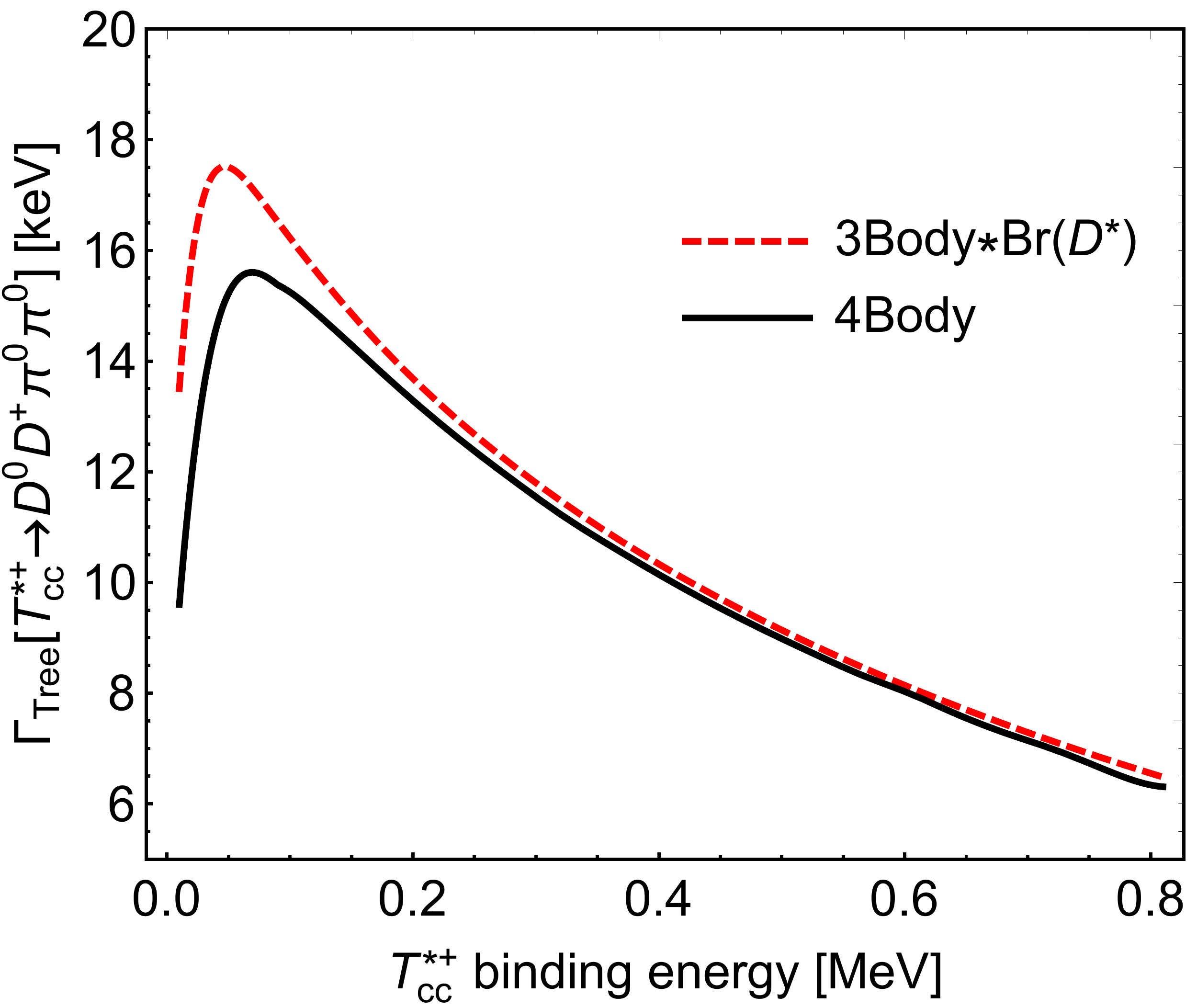} \label{Fig.DcDpipiTree3B4B}
    }
    \subfigure[$T_{cc}^{\ast +}\to D^0 D^+\pi^0 \pi^0 $] {\includegraphics[scale=0.295]{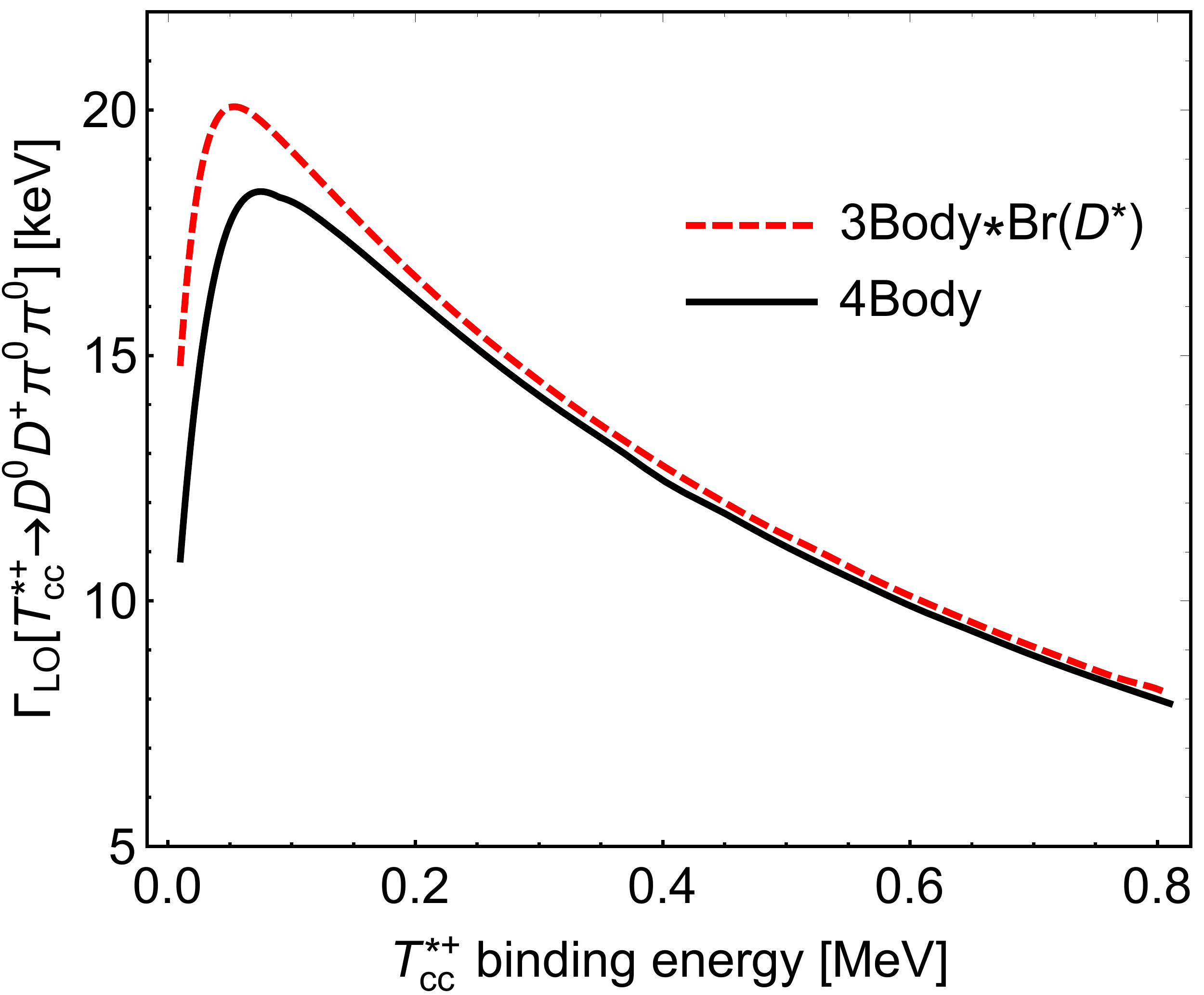} \label{Fig.DcDpipiLO3B4B}
    }
    \subfigure[$T_{cc}^{\ast +}\to D^0 D^+\pi^0 \pi^0 $] {\includegraphics[scale=0.295]{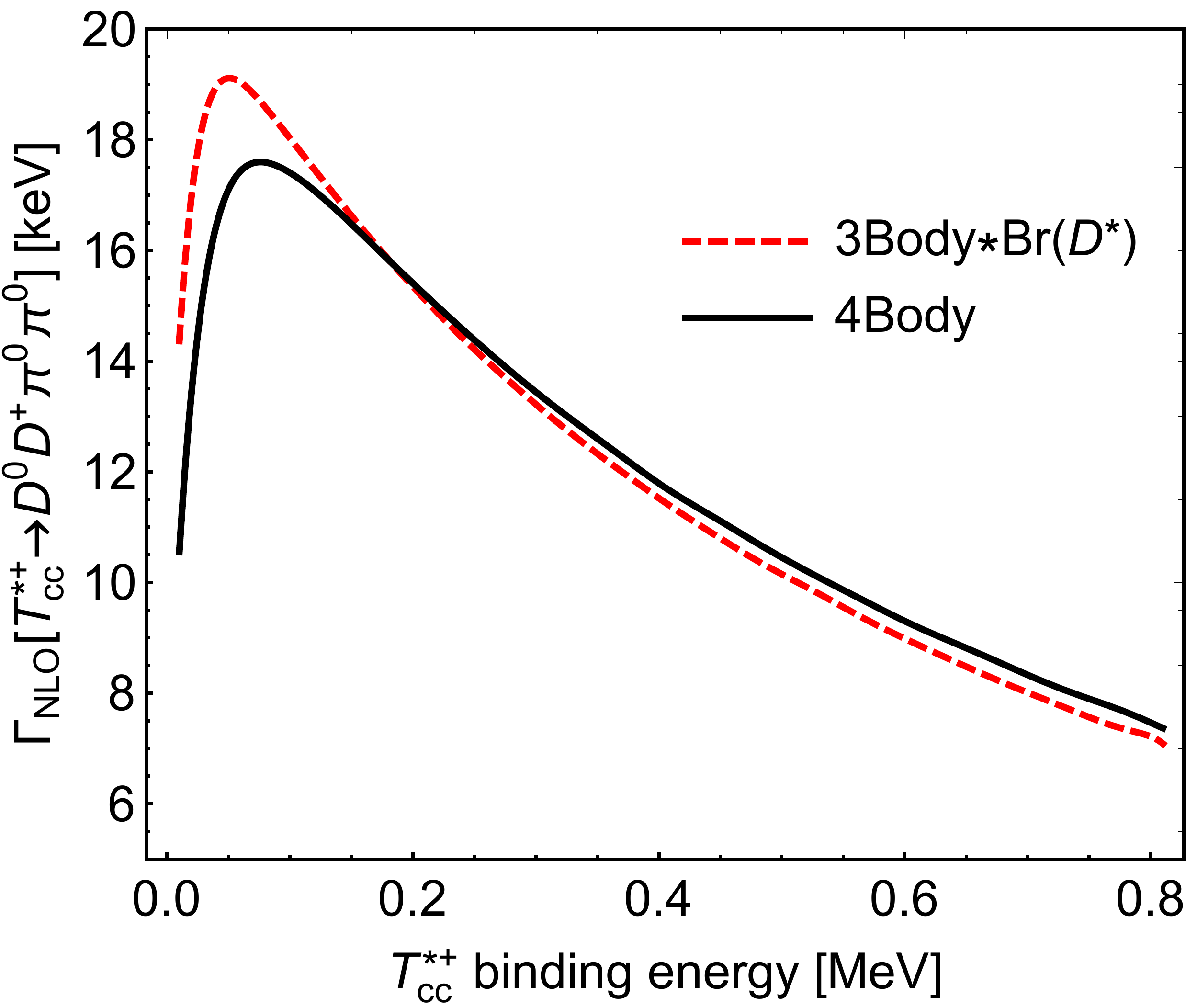} \label{Fig.DcDpipiNLO3B4B}
    }
    \subfigure[$T_{cc}^{\ast +}\to D^0 D^0\pi^0 \pi^+$] {\includegraphics[scale=0.295]{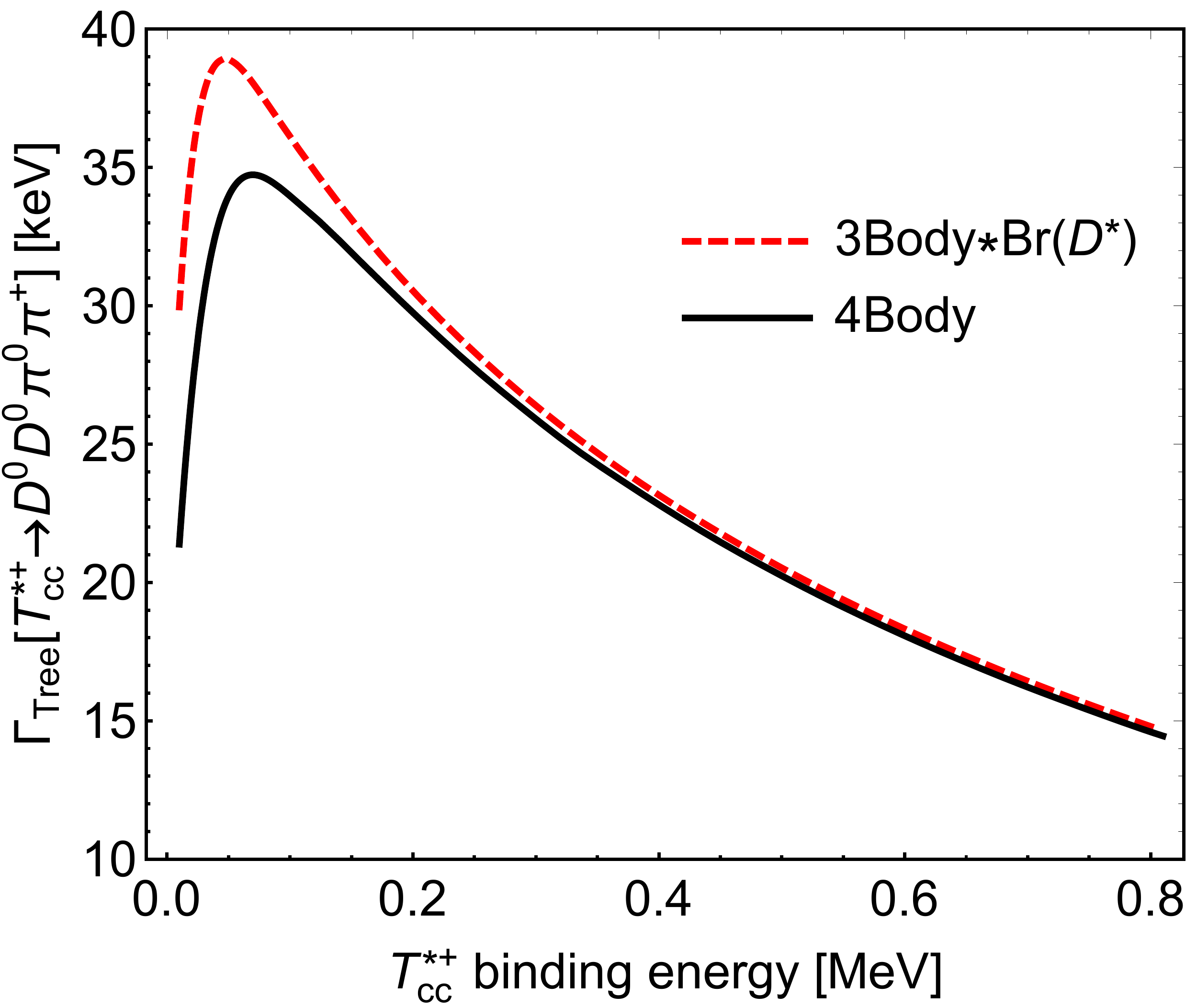} \label{Fig.DDpipicTree3B4B}
    }
    \subfigure[$T_{cc}^{\ast +}\to D^0 D^0\pi^0 \pi^+$] {\includegraphics[scale=0.295]{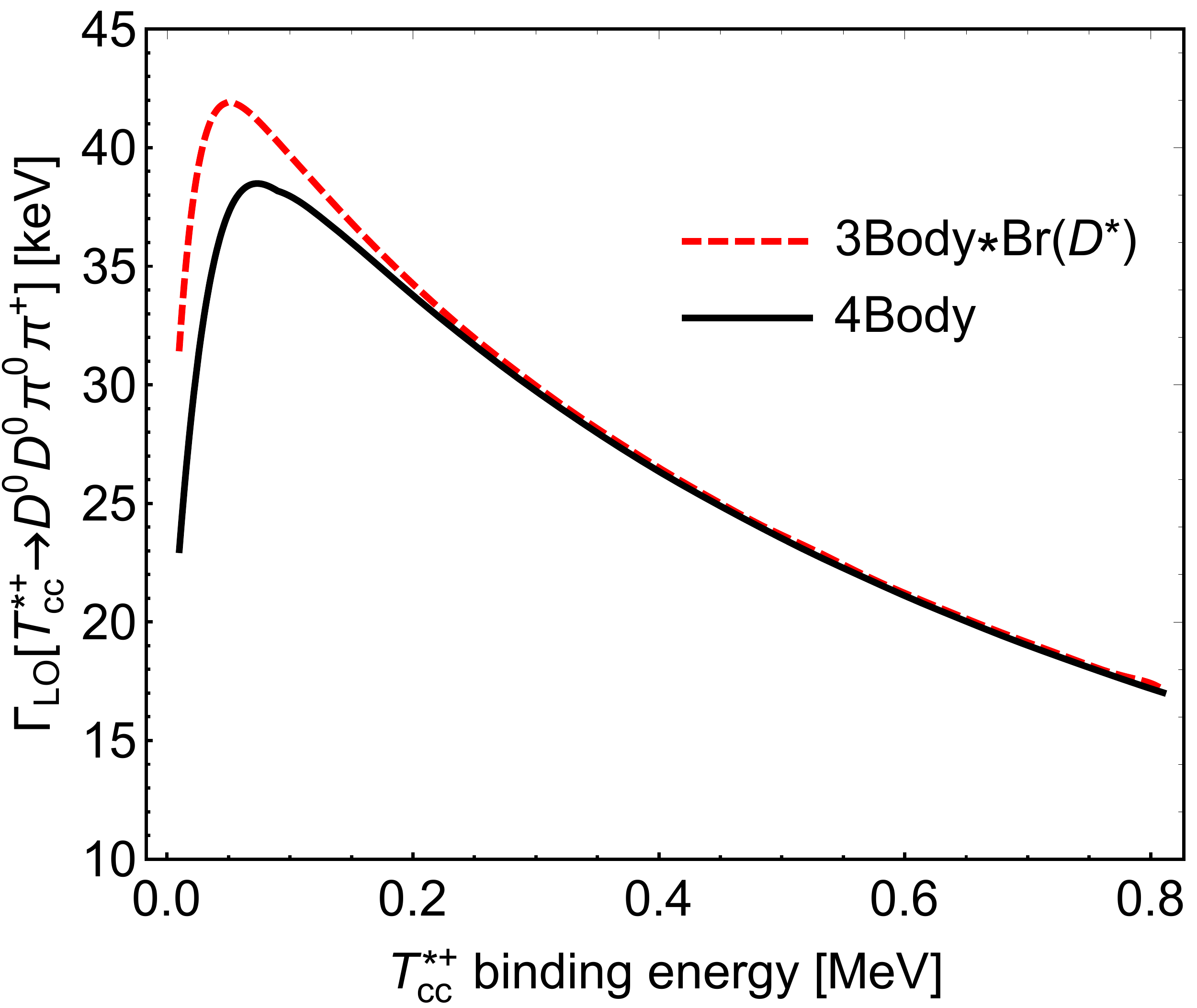} \label{Fig.DDpipicLO3B4B}
    }
    \subfigure[$T_{cc}^{\ast +}\to D^0 D^0\pi^0 \pi^+$] {\includegraphics[scale=0.295]{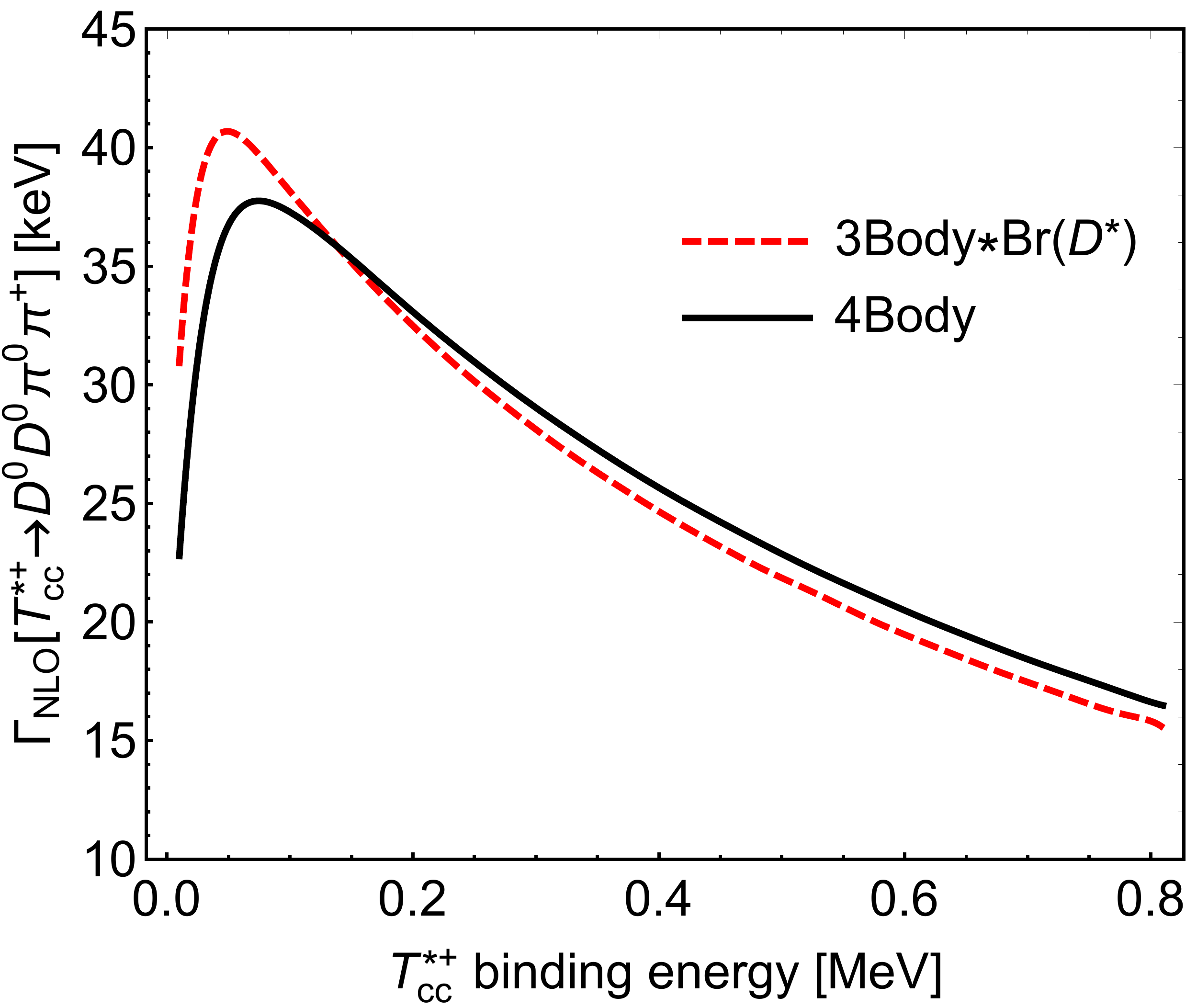} \label{Fig.DDpipicNLO3B4B}
    }
   \caption{Partial decay widths of the $T_{cc}^{\ast }\rightarrow DD\pi\pi$ and $T_{cc}^{*+} \to D^*D\pi$ times the branch ratio of $D^* \to D\pi$ versus the binding energy of the $T_{cc}^{\ast +}$.}
    \label{Fig.Tccstar_DDpipi3B4B}
\end{figure}
\begin{figure}[htbp]
    \subfigure[$T_{cc}^{\ast +}\to D^0 D^+\pi^0 \pi^0 $] {\includegraphics[scale=0.295]{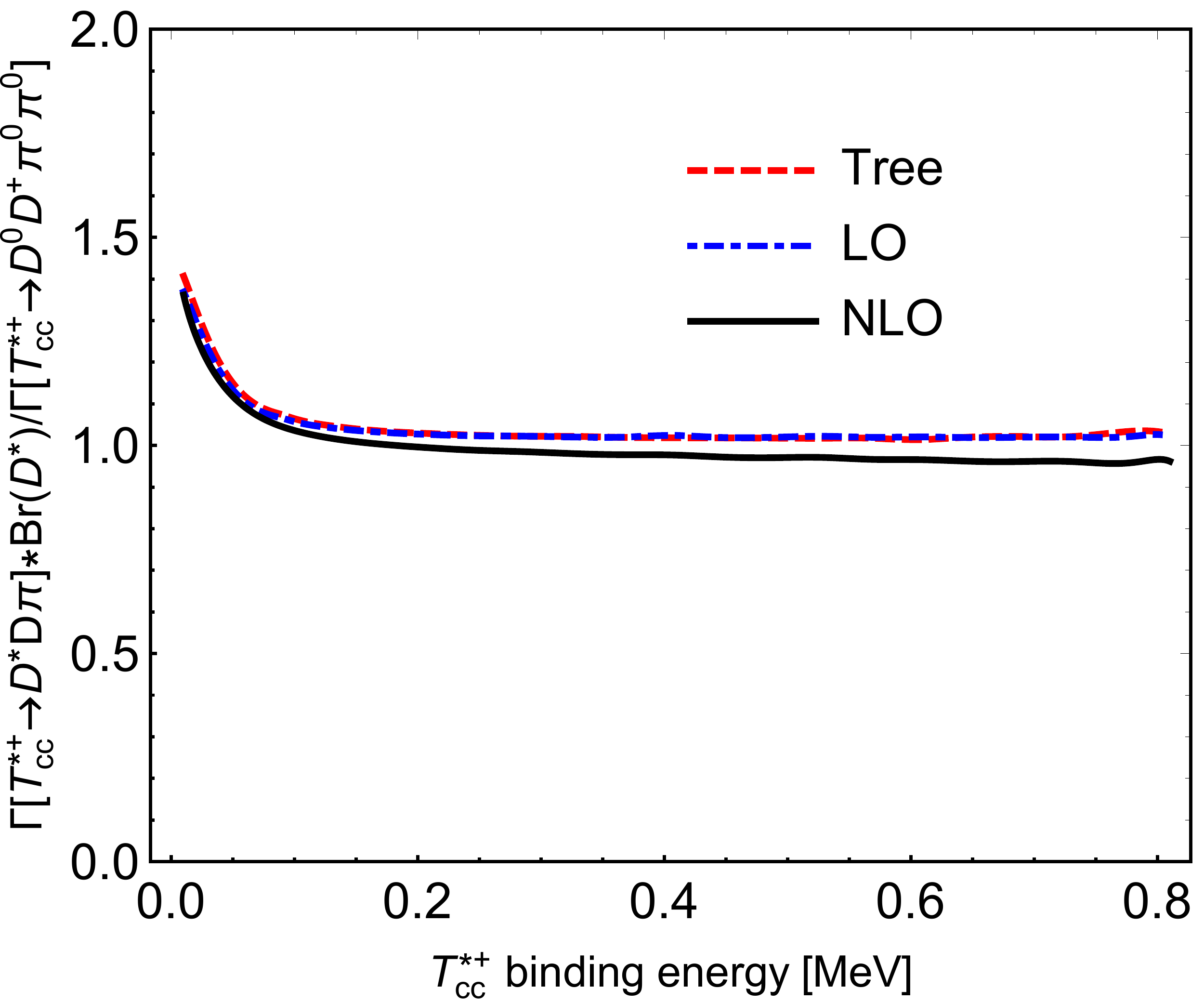} \label{Fig.DcDpipi3BBrvs4B}
    }
    \subfigure[$T_{cc}^{\ast +}\to D^0 D^0\pi^0 \pi^+$] {\includegraphics[scale=0.295]{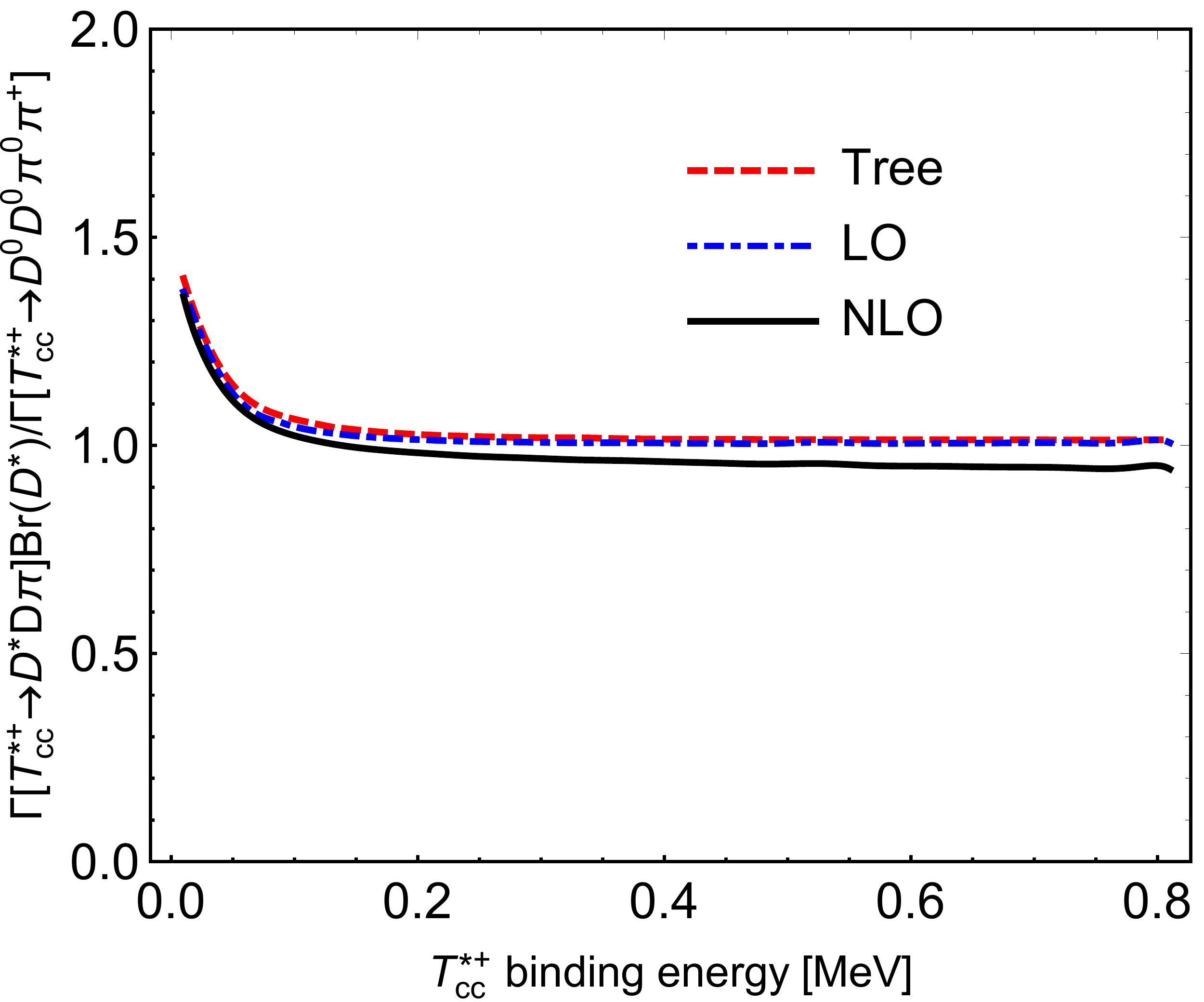} \label{Fig.DDpipic3BBrvs4B}
    }
   \caption{Ratio between the partial decay widths of the $T_{cc}^{\ast }\rightarrow DD\pi\pi$ and those obtained using $\Gamma(T_{cc}^{*+} \to D^*D\pi){\rm Br}(D^*\to D\pi)$ versus the binding energy of the $T_{cc}^{\ast +}$.}
\label{Fig.Tccstar_DDpipi3BBrvs4B}
\end{figure}
 
Table~\ref{Tab:Tccstar+4Body} shows the decay widths with the binding energy of the $T_{cc}^{*+}$ being $\mathcal{B}=(503 \pm 40)~\rm{keV}$. The second column includes only the contribution from the tree-level diagram denoted by $\Gamma_{\text{Tree}}$. The third column lists the LO decay widths, marked by $\Gamma_{\rm{LO}}$, including the tree-level and the $D^*D$ rescattering contribution. The fourth column lists the results up to NLO including corrections from the $D^*\pi$ rescattering. 
For comparison, we also list the results obtained by multiplying the three-body decays into $D^*D\pi$ with the corresponding $D^*\to D\pi$ branching fractions (and thus the interference between different intermediate three-body decays is neglected).
 
One can see that the difference between the results with and without the interference between different intermediate three-body $D^*D\pi$ is marginal. Thus, the $T_{cc}^*$ decay width can be well approximated by summing over the 3-body final state $D^*D\pi$, given in Eq.~\eqref{eq:total}.

As the binding energy of the $T_{cc}^*$ is uncertain, we give the partial widths of $T_{cc}^* \rightarrow DD\pi\pi$ varying the binding energy from $0.01~\rm{MeV}$ to $0.80~\rm{MeV}$ in Fig.~\ref{Fig.Tccstar_DDpipi decay width}. To see the relations between the 3-body decay $T_{cc}^{*} \rightarrow D^* D \pi$ and the 4-body decay $T_{cc}^{*} \rightarrow DD\pi\pi$ more clearly, we compare the partial decay widths $\Gamma[T_{cc}^{*} \to  DD\pi\pi ]$ and $\Gamma[T_{cc}^{*} \rightarrow D^*D\pi]\times\mathrm{Br}[D^* \rightarrow D\pi]$ in Fig.~\ref{Fig.Tccstar_DDpipi3B4B} and give their ratio in Fig.~\ref{Fig.Tccstar_DDpipi3BBrvs4B}. One can see that the difference between the decay widths with and without the interference between the intermediate 3-body $D^*D\pi$ states is marginal for the $T_{cc}^*$ binding energy larger than $200$~keV, and the binding energy $(503 \pm 40)\, \rm{keV}$ predicted in Ref.~\cite{Du:2021zzh} is within this region.   

\section{Summary}\label{sec:SUMMARY}

In this paper, we calculated the contributions of the $D^{\ast} D$ and $D^{\ast} \pi$ rescattering to the partial decay widths of the $T_{cc}^{\ast }\to D^*D\pi$ through the XEFT assuming that the $T_{cc}^{\ast +}$ is a $D^{\ast +} D^{\ast 0}$ shallow bound state. We found that the $I=0$ $D^{\ast+} D^0/D^{\ast0} D^+$ rescattering, which generates a $T_{cc}^+$ pole just below the threshold, can contribute at LO and has a sizeable impact on the partial widths of the $T_{c c}^{*+} \rightarrow D^{*+} D^0 \pi^0$ and $T_{c c}^{*+} \rightarrow D^{* 0} D^{+} \pi^0$. The corrections from the $D^{\ast} \pi$ rescattering to the LO result are marginal, at the level of 10\%.
Being an isoscalar $1^+$ $D^{*+}D^{*0}$ molecular state, the $T_{cc}^{*+}$ should decay dominantly into the three $D^*D\pi$ channels calculated here. 
Since the $D^*$ may be reconstructed from the $D\pi$ final state, we also calculate the four-body decay widths of $T_{cc}^{*+} \rightarrow DD\pi\pi$. We find that the interference effect between different intermediate $D^*D\pi$ states is small and the $T_{cc}^{*}$ width can be well approximated by summing over the $D^*D\pi$ partial widths for the $T_{cc}^*$ binding energy larger than $200$~keV. Taking the binding energy $(503\pm40)$~keV predicted in Ref.~\cite{Du:2021zzh}, the $T_{cc}^{*}$ width is obtained as about 41~keV.
The result reported here should be useful for searching the $T_{cc}^{\ast+}$ state at LHCb in the future.

\section{Acknowledgments}\label{sec: ACKNOWLEDGMENTS}

We would like to thank Xu Zhang for helpful discussions on the $D^{\ast}$ self-energy contribution.  This work is supported in part by the Chinese Academy of Sciences under Grant No. XDB34030000; by the
National Natural Science Foundation of China (NSFC) under Grants No. 12125507, No. 11835015, No. 12047503, No. 12075133, and No. 11961141012; by the China Postdoctoral Science Foundation under Grant No. 2022M713229 and by the NSFC and the Deutsche Forschungsgemeinschaft
(DFG) through the funds provided to the TRR110 “Symmetries and the Emergence of Structure in
QCD” (NSFC Grant No. 12070131001, DFG Project-ID 196253076). This work is also supported by Taishan Scholar Project of Shandong Province under Grant No. tsqn202103062,
the Higher Educational Youth Innovation Science and Technology
Program Shandong Province  under Grant No. 2020KJJ004.

\appendix

\section{ISOSPIN PHASE CONVENTIONS AND CONTACT INTERACTIONS}\label{sec:Contact interactions}

In this section, we give the isospin phase conventions in our calculation and derive the couplings $C_{\frac{3}{2}\pi}$ and $C_{\frac{1}{2}\pi}$ in Eq.~\eqref{eq:XEFTlagrangian} from the $D^{\ast} \pi$ scattering lengths. The isospin phase conventions for $D^*$ and $\pi$ are \cite{Dai:2019hrf}
\begin{align}
&\left|\pi^{+}\right\rangle=-|1,+1\rangle,
\quad\left|\pi^{0}\right\rangle=|1,0\rangle,
\quad\left|\pi^{-}\right\rangle=|1,-1\rangle,\nonumber\\
&\left|D^{\ast +}\right\rangle=\left|D^{ +}\right\rangle=\left|\frac{1}{2},+\frac{1}{2}\right\rangle,
\quad\left|D^{\ast 0}\right\rangle=\left|D^{ 0}\right\rangle=\left|\frac{1}{2},-\frac{1}{2}\right\rangle,
\end{align}
where the right-hand side represents states $|I,I_3\rangle$ in the isospin basis with $I$ and $I_3$ the isospin and its third component, respectively.
For the derivation of the contact interactions between the $D^*$ and $D$, all the couplings can be expressed in terms of two couplings,  $C_{0D}$ with $I=0$ and  $C_{1D}$ with $I=1$.
The $\left|D^{\ast}D\right\rangle$ states can be expressed in terms of the isospin basis as
\begin{align}
&\left|D^{\ast +} D^{0}\right\rangle=\sqrt{\frac{1}{2}}\left|1,0\right\rangle+\sqrt{\frac{1}{2}}\left|0,0\right\rangle,\\
\quad &\left|D^{\ast 0} D^{+}\right\rangle=\sqrt{\frac{1}{2}}\left|1,0\right\rangle-\sqrt{\frac{1}{2}}\left|0,0\right\rangle,\\
\quad &\left|D^{\ast 0} D^{0}\right\rangle=\left|1,-1\right\rangle.
\end{align}
The $D^*D$ amplitude can be written in terms of the amplitudes with the total isospin $I=1$ and $I=0$ as
\begin{align}
   \langle D^{\ast +}D^{0}|T|D^{\ast +}D^{0}\rangle=&\,\frac{{1}}{2}\langle D^{\ast}D|T|D^{\ast}D\rangle_{I=1}+\frac{{1}}{2}\langle D^{\ast}D|T|D^{\ast}D\rangle_{I=0},\\
   \langle D^{\ast +}D^{0}|T|D^{\ast 0}D^{+}\rangle=&\,\frac{{1}}{2}\langle D^{\ast}D|T|D^{\ast}D\rangle_{I=1}-\frac{{1}}{2}\langle D^{\ast}D|T|D^{\ast}D\rangle_{I=0},\\
   \langle D^{\ast 0}D^{+}|T|D^{\ast 0}D^{+}\rangle=&\,\frac{{1}}{2}\langle D^{\ast}D|T|D^{\ast}D\rangle_{I=1}+\frac{{1}}{2}\langle D^{\ast}D|T|D^{\ast}D\rangle_{I=0},\\
   \langle D^{\ast 0}D^{+}|T|D^{\ast +}D^{0}\rangle=&\,\frac{{1}}{2}\langle D^{\ast}D|T|D^{\ast}D\rangle_{I=1}-\frac{{1}}{2}\langle D^{\ast}D|T|D^{\ast}D\rangle_{I=0},\\
   \langle D^{\ast 0}D^{0}|T|D^{\ast 0}D^{0}\rangle=&\,\langle D^{\ast}D|T|D^{\ast}D\rangle_{I=1},
\end{align}
which give the expressions of  $C_{0D1}$, $C_{0D2}$, $C_{D3}$, $C_{0D1\text{ex}}$, and $C_{0D2\text{ex}}$ in terms of $C_{0D}$ and $C_{1D}$ as 
\begin{align}
    C_{0D1}&=\frac{1}{2}C_{1D}+\frac{1}{2}C_{0D},\\
    C_{0D1\text{ex}}&=\frac{1}{2}C_{1D}-\frac{1}{2}C_{0D},\\
    C_{0D2}&=\frac{1}{2}C_{1D}+\frac{1}{2}C_{0D},\\
    C_{0D2\text{ex}}&=\frac{1}{2}C_{1D}-\frac{1}{2}C_{0D},\\
    C_{D3}&=C_{1D}.
\end{align}
Here $C_{1D}$ is the $D^*D$ contact interaction with $I=1$ and is neglected in our calculation as there is no isovector exotic state like the $T_{cc}$ near the $D^{*}D$ threshold, $C_{0D1}$, $C_{0D2}$, and $C_{D3}$ are the contact couplings for $D^{*+}D^0\to D^{*+} D^0$, $D^{*0}D^+\to D^{*0} D^+$, and $D^{*0}D^0\to D^{*0} D^0$, respectively, and $C_{0D1\text{ex}}$ and $C_{0D2\text{ex}}$ are the contact couplings for $D^{*0}D^+\to D^{*+} D^0$ and $D^{*+}D^0\to D^{*0} D^+$, respectively.

For the derivation of the contact interactions between the $D^*$ and $\pi$, all the couplings can be expressed in terms of two couplings, $C_{\frac{3}{2}\pi}$ with $I=\frac{3}{2}$ and $C_{\frac{1}{2}\pi}$ with $I=\frac{1}{2}$, and the two couplings can be obtained by matching the $D^*\pi$ scattering amplitude at the $D^*\pi$ threshold,
\begin{align}
    \sqrt{2m_{D^{\ast}} 2m_{\pi} 2m_{D^{\ast}} 2m_{\pi}}\frac{C_{\pi }^{I}}{2m_{\pi}}=A^I_{D^{\ast} \pi}\left(\sqrt{s}=m_{D^{\ast}}+m_{\pi} \right)=8 \pi\left(m_{D^{\ast}}+m_{\pi} \right) a^I_{D^{\ast} \pi},
\end{align}
where $I=\frac{3}{2},\frac{1}{2}$, and $a^I_{D^{\ast} \pi}$ is the $D^*\pi$ scattering length with isospin $I$. By using the central values of the scattering lengths $a_{D^{\ast} \pi}^{3 / 2}=a_{D \pi}^{3 / 2}=-(0.100 \pm 0.002)\, \mathrm{fm}$ and $a_{D^{\ast} \pi}^{1 / 2}=a_{D \pi}^{1 / 2}=0.37_{-0.02}^{+0.03}\, \mathrm{fm}$ given in Ref.~\cite{Liu:2012zya}, we have 
\begin{align}
    C_{\frac{3}{2}\pi}=-6.8~\mathrm{GeV}^{-1},\\
    C_{\frac{1}{2}\pi}=25.2~\mathrm{GeV}^{-1}.
\end{align}
The $\left|D^{\ast}\pi\right\rangle$ states can be expressed in terms of the isospin basis as
\begin{align}
-&\left|D^{\ast 0} \pi^{+}\right\rangle=\sqrt{\frac{1}{3}}\left|\frac{3}{2},+\frac{1}{2}\right\rangle+\sqrt{\frac{2}{3}}\left|\frac{1}{2},+\frac{1}{2}\right\rangle,\\
&\left|D^{\ast 0} \pi^{0}\right\rangle=\sqrt{\frac{2}{3}}\left|\frac{3}{2},-\frac{1}{2}\right\rangle+\sqrt{\frac{1}{3}}\left|\frac{1}{2},-\frac{1}{2}\right\rangle,\\
&\left|D^{\ast +} \pi^{0}\right\rangle=\sqrt{\frac{2}{3}}\left|\frac{3}{2},+\frac{1}{2}\right\rangle-\sqrt{\frac{1}{3}}\left|\frac{1}{2},+\frac{1}{2}\right\rangle,\\
&\left|D^{\ast +} \pi^{-}\right\rangle=\sqrt{\frac{1}{3}}\left|\frac{3}{2},-\frac{1}{2}\right\rangle-\sqrt{\frac{2}{3}}\left|\frac{1}{2},-\frac{1}{2}\right\rangle.
\end{align}

The $D^*\pi$ amplitude can be written in terms of the amplitudes with total isospin $I=\frac{3}{2}$ and $I=\frac{1}{2}$ as
\begin{align}
   \langle D^{\ast +}\pi^0|T|D^{\ast +}\pi^0\rangle=&\frac{{2}}{3}\langle D^{\ast}\pi|T|D^{\ast}\pi\rangle_{I=\frac{3}{2}}+\frac{{1}}{3}\langle D^{\ast}\pi|T|D^{\ast}\pi\rangle_{I=\frac{1}{2}},\\
   \langle D^{\ast +}\pi^0|T|D^{\ast 0}\pi^+\rangle=&-\frac{\sqrt{2}}{3}\langle D^{\ast}\pi|T|D^{\ast}\pi\rangle_{I=\frac{3}{2}}+\frac{\sqrt{2}}{3}\langle D^{\ast}\pi|T|D^{\ast}\pi\rangle_{I=\frac{1}{2}},\\
   \langle D^{\ast 0}\pi^0|T|D^{\ast 0}\pi^0\rangle=&\frac{2}{3}\langle D^{\ast}\pi|T|D^{\ast}\pi\rangle_{I=\frac{3}{2}}+\frac{1}{3}\langle D^{\ast}\pi|T|D^{\ast}\pi\rangle_{I=\frac{1}{2}},\\
   \langle D^{\ast 0}\pi^0|T|D^{\ast +}\pi^-\rangle=&\frac{\sqrt{2}}{3}\langle D^{\ast}\pi|T|D^{\ast}\pi\rangle_{I=\frac{3}{2}}-\frac{\sqrt{2}}{3}\langle D^{\ast}\pi|T|D^{\ast}\pi\rangle_{I=\frac{1}{2}},\\
   \langle D^{\ast 0}\pi^+|T|D^{\ast 0}\pi^+\rangle=&\frac{1}{3}\langle D^{\ast}\pi|T|D^{\ast}\pi\rangle_{I=\frac{3}{2}}+\frac{2}{3}\langle D^{\ast}\pi|T|D^{\ast}\pi\rangle_{I=\frac{1}{2}},\\
   \langle D^{\ast 0}\pi^+|T|D^{\ast +}\pi^0\rangle=&-\frac{\sqrt{2}}{3}\langle D^{\ast}\pi|T|D^{\ast}\pi\rangle_{I=\frac{3}{2}}+\frac{\sqrt{2}}{3}\langle D^{\ast}\pi|T|D^{\ast}\pi\rangle_{I=\frac{1}{2}},
\end{align}
which give the expressions of  $C_{\pi1}$, $C_{\pi2}$, $C_{\pi3}$, $C_{\pi1\text{ex}}$, $C_{\pi2\text{ex}}$, and $C_{\pi3\text{ex}}$ in terms of $C_{\frac{3}{2}\pi}$ and $C_{\frac{1}{2}\pi}$ as 
\begin{align}
    C_{\pi1}&=\frac{2}{3}C_{\frac{3}{2}\pi}+\frac{1}{3}C_{\frac{1}{2}\pi}=4.1\, \mathrm{GeV}^{-1},\\
    C_{\pi1\text{ex}}&=-\frac{\sqrt{2}}{3}C_{\frac{3}{2}\pi}+\frac{\sqrt{2}}{3}C_{\frac{1}{2}\pi}=15.1\, \mathrm{GeV}^{-1},\\
    C_{\pi2}&=\frac{2}{3}C_{\frac{3}{2}\pi}+\frac{1}{3}C_{\frac{1}{2}\pi}=4.1\, \mathrm{GeV}^{-1},\\
    C_{\pi2\text{ex}}&=\frac{\sqrt{2}}{3}C_{\frac{3}{2}\pi}-\frac{\sqrt{2}}{3}C_{\frac{1}{2}\pi}=-15.1\, \mathrm{GeV}^{-1},\\
    C_{\pi3}&=\frac{1}{3}C_{\frac{3}{2}\pi}+\frac{2}{3}C_{\frac{1}{2}\pi}=14.4\, \mathrm{GeV}^{-1},\\
    C_{\pi3\text{ex}}=&-\frac{\sqrt{2}}{3}C_{\frac{3}{2}\pi}+\frac{\sqrt{2}}{3}C_{\frac{1}{2}\pi}=15.1\, \mathrm{GeV}^{-1}.
\end{align}
Here $C_{\pi1}$, $C_{\pi2}$, and $C_{\pi3}$ are the contact interactions for $D^{*+}\pi^0\to D^{*+}\pi^0$,
$D^{*0}\pi^0\to D^{*0}\pi^0$, and $D^{*0}\pi^+\to D^{*0}\pi^+$, respectively, and $C_{\pi1\text{ex}}$, $C_{\pi2\text{ex}}$, and $C_{\pi3\text{ex}}$ are the contact interactions for $D^{*0}\pi^+\to D^{*+}\pi^0$,
$D^{*+}\pi^-\to D^{*0}\pi^0$, and $D^{*+}\pi^0\to D^{*0}\pi^+$, respectively.

\section{3-POINT LOOP INTEGRALS}\label{sec:Triangal loop}

In the rest frame of the decay particle, the scalar 3-point loop integral is ultraviolet (UV) convergent and can be worked out as~\cite{Guo:2010ak}
\begin{align}
I(q)
&=i\int \frac{d^dl}{(2 \pi)^{d}} \frac{1}{\left(l^0-m_{1}-\frac{\vec{l}^{2}}{2 m_{1}}+i \epsilon\right)\left(M-l^{0}-m_{2}-\frac{\vec{l}^{2}}{2 m_{2}}+i \epsilon\right)\left[l^{0}-q^0-m_{3}-\frac{\left(\vec{l}-\vec{q}\right)^{2}}{2 m_{3}}+i \epsilon\right]}\nonumber \\
&=\int \frac{d^{d-1}l}{(2 \pi)^{d-1}} \frac{1}{\left(b_{12}+\frac{\vec{l}^{2}}{2 \mu_{12}}-i \epsilon\right)\left[b_{23}+\frac{\vec{l}^{2}}{2 m_{2}}+\frac{\left(\vec{l}-\vec{q}\right)^{2}}{2 m_{3}}-i \epsilon\right]}\nonumber \\
&=4 \mu_{12} \mu_{23} \int \frac{d^{d-1} l}{(2 \pi)^{d-1}} \frac{1}{\left(\vec{l}^2+c_1-i \epsilon\right)\left(\vec{l}^2-\frac{2 \mu_{23}}{m_3} \vec{l} \cdot \vec{q}+c_2-i \epsilon\right)}\nonumber\\
&=4\mu_{12}\mu_{23}\int_0^1 d x \int \frac{d^{d-1} l}{(2 \pi)^{d-1}} \frac{1}{\left[\vec{l}^2-a x^2+\left(c_2-c_1\right) x+c_1-i \epsilon\right]^2} \nonumber \\
&=\frac{4 \mu_{12} \mu_{23}}{(4 \pi)^{(d-1) / 2}} \Gamma\left(\frac{5-d}{2}\right) \int_0^1 d x\left[-a x^2+\left(c_2-c_1\right) x+c_1-i \epsilon\right]^{(d-5) / 2} \nonumber \\
&=\frac{\mu_{12} \mu_{23}}{2 \pi } \frac{1}{\sqrt{a}}\left[\tan ^{-1}\left(\frac{c_2-c_1}{2 \sqrt{a c_1}}\right)+\tan ^{-1}\left(\frac{2 a+c_1-c_2}{2 \sqrt{a\left(c_2-a\right)}}\right)\right],
\label{Eq:loop_integral}
\end{align}
where $\mu_{ij}=m_{i} m_{j} /\left(m_{i}+m_{j}\right)$ are the reduced masses, $b_{12}=m_1+m_2-M,b_{23}=m_{2}+m_{3}+q^0-M$, and 
\begin{align}
 a=\left(\frac{\mu_{23}}{m_3}\right)^2 \vec{q}^2, \quad c_1=2 \mu_{12} b_{12}, \quad c_2=2 \mu_{23} b_{23}+\frac{\mu_{23}}{m_3} \vec{q}^2.
\end{align}
There is no pole for the spacetime dimension $d\leq 4$, and we have taken $d=4$ in the last step of Eq. \eqref{Eq:loop_integral}.

We also need the vector integral which is defined as 
\begin{align}
&q^{i} I^{(1)}(q)=\int \frac{d^{3} \vec{l}}{(2 \pi)^{3}}  \frac{l^{i}}{\left(b_{12}+\frac{\vec{l}^{2}}{2 \mu_{12}}-i \epsilon\right)\left[b_{23}+\frac{\vec{l}^{2}}{2 m_{2}}+\frac{\left(\vec{l}-\vec{q}\right)^{2}}{2 m_{3}}-i \epsilon\right]},
\label{Eq.qiqjI02}
\end{align}
and $I^{(1)}(q)$ can be expressed in terms of the scalar 2-point and 3-point loop integrals as 
\begin{align}
    I^{(1)}(q)=\frac{\mu_{23}}{a m_3}\left[B\left(c_2-a\right)-B\left(c_1\right)+\frac{1}{2}\left(c_2-c_1\right) I(q)\right],
\end{align}
where the two-point function $B(c)=2 \mu_{12} \mu_{23} \Sigma(c)$, with $\Sigma(c)$ defined in the power divergence subtraction (PDS) scheme~\cite{Kaplan:1998tg} as 
\begin{align}
\Sigma(c) & \equiv\left(\frac{\Lambda_{\mathrm{PDS}}}{2}\right)^{4-d} \int \frac{d^{d-1} l}{(2 \pi)^{d-1}} \frac{1}{\vec{l}^2+c-i \epsilon} \\
&=\left(\frac{\Lambda_{\mathrm{PDS}}}{2}\right)^{4-d}(4 \pi)^{(1-d) / 2} \Gamma\left(\frac{3-d}{2}\right)(c-i \epsilon)^{(d-3) / 2} \\
&=\frac{1}{4 \pi}\left(\Lambda_{\mathrm{PDS}}-\sqrt{c-i \epsilon}\right)
\end{align}
where $\Lambda_{\rm{PDS}}$ is the sharp cutoff to regulate the UV divergence in the two-point scalar loop integral, and $I^{(1)}(q)$ is also UV convergent.
When the particles in the 3-point integrals are unstable, e.g., considering the $D^*$ self-energy contribution shown in Eq.~\eqref{Eq.Dstar_self_energy}, one needs to include their widths by replacing
\begin{align}
    m_k\to m_k-i\frac{\Gamma_k}{2}, \quad k=1,2,3,
\end{align}
and in the loop integrals $I(q)$ and $I^{(1)}(q)$, one just makes the replacements for $c_1$ and $c_2$ as
\begin{align}
    c_1\to c_1-i\mu_{12}(\Gamma_1+\Gamma_2), c_2\to c_2-i\mu_{23}(\Gamma_2+\Gamma_3).
\end{align}

\section{PHASE SPACE}\label{sec:Phase space}

\subsection{Three-body phase space}\label{sec:Three-body phase space}
In this section, we derive the three-body phase space in Eq. \eqref{Eq.three_body_phase_space} in the rest frame of the initial particle. The three-body phase space can be written as
\begin{align}
\int d\Phi_{3}\left(P; p_1, p_2, p_3\right)=&\int\left(2 \pi\right)^4 \delta^{4}\left(P-p_1-p_2-p_3\right)\frac{d^3 \vec{p}_1}{(2\pi)^3 2E_{1}} \frac{d^3 \vec{p}_2}{(2\pi )^3 2E_{2}} \frac{d^3 \vec{p}_{3}}{(2 \pi)^3 2E_{3}}\nonumber\\
=&\int\frac{1}{(2\pi)^5}\delta(E-E_1-E_2-E_3)\delta^{3}(\vec{p}_1+\vec{p}_2+\vec{p}_3)\frac{d^3 \vec{p}_1}{2E_{1}} \frac{d^3 \vec{p}_2}{2E_{2}} \frac{d^3 \vec{p}_{3}}{2E_{3}}\nonumber\\
=&\int\frac{1}{(2\pi)^5}\delta(E-E_1-E_2-E_3)\frac{\vert \vec{p}_{1} \vert d\vert \vec{p}_{1}\vert^{2}d\Omega_{1}}{4 E_1}\frac{\vert \vec{p}_{2} \vert d\vert \vec{p}_{2}\vert^{2}d\Omega_{12}}{4 E_2} \frac{1}{2E_3},
\label{eq:three body phase space}
\end{align}
where $P=(M, \vec{0})$, $d^3 \vec{p}_1=\vert \vec{p}_1 \vert^2 d\vert \vec{p}_1 \vert d\Omega_1$ and $d^3 \vec{p}_2=\vert \vec{p}_2 \vert^2 d\vert \vec{p}_2 \vert d\Omega_{12}$. Here $\Omega_{12}$ is the solid angle between the moving directions of particle 1 and particle 2, $ d\Omega_{12}=\sin\theta_{12} d\theta_{12}d\varphi_{12}$, where $\theta_{12}$ is the angle between particles 1 and 2 and is related to the three-momenta as
\begin{align}
\vert \vec{p}_3 \vert^{2}
=\vert \vec{p}_1 \vert^{2}+\vert \vec{p}_2 \vert^{2}-2\cos \theta_{12}\vert \vec{p}_1 \vert \vert \vec{p}_2 \vert.
\label{eq:p32=Costheta12}
\end{align}
The integration over $\theta_{12}$ in Eq.~\eqref{eq:three body phase space} can be changed to the integration over $E_3$ through
\begin{align}
d\vert \vec{p}_3 \vert^{2}= 2\vert \vec{p}_1\vert \vert \vec{p}_2\vert \sin\theta_{12} d\theta_{12}=2E_{3}dE_{3},
\label{eq:dp32}
\end{align}
and the three-body phase space reads
\begin{align}
\int d\Phi_{3}\left(P; p_1, p_2, p_3\right)=&\int\frac{1}{(2\pi)^5} \delta(E-E_1-E_2-E_3) \frac{\vert \vec{p}_1 \vert \vert \vec{p}_2 \vert}{16 E_1 E_2} d\Omega_1 \frac{d \vert \vec{p}_3 \vert^2}{2 \vert \vec{p}_1 \vert \vert \vec{p}_2 \vert} d\varphi_{12} \frac{1}{2 E_3} d\vert \vec{p}_{1}\vert^{2} d\vert \vec{p}_{2}\vert^{2}\nonumber\\
=&\int\frac{1}{(2\pi)^5} \delta(E-E_1-E_2-E_3) \frac{\vert \vec{p}_1 \vert \vert \vec{p}_2 \vert}{16 E_1 E_2} d\Omega_1 \frac{2 E_3 dE_3}{2 \vert \vec{p}_1 \vert \vert \vec{p}_2 \vert} d\varphi_{12} \frac{1}{2 E_3} d\vert \vec{p}_{1}\vert^{2} d\vert \vec{p}_{2}\vert^{2}\nonumber\\
=&\int\frac{1}{32 \pi^3} \frac{1}{4 E_1 E_2} d\vert \vec{p}_1 \vert^2 d\vert \vec{p}_2 \vert^2,
\end{align}

\subsection{Four-body phase space} \label{sec:Four-body phase space}

Here, we derive the four-body phase space using the graphic method of Ref.~\cite{Jing:2020tth}.
\begin{figure}[tb] {
    \includegraphics[scale=0.5]{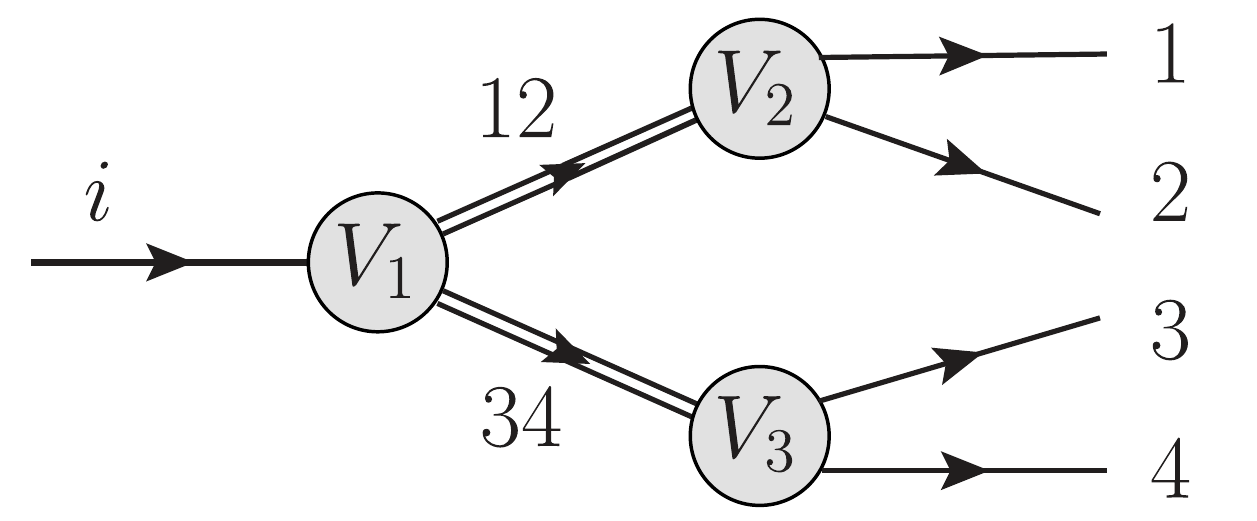}
    }
 \caption{Complete expanded graph for the four-body phase space in terms of integrals over invariant masses squared $s_{12}$ and $s_{34}$ using the graphic method proposed in Ref.~\cite{Jing:2020tth}. The double lines denote the invariant masses, the single lines denote the particles and the vertices are the two-body phase spaces. }
\label{Fig.1to4Dec_tree}
\end{figure}
In the graphic method, an arbitrary $n$-body phase space can be easily reduced to a product of two-body phase spaces by drawing intuitive phase space graphs. Then it can be expressed in integrations over any allowed invariant masses of interest with the involved momenta being in any reference frame.
The graph for expressing the four-body phase space integration over $s_{12} = p_{12}^2$ and $s_{34}=p_{34}^2$, where $p_{12}=p_1+p_2$ and $p_{34}=p_3+p_4$, is shown in Fig.~\ref{Fig.1to4Dec_tree}. 
The phase space expression can be easily obtained by multiplying together the following building blocks:
\begin{align}
    &V_1:~ d\Phi_2(P;p_{12},p_{34}),\hspace{.5cm} V_2:~ d\Phi_2(p_{12};p_1,p_2),\hspace{.5cm}
    V_{3}:~ d\Phi_2(p_{34};p_3,p_4),\notag\\ 
    &``12\text{''}:~\frac{d s_{12}}{2\pi},\hspace{.5cm}``34\text{''}:~\frac{d s_{34}}{2\pi}.
    \label{Eq.building_blocks}
\end{align}
Here we use the following two-body phase space in the c.m. frame for simplicity.
\begin{equation}
    d\Phi_2(p;p_1,p_2) = d\Omega_1\frac{|\vec{p}_1|}{(2\pi)^2 4m},
    \label{eq_1to2PSFcm}
\end{equation}
where $|\vec{p}_1|$ is the magnitude of the three-momentum of particle 1 in the c.m. frame of the initial state, $d\Omega_1=d\cos\theta_1d\varphi_1$ is the solid angle of particle 1 in the c.m. frame of particles 1 and 2, and the integration region is given by $\cos{\theta_1}\in[-1,1]$ and $\varphi_1\in[0,2\pi)$.

With all the building blocks in Eq.~\eqref{Eq.building_blocks} and the two-body phase space in Eq.~\eqref{eq_1to2PSFcm}, one obtains the four-body phase space,
\begin{align}
    d\Phi_4(P;p_1,p_2,p_3,p_4)
    =\frac{|\vec{p}^{\,\prime\prime}_{12}||\vec{p}^*_1||\vec{p}^{\,\prime}_3|}{(8\pi^2)^{4} M}d \sqrt{s_{12}} d \sqrt{s_{34}} d\Omega_{12}''d\Omega_1^*d\Omega_3',
    \label{eq_1to4Tree}
\end{align}
where $(|\vec{p}^{\,\prime\prime}_{12}|, \Omega_{12}'')\equiv (|\vec{q}|, \Omega) $ is the three-momentum of the final-state (1,2) particle system in the c.m. frame of the initial state, $(|\vec{p}^*_{1}|, \Omega_{1}^*)$ is the three-momentum of particle 1 in the c.m. frame of particles 1 and 2, and $(|\vec{p}^{\,\prime}_3|, \Omega_3')$ is the three-momentum of particle 3 in the c.m. frame of particles 3 and 4. The integration regions of $\sqrt{s_{12}}$ and $\sqrt{s_{34}}$ are $[m_1+m_2,M-m_3-m_4]$ and $[m_3+m_4,M-\sqrt{s_{12}}]$, respectively.

\section{FOUR-BODY DECAY AMPLITUDES\label{Appendix:4body_amplitudes}}

In this section, we show all the amplitudes for the diagrams in Figs.~\ref{Fig.4bodypi0} and \ref{Fig.4bodypic} of the four-body $T_{cc}^{*} \to \pi D\pi D$ decays. 

\subsection{$T_{cc}^{*+}\to \pi^0 D^0\pi^0 D^+$ amplitudes}

We first consider the decay $T_{cc}^{*+}  \to \pi^0 D^0\pi^0 D^+$. The LO amplitude from the tree diagram in Fig.~\ref{Fig.4bodypi0Tree} reads
\begin{align}
i \mathcal{A}_{\text{Tree}}[T_{cc}^{*+} \to  \pi^0(p_1)D^0(p_2)\pi^0(p_3)D^+(p_4)]=&\, \frac{-i g_0 \bar{g}^2}{2\sqrt{2}F_{\pi}^2 m_{\pi^0}}\frac{1}{q^0-m_{D^{*0}}-\frac{\vec{q}^2}{2m_{D^{*0}}}+i\frac{\Gamma_{D^{*0}}}{2}}\nonumber\\
&\, \times \frac{1}{k^0-m_{D^{*+}}-\frac{\vec{k}^2}{2m_{D^{*+}}}+i\frac{\Gamma_{D^{*+}}}{2}}\epsilon_{ijk}\epsilon^i(T_{cc}^{*+})p_1^jp_3^k,
\end{align}
with $s_{12}=q^2$, $s_{34}=k^2$, $s_{ij}=(p_i+p_j)^2$, $i,j=1,...,4$. Here $p_1^j$ and $p_3^k$ are the three-momenta of the two $\pi^0$ in the final states in the $T_{cc}^{*}$ rest frame, respectively, and $q^{\mu}=(q^0, \vec{q})$, $k^{\mu}=(k^0, \vec{k})$ are the four-momenta of the 1,2 and 3,4 two-particle systems in the $T_{cc}^{*}$ rest frame, respectively.
Considering the crossed-channel effects of the two identical $\pi^0$ in the final states, we also have
\begin{align}
i \mathcal{A}_{\text{Tree}}[T_{cc}^{*+} \to \pi^0(p_3)D^0(p_2)\pi^0(p_1)D^+(p_4)]=&\, \frac{i g_0 \bar{g}^2}{2\sqrt{2}F_{\pi}^2 m_{\pi^0}}\frac{1}{l^0-m_{D^{*0}}-\frac{\vec{l}^2}{2m_{D^{*0}}}+i\frac{\Gamma_{D^{*0}}}{2}}\nonumber\\
&\, \times\frac{1}{t^0-m_{D^{*+}}-\frac{\vec{t}^2}{2m_{D^{*+}}}+i\frac{\Gamma_{D^{*+}}}{2}} \epsilon_{ijk}\epsilon^i(T_{cc}^{*+})p_1^jp_3^k,
\end{align}
where $s_{23}=l^2$, $s_{14}=t^2$, and $l^{\mu}=(l^0, \vec{l})$, $t^{\mu}=(t^0, \vec{t})$ are the four-momenta of the 2,3 and 1,4 two-particle systems in the $T_{cc}^{*}$ rest frame, respectively.

The LO amplitudes from the $D^{*+}D^0/D^{*0}D^+$ rescattering diagrams in Figs.~\ref{Fig.4bodypi0C0D1}, \ref{Fig.4bodypi0C0D1ex}, \ref{Fig.4bodypi0C0D2} and \ref{Fig.4bodypi0C0D2ex} read
\begin{align}
i \mathcal{A}_{b}[T_{cc}^{*+} \to \pi^0(p_1)D^0(p_2)\pi^0(p_3)D^+(p_4)]=&\, \frac{i g_0 \bar{g}^2C_{0D1}}{2\sqrt{2}F_{\pi}^2 m_{\pi^0}}\frac{1}{(p_3^0+p_4^0)-m_{D^{*+}}-\frac{(\vec{p}_3+\vec{p}_4)^2}{2m_{D^{*+}}}+i\frac{\Gamma_{D^{*+}}}{2}}\nonumber\\
&\, \times \epsilon_{ijk}\epsilon^i(T_{cc}^{*+})p_1^jp_3^kI(p_1),
\end{align}
\begin{align}
i \mathcal{A}_{b}[T_{cc}^{*+} \to \pi^0(p_3)D^0(p_2)\pi^0(p_1)D^+(p_4)]=&\, \frac{-i g_0 \bar{g}^2C_{0D1}}{2\sqrt{2}F_{\pi}^2 m_{\pi^0}}\frac{1}{(p_1^0+p_4^0)-m_{D^{*+}}-\frac{(\vec{p}_1+\vec{p}_4)^2}{2m_{D^{*+}}}+i\frac{\Gamma_{D^{*+}}}{2}}\nonumber\\
&\, \times \epsilon_{ijk}\epsilon^i(T_{cc}^{*+})p_1^jp_3^kI(p_3),
\end{align}
\begin{align}
i \mathcal{A}_{c}[T_{cc}^{*+} \to \pi^0(p_1)D^0(p_2)\pi^0(p_3)D^+(p_4)]=&\, \frac{i g_0 \bar{g}^2C_{0D1{\rm ex}}}{2\sqrt{2}F_{\pi}^2 m_{\pi^0}}\frac{1}{(p_3^0+p_4^0)-m_{D^{*+}}-\frac{(\vec{p}_3+\vec{p}_4)^2}{2m_{D^{*+}}}+i\frac{\Gamma_{D^{*+}}}{2}}\nonumber\\
&\, \times \epsilon_{ijk}\epsilon^i(T_{cc}^{*+})p_1^jp_3^kI(p_1),
\end{align}
\begin{align}
i \mathcal{A}_{c}[T_{cc}^{*+} \to \pi^0(p_3)D^0(p_2)\pi^0(p_1)D^+(p_4)]=&\, \frac{-i g_0 \bar{g}^2C_{0D1{\rm ex}}}{2\sqrt{2}F_{\pi}^2 m_{\pi^0}}\frac{1}{(p_1^0+p_4^0)-m_{D^{*+}}-\frac{(\vec{p}_1+\vec{p}_4)^2}{2m_{D^{*+}}}+i\frac{\Gamma_{D^{*+}}}{2}}\nonumber\\
&\, \times \epsilon_{ijk}\epsilon^i(T_{cc}^{*+})p_1^jp_3^kI(p_3),
\end{align}
\begin{align}
i \mathcal{A}_{d}[T_{cc}^{*+} \to \pi^0(p_1)D^0(p_2)\pi^0(p_3)D^+(p_4)]=&\, \frac{i g_0 \bar{g}^2C_{0D2}}{2\sqrt{2}F_{\pi}^2 m_{\pi^0}}\frac{1}{(p_1^0+p_2^0)-m_{D^{*0}}-\frac{(\vec{p}_1+\vec{p}_2)^2}{2m_{D^{*0}}}+i\frac{\Gamma_{D^{*0}}}{2}}\nonumber\\
&\, \times \epsilon_{ijk}\epsilon^i(T_{cc}^{*+})p_1^jp_3^kI(p_3),
\end{align}
\begin{align}
i \mathcal{A}_{d}[T_{cc}^{*+} \to \pi^0(p_3)D^0(p_2)\pi^0(p_1)D^+(p_4)]=&\, \frac{-i g_0 \bar{g}^2C_{0D2}}{2\sqrt{2}F_{\pi}^2 m_{\pi^0}}\frac{1}{(p_2^0+p_3^0)-m_{D^{*0}}-\frac{(\vec{p}_2+\vec{p}_3)^2}{2m_{D^{*0}}}+i\frac{\Gamma_{D^{*0}}}{2}}\nonumber\\
&\, \times \epsilon_{ijk}\epsilon^i(T_{cc}^{*+})p_1^jp_3^kI(p_1),
\end{align}
\begin{align}
i \mathcal{A}_{e}[T_{cc}^{*+} \to \pi^0(p_1)D^0(p_2)\pi^0(p_3)D^+(p_4)]=&\, \frac{i g_0 \bar{g}^2C_{0D2{\rm ex}}}{2\sqrt{2}F_{\pi}^2 m_{\pi^0}}\frac{1}{(p_1^0+p_2^0)-m_{D^{*0}}-\frac{(\vec{p}_1+\vec{p}_2)^2}{2m_{D^{*0}}}+i\frac{\Gamma_{D^{*0}}}{2}}\nonumber\\
&\, \times \epsilon_{ijk}\epsilon^i(T_{cc}^{*+})p_1^jp_3^kI(p_3),
\end{align}
\begin{align}
i \mathcal{A}_{e}[T_{cc}^{*+} \to \pi^0(p_3)D^0(p_2)\pi^0(p_1)D^+(p_4)]=&\, \frac{-i g_0 \bar{g}^2C_{0D2{\rm ex}}}{2\sqrt{2}F_{\pi}^2 m_{\pi^0}}\frac{1}{(p_2^0+p_3^0)-m_{D^{*0}}-\frac{(\vec{p}_2+\vec{p}_3)^2}{2m_{D^{*0}}}+i\frac{\Gamma_{D^{*0}}}{2}}\nonumber\\
&\, \times \epsilon_{ijk}\epsilon^i(T_{cc}^{*+})p_1^jp_3^kI(p_1).
\end{align}
 where $p_1^{\mu}=(p_1^0, \vec{p}_1)$, $p_2^{\mu}=(p_2^0, \vec{p}_2)$, $p_3^{\mu}=(p_3^0, \vec{p}_3)$, and $p_4^{\mu}=(p_4^0, \vec{p}_4)$ are the four-momenta of the four final-state particles in the rest frame of the $T_{cc}^{*}$, respectively. 
The masses $m_1$, $m_2$, and $m_3$ in the loop integrals $I(p_i)$ are taken to be the masses of $D^{*0}$, $D^{*+}$, and $D^0$ in Figs.~\ref{Fig.4bodypi0C0D1} and \ref{Fig.4bodypi0C0D2ex}, and the masses of $D^{*+}$, $D^{*0}$, and $D^+$ in Figs.~\ref{Fig.4bodypi0C0D1ex} and \ref{Fig.4bodypi0C0D2}, respectively.

The NLO amplitudes from the $D^*\pi$ rescattering diagrams considering the crossed-channel effects in Figs.~\ref{Fig.4bodypi0Cpi1}, \ref{Fig.4bodypi0Cpi1ex}, \ref{Fig.4bodypi0Cpi2}, and \ref{Fig.4bodypi0Cpi2ex} are
\begin{align}
i \mathcal{A}_{f}[T_{cc}^{*+} \to \pi^0(p_1)D^0(p_2)\pi^0(p_3)D^+(p_4)]=&\, \frac{-i g_0 \bar{g}^2C_{\pi1}}{4\sqrt{2}F_{\pi}^2 m_{\pi^0}^2}\frac{1}{(p_3^0+p_4^0)-m_{D^{*+}}-\frac{(\vec{p}_3+\vec{p}_4)^2}{2m_{D^{*+}}}+i\frac{\Gamma_{D^{*+}}}{2}}\nonumber\\
&\, \times \epsilon_{ijk}\epsilon^i(T_{cc}^{*+})p_2^jp_3^k\left[I^{(1)}(p_2)+I(p_2)\right],
\end{align}
\begin{align}
i \mathcal{A}_{f}[T_{cc}^{*+} \to \pi^0(p_3)D^0(p_2)\pi^0(p_1)D^+(p_4)]=&\, \frac{i g_0 \bar{g}^2C_{\pi1}}{4\sqrt{2}F_{\pi}^2 m_{\pi^0}^2}\frac{1}{(p_1^0+p_4^0)-m_{D^{*+}}-\frac{(\vec{p}_1+\vec{p}_4)^2}{2m_{D^{*+}}}+i\frac{\Gamma_{D^{*+}}}{2}}\nonumber\\
&\, \times \epsilon_{ijk}\epsilon^i(T_{cc}^{*+})p_1^jp_2^k\left[I^{(1)}(p_2)+I(p_2)\right],
\end{align}
\begin{align}
i \mathcal{A}_{g}[T_{cc}^{*+} \to \pi^0(p_1)D^0(p_2)\pi^0(p_3)D^+(p_4)]=&\, \frac{-\sqrt{2}i g_0 \bar{g}^2C_{\pi1{\rm ex}}}{4\sqrt{2}F_{\pi}^2 m_{\pi^0} m_{\pi^+}}\frac{1}{(p_3^0+p_4^0)-m_{D^{*+}}-\frac{(\vec{p}_3+\vec{p}_4)^2}{2m_{D^{*+}}}+i\frac{\Gamma_{D^{*+}}}{2}}\nonumber\\
&\, \times \epsilon_{ijk}\epsilon^i(T_{cc}^{*+})p_2^jp_3^k\left[I^{(1)}(p_2)-I(p_2)\right],
\end{align}
\begin{align}
i \mathcal{A}_{g}[T_{cc}^{*+} \to \pi^0(p_3)D^0(p_2)\pi^0(p_1)D^+(p_4)]=&\, \frac{\sqrt{2}i g_0 \bar{g}^2C_{\pi1{\rm ex}}}{4\sqrt{2}F_{\pi}^2 m_{\pi^0} m_{\pi^+}}\frac{1}{(p_1^0+p_4^0)-m_{D^{*+}}-\frac{(\vec{p}_1+\vec{p}_4)^2}{2m_{D^{*+}}}+i\frac{\Gamma_{D^{*+}}}{2}}\nonumber\\
&\, \times \epsilon_{ijk}\epsilon^i(T_{cc}^{*+})p_1^jp_2^k\left[I^{(1)}(p_2)-I(p_2)\right],
\end{align}
\begin{align}
i \mathcal{A}_{h}[T_{cc}^{*+} \to \pi^0(p_1)D^0(p_2)\pi^0(p_3)D^+(p_4)]=&\, \frac{i g_0 \bar{g}^2C_{\pi2}}{4\sqrt{2}F_{\pi}^2 m_{\pi^0}^2}\frac{1}{(p_1^0+p_2^0)-m_{D^{*0}}-\frac{(\vec{p}_1+\vec{p}_2)^2}{2m_{D^{*0}}}+i\frac{\Gamma_{D^{*0}}}{2}}\nonumber\\
&\, \times \epsilon_{ijk}\epsilon^i(T_{cc}^{*+})p_1^jp_4^k\left[I^{(1)}(p_4)-I(p_4)\right],
\end{align}
\begin{align}
i \mathcal{A}_{h}[T_{cc}^{*+} \to \pi^0(p_3)D^0(p_2)\pi^0(p_1)D^+(p_4)]=&\, \frac{i g_0 \bar{g}^2C_{\pi2}}{4\sqrt{2}F_{\pi}^2 m_{\pi^0}^2}\frac{1}{(p_2^0+p_3^0)-m_{D^{*0}}-\frac{(\vec{p}_2+\vec{p}_3)^2}{2m_{D^{*0}}}+i\frac{\Gamma_{D^{*0}}}{2}}\nonumber\\
&\, \times \epsilon_{ijk}\epsilon^i(T_{cc}^{*+})p_3^jp_4^k\left[I^{(1)}(p_4)-I(p_4)\right],
\end{align}
\begin{align}
i \mathcal{A}_{i}[T_{cc}^{*+} \to \pi^0(p_1)D^0(p_2)\pi^0(p_3)D^+(p_4)]=&\, \frac{-\sqrt{2}i g_0 \bar{g}^2C_{\pi2{\rm ex}}}{4\sqrt{2}F_{\pi}^2 m_{\pi^-} m_{\pi^0}}\frac{1}{(p_1^0+p_2^0)-m_{D^{*0}}-\frac{(\vec{p}_1+\vec{p}_2)^2}{2m_{D^{*0}}}+i\frac{\Gamma_{D^{*0}}}{2}}\nonumber\\
&\, \times \epsilon_{ijk}\epsilon^i(T_{cc}^{*+})p_1^jp_4^k\left[I^{(1)}(p_4)+I(p_4)\right],
\end{align}
\begin{align}
i \mathcal{A}_{i}[T_{cc}^{*+} \to \pi^0(p_3)D^0(p_2)\pi^0(p_1)D^+(p_4)]=&\, \frac{-\sqrt{2}i g_0 \bar{g}^2C_{\pi2{\rm ex}}}{4\sqrt{2}F_{\pi}^2 m_{\pi^-} m_{\pi^0}}\frac{1}{(p_2^0+p_3^0)-m_{D^{*0}}-\frac{(\vec{p}_2+\vec{p}_3)^2}{2m_{D^{*0}}}+i\frac{\Gamma_{D^{*0}}}{2}}\nonumber\\
&\, \times \epsilon_{ijk}\epsilon^i(T_{cc}^{*+})p_3^jp_4^k\left[I^{(1)}(p_4)+I(p_4)\right],
\end{align}
where the masses $m_1$, $m_2$, and $m_3$ in the loop integrals $I^{(1)}(p_i)$ and $I(p_i)$ are taken to be the masses of $D^{*0}$, $D^{*+}$, and $\pi^0$ in Fig.~\ref{Fig.4bodypi0Cpi1}, the masses of $D^{*+}$, $D^{*0}$, and $\pi^+$ in Fig.~\ref{Fig.4bodypi0Cpi1ex}, the masses of $D^{*+}$, $D^{*0}$, and $\pi^0$ in Fig.~\ref{Fig.4bodypi0Cpi2}, and the masses of $D^{*0}$, $D^{*+}$, and $\pi^-$ in Fig.~\ref{Fig.4bodypi0Cpi2ex}, respectively.

\subsection{$T_{cc}^{*+} \to \pi^0 D^0 \pi^+ D^0$ amplitudes}

For the decay $T_{cc}^{*+} \to D^0 D^0\pi^0 \pi^+$, the LO amplitude from the tree diagram in Fig.~\ref{Fig.4bodypi+Tree} reads
\begin{align}
i \mathcal{A}_{a}[T_{cc}^{*+} \to \pi^0(p_1)D^0(p_2)\pi^+(p_3)D^0(p_4)]=&\, \frac{i g_0 \bar{g}^2}{2F_{\pi}^2 \sqrt{m_{\pi^0} m_{\pi^+}}}\frac{1}{q^0-m_{D^{*0}}-\frac{\vec{q}^2}{2m_{D^{*0}}}+i\frac{\Gamma_{D^{*0}}}{2}}\nonumber\\
&\, \times \frac{1}{k^0-m_{D^{*+}}-\frac{\vec{k}^2}{2m_{D^{*+}}}+i\frac{\Gamma_{D^{*+}}}{2}}\epsilon_{ijk}\epsilon^i(T_{cc}^{*+})p_1^jp_3^k,
\end{align}
and the other amplitude from the crossed-channel effects of the final-state identical $D^0$ particles is
\begin{align}
i \mathcal{A}_{a}[T_{cc}^{*+} \to \pi^0(p_1)D^0(p_4)\pi^+(p_3)D^0(p_2)]=&\, \frac{i g_0 \bar{g}^2}{2F_{\pi}^2 \sqrt{m_{\pi^0} m_{\pi^+}}}\frac{1}{t^0-m_{D^{*0}}-\frac{\vec{t}^2}{2m_{D^{*0}}}+i\frac{\Gamma_{D^{*0}}}{2}}\nonumber\\
&\, \times \frac{1}{l^0-m_{D^{*+}}-\frac{\vec{l}^2}{2m_{D^{*+}}}+i\frac{\Gamma_{D^{*+}}}{2}}\epsilon_{ijk}\epsilon^i(T_{cc}^{*+})p_1^jp_3^k. 
\end{align}

The LO amplitudes from the $D^{*+}D^0/D^{*0}D^+$ rescattering diagrams including the crossed-channel contributions in Figs.~\ref{Fig.4bodypi+C0D1} and \ref{Fig.4bodypi+C0D1ex} are
\begin{align}
i \mathcal{A}_{b}[T_{cc}^{*+} \to \pi^0(p_1)D^0(p_2)\pi^+(p_3)D^0(p_4)]=&\, \frac{-i g_0 \bar{g}^2 C_{0D1}}{2F_{\pi}^2 \sqrt{m_{\pi^0} m_{\pi^+}}}\frac{1}{(p_3^0+p_4^0)-m_{D^{*+}}-\frac{(\vec{p}_3+\vec{p}_4)^2}{2m_{D^{*+}}}+i\frac{\Gamma_{D^{*+}}}{2}}\nonumber\\
&\, \times \epsilon_{ijk}\epsilon^i(T_{cc}^{*+})p_1^jp_3^kI(p_1),
\end{align}
\begin{align}
i \mathcal{A}_{b}[T_{cc}^{*+} \to \pi^0(p_1)D^0(p_4)\pi^+(p_3)D^0(p_2)]=&\, \frac{-i g_0 \bar{g}^2 C_{0D1}}{2F_{\pi}^2 \sqrt{m_{\pi^0} m_{\pi^+}}}\frac{1}{(p_2^0+p_3^0)-m_{D^{*+}}-\frac{(\vec{p}_2+\vec{p}_3)^2}{2m_{D^{*+}}}+i\frac{\Gamma_{D^{*+}}}{2}}\nonumber\\
&\, \times \epsilon_{ijk}\epsilon^i(T_{cc}^{*+})p_1^jp_3^kI(p_1),
\end{align}
\begin{align}
i \mathcal{A}_{c}[T_{cc}^{*+} \to \pi^0(p_1)D^0(p_2)\pi^+(p_3)D^0(p_4)]=&\, \frac{-i g_0 \bar{g}^2 C_{0D1{\rm ex}}}{2F_{\pi}^2 \sqrt{m_{\pi^0} m_{\pi^+}}}\frac{1}{(p_3^0+p_4^0)-m_{D^{*+}}-\frac{(\vec{p}_3+\vec{p}_4)^2}{2m_{D^{*+}}}+i\frac{\Gamma_{D^{*+}}}{2}}\nonumber\\
&\, \times \epsilon_{ijk}\epsilon^i(T_{cc}^{*+})p_1^jp_3^kI(p_1),
\end{align}
\begin{align}
i \mathcal{A}_{c}[T_{cc}^{*+} \to \pi^0(p_1)D^0(p_4)\pi^+(p_3)D^0(p_2)]=&\, \frac{-i g_0 \bar{g}^2 C_{0D1{\rm ex}}}{2F_{\pi}^2 \sqrt{m_{\pi^0} m_{\pi^+}}}\frac{1}{(p_2^0+p_3^0)-m_{D^{*+}}-\frac{(\vec{p}_2+\vec{p}_3)^2}{2m_{D^{*+}}}+i\frac{\Gamma_{D^{*+}}}{2}}\nonumber\\
&\, \times \epsilon_{ijk}\epsilon^i(T_{cc}^{*+})p_1^jp_3^kI(p_1).
\end{align}
Since the isospin of $D^{\ast 0}D^0$ is 1 and no isospin vector double-charm tetraquark around the $D^*D$ threshold has been found,  the $D^{\ast 0}D^0$ rescattering in Figs.~\ref{Fig.4bodypi+C0D3} is not promoted to LO.

The NLO amplitudes from the $D^*\pi$ rescattering diagrams considering the crossed-channel effects in Figs.~\ref{Fig.4bodypi+Cpi1}, \ref{Fig.4bodypi+Cpi1ex}, \ref{Fig.4bodypi+Cpi3} and \ref{Fig.4bodypi+Cpi3ex} are
\begin{align}
i \mathcal{A}_{e}[T_{cc}^{*+} \to \pi^0(p_1)D^0(p_2)\pi^+(p_3)D^0(p_4)]=&\, \frac{\sqrt{2}i g_0 \bar{g}^2C_{\pi1}}{4\sqrt{2}F_{\pi}^2 m_{\pi^0}\sqrt{m_{\pi^0}m_{\pi^+}}}\frac{1}{(p_3^0+p_4^0)-m_{D^{*+}}-\frac{(\vec{p}_3+\vec{p}_4)^2}{2m_{D^{*+}}}+i\frac{\Gamma_{D^{*+}}}{2}}\nonumber\\
&\, \times \epsilon_{ijk}\epsilon^i(T_{cc}^{*+})p_2^jp_3^k\left[I^{(1)}(p_2)+I(p_2)\right],
\end{align}
\begin{align}
i \mathcal{A}_{e}[T_{cc}^{*+} \to \pi^0(p_1)D^0(p_4)\pi^+(p_3)D^0(p_2)]=&\, \frac{-\sqrt{2}i g_0 \bar{g}^2C_{\pi1}}{4\sqrt{2}F_{\pi}^2 m_{\pi^0}\sqrt{m_{\pi^0}m_{\pi^+}}}\frac{1}{(p_2^0+p_3^0)-m_{D^{*+}}-\frac{(\vec{p}_2+\vec{p}_3)^2}{2m_{D^{*+}}}+i\frac{\Gamma_{D^{*+}}}{2}}\nonumber\\
&\, \times \epsilon_{ijk}\epsilon^i(T_{cc}^{*+})p_3^jp_4^k\left[I^{(1)}(p_4)+I(p_4)\right],
\end{align}
\begin{align}
i \mathcal{A}_{f}[T_{cc}^{*+} \to \pi^0(p_1)D^0(p_2)\pi^+(p_3)D^0(p_4)]=&\, \frac{2 i g_0 \bar{g}^2C_{\pi1{\rm ex}}}{4\sqrt{2}F_{\pi}^2 m_{\pi^+}\sqrt{m_{\pi^0}m_{\pi^+}}}\frac{1}{(p_3^0+p_4^0)-m_{D^{*+}}-\frac{(\vec{p}_3+\vec{p}_4)^2}{2m_{D^{*+}}}+i\frac{\Gamma_{D^{*+}}}{2}}\nonumber\\
&\, \times \epsilon_{ijk}\epsilon^i(T_{cc}^{*+})p_2^jp_3^k\left[I^{(1)}(p_2)-I(p_2)\right],
\end{align}
\begin{align}
i \mathcal{A}_{f}[T_{cc}^{*+} \to \pi^0(p_1)D^0(p_4)\pi^+(p_3)D^0(p_2)]=&\, \frac{-2 i g_0 \bar{g}^2C_{\pi1{\rm ex}}}{4\sqrt{2}F_{\pi}^2 m_{\pi^+}\sqrt{m_{\pi^0}m_{\pi^+}}}\frac{1}{(p_2^0+p_3^0)-m_{D^{*+}}-\frac{(\vec{p}_2+\vec{p}_3)^2}{2m_{D^{*+}}}+i\frac{\Gamma_{D^{*+}}}{2}}\nonumber\\
&\, \times \epsilon_{ijk}\epsilon^i(T_{cc}^{*+})p_3^jp_4^k\left[I^{(1)}(p_4)-I(p_4)\right],
\end{align}
\begin{align}
i \mathcal{A}_{g}[T_{cc}^{*+} \to \pi^0(p_1)D^0(p_2)\pi^+(p_3)D^0(p_4)]=&\, \frac{-\sqrt{2}i g_0 \bar{g}^2C_{\pi3}}{4\sqrt{2}F_{\pi}^2 m_{\pi^+}\sqrt{m_{\pi^0}m_{\pi^+}}}\frac{1}{(p_1^0+p_2^0)-m_{D^{*0}}-\frac{(\vec{p}_1+\vec{p}_2)^2}{2m_{D^{*0}}}+i\frac{\Gamma_{D^{*0}}}{2}}\nonumber\\
&\, \times \epsilon_{ijk}\epsilon^i(T_{cc}^{*+})p_1^jp_4^k\left[I^{(1)}(p_4)-I(p_4)\right],
\end{align}
\begin{align}
i \mathcal{A}_{g}[T_{cc}^{*+} \to \pi^0(p_1)D^0(p_4)\pi^+(p_3)D^0(p_2)]=&\, \frac{-\sqrt{2}i g_0 \bar{g}^2C_{\pi3}}{4\sqrt{2}F_{\pi}^2 m_{\pi^+}\sqrt{m_{\pi^0}m_{\pi^+}}}\frac{1}{(p_1^0+p_4^0)-m_{D^{*0}}-\frac{(\vec{p}_1+\vec{p}_4)^2}{2m_{D^{*0}}}+i\frac{\Gamma_{D^{*0}}}{2}}\nonumber\\
&\, \times \epsilon_{ijk}\epsilon^i(T_{cc}^{*+})p_1^jp_2^k\left[I^{(1)}(p_2)-I(p_2)\right],
\end{align}
\begin{align}
i \mathcal{A}_{h}[T_{cc}^{*+} \to \pi^0(p_1)D^0(p_2)\pi^+(p_3)D^0(p_4)]=&\, \frac{-i g_0 \bar{g}^2C_{\pi3ex}}{4\sqrt{2}F_{\pi}^2 m_{\pi^0}\sqrt{m_{\pi^0}m_{\pi^+}}}\frac{1}{(p_1^0+p_2^0)-m_{D^{*0}}-\frac{(\vec{p}_1+\vec{p}_2)^2}{2m_{D^{*0}}}+i\frac{\Gamma_{D^{*0}}}{2}}\nonumber\\
&\, \times \epsilon_{ijk}\epsilon^i(T_{cc}^{*+})p_1^jp_4^k\left[I^{(1)}(p_4)+I(p_4)\right],
\end{align}
\begin{align}
i \mathcal{A}_{h}[T_{cc}^{*+} \to \pi^0(p_1)D^0(p_4)\pi^+(p_3)D^0(p_2)]=&\, \frac{-i g_0 \bar{g}^2C_{\pi3ex}}{4\sqrt{2}F_{\pi}^2 m_{\pi^0}\sqrt{m_{\pi^0}m_{\pi^+}}}\frac{1}{(p_1^0+p_4^0)-m_{D^{*0}}-\frac{(\vec{p}_1+\vec{p}_4)^2}{2m_{D^{*0}}}+i\frac{\Gamma_{D^{*0}}}{2}}\nonumber\\
&\, \times \epsilon_{ijk}\epsilon^i(T_{cc}^{*+})p_1^jp_2^k\left[I^{(1)}(p_2)+I(p_2)\right].
\end{align}

\bibliography{TccStar.bib}

\begin{thebibliography}{46}%
\makeatletter
\providecommand \@ifxundefined [1]{%
 \@ifx{#1\undefined}
}%
\providecommand \@ifnum [1]{%
 \ifnum #1\expandafter \@firstoftwo
 \else \expandafter \@secondoftwo
 \fi
}%
\providecommand \@ifx [1]{%
 \ifx #1\expandafter \@firstoftwo
 \else \expandafter \@secondoftwo
 \fi
}%
\providecommand \natexlab [1]{#1}%
\providecommand \enquote  [1]{``#1''}%
\providecommand \bibnamefont  [1]{#1}%
\providecommand \bibfnamefont [1]{#1}%
\providecommand \citenamefont [1]{#1}%
\providecommand \href@noop [0]{\@secondoftwo}%
\providecommand \href [0]{\begingroup \@sanitize@url \@href}%
\providecommand \@href[1]{\@@startlink{#1}\@@href}%
\providecommand \@@href[1]{\endgroup#1\@@endlink}%
\providecommand \@sanitize@url [0]{\catcode `\\12\catcode `\$12\catcode `\&12\catcode `\#12\catcode `\^12\catcode `\_12\catcode `\%12\relax}%
\providecommand \@@startlink[1]{}%
\providecommand \@@endlink[0]{}%
\providecommand \url  [0]{\begingroup\@sanitize@url \@url }%
\providecommand \@url [1]{\endgroup\@href {#1}{\urlprefix }}%
\providecommand \urlprefix  [0]{URL }%
\providecommand \Eprint [0]{\href }%
\providecommand \doibase [0]{http://dx.doi.org/}%
\providecommand \selectlanguage [0]{\@gobble}%
\providecommand \bibinfo  [0]{\@secondoftwo}%
\providecommand \bibfield  [0]{\@secondoftwo}%
\providecommand \translation [1]{[#1]}%
\providecommand \BibitemOpen [0]{}%
\providecommand \bibitemStop [0]{}%
\providecommand \bibitemNoStop [0]{.\EOS\space}%
\providecommand \EOS [0]{\spacefactor3000\relax}%
\providecommand \BibitemShut  [1]{\csname bibitem#1\endcsname}%
\let\auto@bib@innerbib\@empty
\bibitem [{\citenamefont {Aaij}\ \emph {et~al.}(2022{\natexlab{a}})\citenamefont {Aaij} \emph {et~al.}}]{LHCb:2021vvq}%
  \BibitemOpen
  \bibfield  {author} {\bibinfo {author} {\bibfnamefont {R.}~\bibnamefont {Aaij}} \emph {et~al.} (\bibinfo {collaboration} {LHCb}),\ }\href {\doibase 10.1038/s41567-022-01614-y} {\bibfield  {journal} {\bibinfo  {journal} {Nature Phys.}\ }\textbf {\bibinfo {volume} {18}},\ \bibinfo {pages} {751} (\bibinfo {year} {2022}{\natexlab{a}})},\ \Eprint {http://arxiv.org/abs/2109.01038} {arXiv:2109.01038 [hep-ex]} \BibitemShut {NoStop}%
\bibitem [{\citenamefont {Aaij}\ \emph {et~al.}(2022{\natexlab{b}})\citenamefont {Aaij} \emph {et~al.}}]{LHCb:2021auc}%
  \BibitemOpen
  \bibfield  {author} {\bibinfo {author} {\bibfnamefont {R.}~\bibnamefont {Aaij}} \emph {et~al.} (\bibinfo {collaboration} {LHCb}),\ }\href {\doibase 10.1038/s41467-022-30206-w} {\bibfield  {journal} {\bibinfo  {journal} {Nature Commun.}\ }\textbf {\bibinfo {volume} {13}},\ \bibinfo {pages} {3351} (\bibinfo {year} {2022}{\natexlab{b}})},\ \Eprint {http://arxiv.org/abs/2109.01056} {arXiv:2109.01056 [hep-ex]} \BibitemShut {NoStop}%
\bibitem [{\citenamefont {Du}\ \emph {et~al.}(2022)\citenamefont {Du}, \citenamefont {Baru}, \citenamefont {Dong}, \citenamefont {Filin}, \citenamefont {Guo}, \citenamefont {Hanhart}, \citenamefont {Nefediev}, \citenamefont {Nieves},\ and\ \citenamefont {Wang}}]{Du:2021zzh}%
  \BibitemOpen
  \bibfield  {author} {\bibinfo {author} {\bibfnamefont {M.-L.}\ \bibnamefont {Du}}, \bibinfo {author} {\bibfnamefont {V.}~\bibnamefont {Baru}}, \bibinfo {author} {\bibfnamefont {X.-K.}\ \bibnamefont {Dong}}, \bibinfo {author} {\bibfnamefont {A.}~\bibnamefont {Filin}}, \bibinfo {author} {\bibfnamefont {F.-K.}\ \bibnamefont {Guo}}, \bibinfo {author} {\bibfnamefont {C.}~\bibnamefont {Hanhart}}, \bibinfo {author} {\bibfnamefont {A.}~\bibnamefont {Nefediev}}, \bibinfo {author} {\bibfnamefont {J.}~\bibnamefont {Nieves}}, \ and\ \bibinfo {author} {\bibfnamefont {Q.}~\bibnamefont {Wang}},\ }\href {\doibase 10.1103/PhysRevD.105.014024} {\bibfield  {journal} {\bibinfo  {journal} {Phys. Rev. D}\ }\textbf {\bibinfo {volume} {105}},\ \bibinfo {pages} {014024} (\bibinfo {year} {2022})},\ \Eprint {http://arxiv.org/abs/2110.13765} {arXiv:2110.13765 [hep-ph]} \BibitemShut {NoStop}%
\bibitem [{\citenamefont {Dong}\ \emph {et~al.}(2021)\citenamefont {Dong}, \citenamefont {Guo},\ and\ \citenamefont {Zou}}]{Dong:2021bvy}%
  \BibitemOpen
  \bibfield  {author} {\bibinfo {author} {\bibfnamefont {X.-K.}\ \bibnamefont {Dong}}, \bibinfo {author} {\bibfnamefont {F.-K.}\ \bibnamefont {Guo}}, \ and\ \bibinfo {author} {\bibfnamefont {B.-S.}\ \bibnamefont {Zou}},\ }\href {\doibase 10.1088/1572-9494/ac27a2} {\bibfield  {journal} {\bibinfo  {journal} {Commun. Theor. Phys.}\ }\textbf {\bibinfo {volume} {73}},\ \bibinfo {pages} {125201} (\bibinfo {year} {2021})},\ \Eprint {http://arxiv.org/abs/2108.02673} {arXiv:2108.02673 [hep-ph]} \BibitemShut {NoStop}%
\bibitem [{\citenamefont {Chen}\ \emph {et~al.}(2022)\citenamefont {Chen}, \citenamefont {Chen}, \citenamefont {Liu}, \citenamefont {Liu},\ and\ \citenamefont {Zhu}}]{Chen:2022asf}%
  \BibitemOpen
  \bibfield  {author} {\bibinfo {author} {\bibfnamefont {H.-X.}\ \bibnamefont {Chen}}, \bibinfo {author} {\bibfnamefont {W.}~\bibnamefont {Chen}}, \bibinfo {author} {\bibfnamefont {X.}~\bibnamefont {Liu}}, \bibinfo {author} {\bibfnamefont {Y.-R.}\ \bibnamefont {Liu}}, \ and\ \bibinfo {author} {\bibfnamefont {S.-L.}\ \bibnamefont {Zhu}},\ }\href@noop {} {\  (\bibinfo {year} {2022})},\ \Eprint {http://arxiv.org/abs/2204.02649} {arXiv:2204.02649 [hep-ph]} \BibitemShut {NoStop}%
\bibitem [{\citenamefont {Feijoo}\ \emph {et~al.}(2021)\citenamefont {Feijoo}, \citenamefont {Liang},\ and\ \citenamefont {Oset}}]{Feijoo:2021ppq}%
  \BibitemOpen
  \bibfield  {author} {\bibinfo {author} {\bibfnamefont {A.}~\bibnamefont {Feijoo}}, \bibinfo {author} {\bibfnamefont {W.~H.}\ \bibnamefont {Liang}}, \ and\ \bibinfo {author} {\bibfnamefont {E.}~\bibnamefont {Oset}},\ }\href {\doibase 10.1103/PhysRevD.104.114015} {\bibfield  {journal} {\bibinfo  {journal} {Phys. Rev. D}\ }\textbf {\bibinfo {volume} {104}},\ \bibinfo {pages} {114015} (\bibinfo {year} {2021})},\ \Eprint {http://arxiv.org/abs/2108.02730} {arXiv:2108.02730 [hep-ph]} \BibitemShut {NoStop}%
\bibitem [{\citenamefont {Meng}\ \emph {et~al.}(2021)\citenamefont {Meng}, \citenamefont {Wang}, \citenamefont {Wang},\ and\ \citenamefont {Zhu}}]{Meng:2021jnw}%
  \BibitemOpen
  \bibfield  {author} {\bibinfo {author} {\bibfnamefont {L.}~\bibnamefont {Meng}}, \bibinfo {author} {\bibfnamefont {G.-J.}\ \bibnamefont {Wang}}, \bibinfo {author} {\bibfnamefont {B.}~\bibnamefont {Wang}}, \ and\ \bibinfo {author} {\bibfnamefont {S.-L.}\ \bibnamefont {Zhu}},\ }\href {\doibase 10.1103/PhysRevD.104.L051502} {\bibfield  {journal} {\bibinfo  {journal} {Phys. Rev. D}\ }\textbf {\bibinfo {volume} {104}},\ \bibinfo {pages} {051502} (\bibinfo {year} {2021})},\ \Eprint {http://arxiv.org/abs/2107.14784} {arXiv:2107.14784 [hep-ph]} \BibitemShut {NoStop}%
\bibitem [{\citenamefont {Ling}\ \emph {et~al.}(2022)\citenamefont {Ling}, \citenamefont {Liu}, \citenamefont {Geng}, \citenamefont {Wang},\ and\ \citenamefont {Xie}}]{Ling:2021bir}%
  \BibitemOpen
  \bibfield  {author} {\bibinfo {author} {\bibfnamefont {X.-Z.}\ \bibnamefont {Ling}}, \bibinfo {author} {\bibfnamefont {M.-Z.}\ \bibnamefont {Liu}}, \bibinfo {author} {\bibfnamefont {L.-S.}\ \bibnamefont {Geng}}, \bibinfo {author} {\bibfnamefont {E.}~\bibnamefont {Wang}}, \ and\ \bibinfo {author} {\bibfnamefont {J.-J.}\ \bibnamefont {Xie}},\ }\href {\doibase 10.1016/j.physletb.2022.136897} {\bibfield  {journal} {\bibinfo  {journal} {Phys. Lett. B}\ }\textbf {\bibinfo {volume} {826}},\ \bibinfo {pages} {136897} (\bibinfo {year} {2022})},\ \Eprint {http://arxiv.org/abs/2108.00947} {arXiv:2108.00947 [hep-ph]} \BibitemShut {NoStop}%
\bibitem [{\citenamefont {Isgur}\ and\ \citenamefont {Wise}(1989)}]{Isgur:1989vq}%
  \BibitemOpen
  \bibfield  {author} {\bibinfo {author} {\bibfnamefont {N.}~\bibnamefont {Isgur}}\ and\ \bibinfo {author} {\bibfnamefont {M.~B.}\ \bibnamefont {Wise}},\ }\href {\doibase 10.1016/0370-2693(89)90566-2} {\bibfield  {journal} {\bibinfo  {journal} {Phys. Lett. B}\ }\textbf {\bibinfo {volume} {232}},\ \bibinfo {pages} {113} (\bibinfo {year} {1989})}\BibitemShut {NoStop}%
\bibitem [{\citenamefont {Neubert}(1994)}]{Neubert:1993mb}%
  \BibitemOpen
  \bibfield  {author} {\bibinfo {author} {\bibfnamefont {M.}~\bibnamefont {Neubert}},\ }\href {\doibase 10.1016/0370-1573(94)90091-4} {\bibfield  {journal} {\bibinfo  {journal} {Phys. Rept.}\ }\textbf {\bibinfo {volume} {245}},\ \bibinfo {pages} {259} (\bibinfo {year} {1994})},\ \Eprint {http://arxiv.org/abs/hep-ph/9306320} {arXiv:hep-ph/9306320} \BibitemShut {NoStop}%
\bibitem [{\citenamefont {Manohar}\ and\ \citenamefont {Wise}(2000)}]{Manohar:2000dt}%
  \BibitemOpen
  \bibfield  {author} {\bibinfo {author} {\bibfnamefont {A.~V.}\ \bibnamefont {Manohar}}\ and\ \bibinfo {author} {\bibfnamefont {M.~B.}\ \bibnamefont {Wise}},\ }\href@noop {} {\emph {\bibinfo {title} {{Heavy quark physics}}}},\ Vol.~\bibinfo {volume} {10}\ (\bibinfo {year} {2000})\BibitemShut {NoStop}%
\bibitem [{\citenamefont {Albaladejo}(2022)}]{Albaladejo:2021vln}%
  \BibitemOpen
  \bibfield  {author} {\bibinfo {author} {\bibfnamefont {M.}~\bibnamefont {Albaladejo}},\ }\href {\doibase 10.1016/j.physletb.2022.137052} {\bibfield  {journal} {\bibinfo  {journal} {Phys. Lett. B}\ }\textbf {\bibinfo {volume} {829}},\ \bibinfo {pages} {137052} (\bibinfo {year} {2022})},\ \Eprint {http://arxiv.org/abs/2110.02944} {arXiv:2110.02944 [hep-ph]} \BibitemShut {NoStop}%
\bibitem [{\citenamefont {Fleming}\ \emph {et~al.}(2007)\citenamefont {Fleming}, \citenamefont {Kusunoki}, \citenamefont {Mehen},\ and\ \citenamefont {van Kolck}}]{Fleming:2007rp}%
  \BibitemOpen
  \bibfield  {author} {\bibinfo {author} {\bibfnamefont {S.}~\bibnamefont {Fleming}}, \bibinfo {author} {\bibfnamefont {M.}~\bibnamefont {Kusunoki}}, \bibinfo {author} {\bibfnamefont {T.}~\bibnamefont {Mehen}}, \ and\ \bibinfo {author} {\bibfnamefont {U.}~\bibnamefont {van Kolck}},\ }\href {\doibase 10.1103/PhysRevD.76.034006} {\bibfield  {journal} {\bibinfo  {journal} {Phys. Rev. D}\ }\textbf {\bibinfo {volume} {76}},\ \bibinfo {pages} {034006} (\bibinfo {year} {2007})},\ \Eprint {http://arxiv.org/abs/hep-ph/0703168} {arXiv:hep-ph/0703168} \BibitemShut {NoStop}%
\bibitem [{\citenamefont {Fleming}\ and\ \citenamefont {Mehen}(2008)}]{Fleming:2008yn}%
  \BibitemOpen
  \bibfield  {author} {\bibinfo {author} {\bibfnamefont {S.}~\bibnamefont {Fleming}}\ and\ \bibinfo {author} {\bibfnamefont {T.}~\bibnamefont {Mehen}},\ }\href {\doibase 10.1103/PhysRevD.78.094019} {\bibfield  {journal} {\bibinfo  {journal} {Phys. Rev. D}\ }\textbf {\bibinfo {volume} {78}},\ \bibinfo {pages} {094019} (\bibinfo {year} {2008})},\ \Eprint {http://arxiv.org/abs/0807.2674} {arXiv:0807.2674 [hep-ph]} \BibitemShut {NoStop}%
\bibitem [{\citenamefont {Fleming}\ and\ \citenamefont {Mehen}(2012)}]{Fleming:2011xa}%
  \BibitemOpen
  \bibfield  {author} {\bibinfo {author} {\bibfnamefont {S.}~\bibnamefont {Fleming}}\ and\ \bibinfo {author} {\bibfnamefont {T.}~\bibnamefont {Mehen}},\ }\href {\doibase 10.1103/PhysRevD.85.014016} {\bibfield  {journal} {\bibinfo  {journal} {Phys. Rev. D}\ }\textbf {\bibinfo {volume} {85}},\ \bibinfo {pages} {014016} (\bibinfo {year} {2012})},\ \Eprint {http://arxiv.org/abs/1110.0265} {arXiv:1110.0265 [hep-ph]} \BibitemShut {NoStop}%
\bibitem [{\citenamefont {Mehen}\ and\ \citenamefont {Springer}(2011)}]{Mehen:2011ds}%
  \BibitemOpen
  \bibfield  {author} {\bibinfo {author} {\bibfnamefont {T.}~\bibnamefont {Mehen}}\ and\ \bibinfo {author} {\bibfnamefont {R.}~\bibnamefont {Springer}},\ }\href {\doibase 10.1103/PhysRevD.83.094009} {\bibfield  {journal} {\bibinfo  {journal} {Phys. Rev. D}\ }\textbf {\bibinfo {volume} {83}},\ \bibinfo {pages} {094009} (\bibinfo {year} {2011})},\ \Eprint {http://arxiv.org/abs/1101.5175} {arXiv:1101.5175 [hep-ph]} \BibitemShut {NoStop}%
\bibitem [{\citenamefont {Margaryan}\ and\ \citenamefont {Springer}(2013)}]{Margaryan:2013tta}%
  \BibitemOpen
  \bibfield  {author} {\bibinfo {author} {\bibfnamefont {A.}~\bibnamefont {Margaryan}}\ and\ \bibinfo {author} {\bibfnamefont {R.~P.}\ \bibnamefont {Springer}},\ }\href {\doibase 10.1103/PhysRevD.88.014017} {\bibfield  {journal} {\bibinfo  {journal} {Phys. Rev. D}\ }\textbf {\bibinfo {volume} {88}},\ \bibinfo {pages} {014017} (\bibinfo {year} {2013})},\ \Eprint {http://arxiv.org/abs/1304.8101} {arXiv:1304.8101 [hep-ph]} \BibitemShut {NoStop}%
\bibitem [{\citenamefont {Braaten}\ \emph {et~al.}(2010)\citenamefont {Braaten}, \citenamefont {Hammer},\ and\ \citenamefont {Mehen}}]{Braaten:2010mg}%
  \BibitemOpen
  \bibfield  {author} {\bibinfo {author} {\bibfnamefont {E.}~\bibnamefont {Braaten}}, \bibinfo {author} {\bibfnamefont {H.~W.}\ \bibnamefont {Hammer}}, \ and\ \bibinfo {author} {\bibfnamefont {T.}~\bibnamefont {Mehen}},\ }\href {\doibase 10.1103/PhysRevD.82.034018} {\bibfield  {journal} {\bibinfo  {journal} {Phys. Rev. D}\ }\textbf {\bibinfo {volume} {82}},\ \bibinfo {pages} {034018} (\bibinfo {year} {2010})},\ \Eprint {http://arxiv.org/abs/1005.1688} {arXiv:1005.1688 [hep-ph]} \BibitemShut {NoStop}%
\bibitem [{\citenamefont {Canham}\ \emph {et~al.}(2009)\citenamefont {Canham}, \citenamefont {Hammer},\ and\ \citenamefont {Springer}}]{Canham:2009zq}%
  \BibitemOpen
  \bibfield  {author} {\bibinfo {author} {\bibfnamefont {D.~L.}\ \bibnamefont {Canham}}, \bibinfo {author} {\bibfnamefont {H.~W.}\ \bibnamefont {Hammer}}, \ and\ \bibinfo {author} {\bibfnamefont {R.~P.}\ \bibnamefont {Springer}},\ }\href {\doibase 10.1103/PhysRevD.80.014009} {\bibfield  {journal} {\bibinfo  {journal} {Phys. Rev. D}\ }\textbf {\bibinfo {volume} {80}},\ \bibinfo {pages} {014009} (\bibinfo {year} {2009})},\ \Eprint {http://arxiv.org/abs/0906.1263} {arXiv:0906.1263 [hep-ph]} \BibitemShut {NoStop}%
\bibitem [{\citenamefont {Jansen}\ \emph {et~al.}(2014)\citenamefont {Jansen}, \citenamefont {Hammer},\ and\ \citenamefont {Jia}}]{Jansen:2013cba}%
  \BibitemOpen
  \bibfield  {author} {\bibinfo {author} {\bibfnamefont {M.}~\bibnamefont {Jansen}}, \bibinfo {author} {\bibfnamefont {H.~W.}\ \bibnamefont {Hammer}}, \ and\ \bibinfo {author} {\bibfnamefont {Y.}~\bibnamefont {Jia}},\ }\href {\doibase 10.1103/PhysRevD.89.014033} {\bibfield  {journal} {\bibinfo  {journal} {Phys. Rev. D}\ }\textbf {\bibinfo {volume} {89}},\ \bibinfo {pages} {014033} (\bibinfo {year} {2014})},\ \Eprint {http://arxiv.org/abs/1310.6937} {arXiv:1310.6937 [hep-ph]} \BibitemShut {NoStop}%
\bibitem [{\citenamefont {Mehen}(2015)}]{Mehen:2015efa}%
  \BibitemOpen
  \bibfield  {author} {\bibinfo {author} {\bibfnamefont {T.}~\bibnamefont {Mehen}},\ }\href {\doibase 10.1103/PhysRevD.92.034019} {\bibfield  {journal} {\bibinfo  {journal} {Phys. Rev. D}\ }\textbf {\bibinfo {volume} {92}},\ \bibinfo {pages} {034019} (\bibinfo {year} {2015})},\ \Eprint {http://arxiv.org/abs/1503.02719} {arXiv:1503.02719 [hep-ph]} \BibitemShut {NoStop}%
\bibitem [{\citenamefont {Jansen}\ \emph {et~al.}(2015)\citenamefont {Jansen}, \citenamefont {Hammer},\ and\ \citenamefont {Jia}}]{Jansen:2015lha}%
  \BibitemOpen
  \bibfield  {author} {\bibinfo {author} {\bibfnamefont {M.}~\bibnamefont {Jansen}}, \bibinfo {author} {\bibfnamefont {H.~W.}\ \bibnamefont {Hammer}}, \ and\ \bibinfo {author} {\bibfnamefont {Y.}~\bibnamefont {Jia}},\ }\href {\doibase 10.1103/PhysRevD.92.114031} {\bibfield  {journal} {\bibinfo  {journal} {Phys. Rev. D}\ }\textbf {\bibinfo {volume} {92}},\ \bibinfo {pages} {114031} (\bibinfo {year} {2015})},\ \Eprint {http://arxiv.org/abs/1505.04099} {arXiv:1505.04099 [hep-ph]} \BibitemShut {NoStop}%
\bibitem [{\citenamefont {Alhakami}\ and\ \citenamefont {Birse}(2015)}]{Alhakami:2015uea}%
  \BibitemOpen
  \bibfield  {author} {\bibinfo {author} {\bibfnamefont {M.~H.}\ \bibnamefont {Alhakami}}\ and\ \bibinfo {author} {\bibfnamefont {M.~C.}\ \bibnamefont {Birse}},\ }\href {\doibase 10.1103/PhysRevD.91.054019} {\bibfield  {journal} {\bibinfo  {journal} {Phys. Rev. D}\ }\textbf {\bibinfo {volume} {91}},\ \bibinfo {pages} {054019} (\bibinfo {year} {2015})},\ \Eprint {http://arxiv.org/abs/1501.06750} {arXiv:1501.06750 [hep-ph]} \BibitemShut {NoStop}%
\bibitem [{\citenamefont {Braaten}(2015)}]{Braaten:2015tga}%
  \BibitemOpen
  \bibfield  {author} {\bibinfo {author} {\bibfnamefont {E.}~\bibnamefont {Braaten}},\ }\href {\doibase 10.1103/PhysRevD.91.114007} {\bibfield  {journal} {\bibinfo  {journal} {Phys. Rev. D}\ }\textbf {\bibinfo {volume} {91}},\ \bibinfo {pages} {114007} (\bibinfo {year} {2015})},\ \Eprint {http://arxiv.org/abs/1503.04791} {arXiv:1503.04791 [hep-ph]} \BibitemShut {NoStop}%
\bibitem [{\citenamefont {Dai}\ \emph {et~al.}(2020)\citenamefont {Dai}, \citenamefont {Guo},\ and\ \citenamefont {Mehen}}]{Dai:2019hrf}%
  \BibitemOpen
  \bibfield  {author} {\bibinfo {author} {\bibfnamefont {L.}~\bibnamefont {Dai}}, \bibinfo {author} {\bibfnamefont {F.-K.}\ \bibnamefont {Guo}}, \ and\ \bibinfo {author} {\bibfnamefont {T.}~\bibnamefont {Mehen}},\ }\href {\doibase 10.1103/PhysRevD.101.054024} {\bibfield  {journal} {\bibinfo  {journal} {Phys. Rev. D}\ }\textbf {\bibinfo {volume} {101}},\ \bibinfo {pages} {054024} (\bibinfo {year} {2020})},\ \Eprint {http://arxiv.org/abs/1912.04317} {arXiv:1912.04317 [hep-ph]} \BibitemShut {NoStop}%
\bibitem [{\citenamefont {Braaten}\ \emph {et~al.}(2021{\natexlab{a}})\citenamefont {Braaten}, \citenamefont {He},\ and\ \citenamefont {Jiang}}]{Braaten:2020nmc}%
  \BibitemOpen
  \bibfield  {author} {\bibinfo {author} {\bibfnamefont {E.}~\bibnamefont {Braaten}}, \bibinfo {author} {\bibfnamefont {L.-P.}\ \bibnamefont {He}}, \ and\ \bibinfo {author} {\bibfnamefont {J.}~\bibnamefont {Jiang}},\ }\href {\doibase 10.1103/PhysRevD.103.036014} {\bibfield  {journal} {\bibinfo  {journal} {Phys. Rev. D}\ }\textbf {\bibinfo {volume} {103}},\ \bibinfo {pages} {036014} (\bibinfo {year} {2021}{\natexlab{a}})},\ \Eprint {http://arxiv.org/abs/2010.05801} {arXiv:2010.05801 [hep-ph]} \BibitemShut {NoStop}%
\bibitem [{\citenamefont {Braaten}\ \emph {et~al.}(2020)\citenamefont {Braaten}, \citenamefont {He}, \citenamefont {Ingles},\ and\ \citenamefont {Jiang}}]{Braaten:2020iye}%
  \BibitemOpen
  \bibfield  {author} {\bibinfo {author} {\bibfnamefont {E.}~\bibnamefont {Braaten}}, \bibinfo {author} {\bibfnamefont {L.-P.}\ \bibnamefont {He}}, \bibinfo {author} {\bibfnamefont {K.}~\bibnamefont {Ingles}}, \ and\ \bibinfo {author} {\bibfnamefont {J.}~\bibnamefont {Jiang}},\ }\href {\doibase 10.1103/PhysRevD.101.096020} {\bibfield  {journal} {\bibinfo  {journal} {Phys. Rev. D}\ }\textbf {\bibinfo {volume} {101}},\ \bibinfo {pages} {096020} (\bibinfo {year} {2020})},\ \Eprint {http://arxiv.org/abs/2004.12841} {arXiv:2004.12841 [hep-ph]} \BibitemShut {NoStop}%
\bibitem [{\citenamefont {Braaten}\ \emph {et~al.}(2021{\natexlab{b}})\citenamefont {Braaten}, \citenamefont {He}, \citenamefont {Ingles},\ and\ \citenamefont {Jiang}}]{Braaten:2020iqw}%
  \BibitemOpen
  \bibfield  {author} {\bibinfo {author} {\bibfnamefont {E.}~\bibnamefont {Braaten}}, \bibinfo {author} {\bibfnamefont {L.-P.}\ \bibnamefont {He}}, \bibinfo {author} {\bibfnamefont {K.}~\bibnamefont {Ingles}}, \ and\ \bibinfo {author} {\bibfnamefont {J.}~\bibnamefont {Jiang}},\ }\href {\doibase 10.1103/PhysRevD.103.L071901} {\bibfield  {journal} {\bibinfo  {journal} {Phys. Rev. D}\ }\textbf {\bibinfo {volume} {103}},\ \bibinfo {pages} {L071901} (\bibinfo {year} {2021}{\natexlab{b}})},\ \Eprint {http://arxiv.org/abs/2012.13499} {arXiv:2012.13499 [hep-ph]} \BibitemShut {NoStop}%
\bibitem [{\citenamefont {Choi}\ \emph {et~al.}(2003)\citenamefont {Choi} \emph {et~al.}}]{Belle:2003nnu}%
  \BibitemOpen
  \bibfield  {author} {\bibinfo {author} {\bibfnamefont {S.~K.}\ \bibnamefont {Choi}} \emph {et~al.} (\bibinfo {collaboration} {Belle}),\ }\href {\doibase 10.1103/PhysRevLett.91.262001} {\bibfield  {journal} {\bibinfo  {journal} {Phys. Rev. Lett.}\ }\textbf {\bibinfo {volume} {91}},\ \bibinfo {pages} {262001} (\bibinfo {year} {2003})},\ \Eprint {http://arxiv.org/abs/hep-ex/0309032} {arXiv:hep-ex/0309032} \BibitemShut {NoStop}%
\bibitem [{\citenamefont {Workman}\ \emph {et~al.}(2022)\citenamefont {Workman} \emph {et~al.}}]{ParticleDataGroup:2022pth}%
  \BibitemOpen
  \bibfield  {author} {\bibinfo {author} {\bibfnamefont {R.~L.}\ \bibnamefont {Workman}} \emph {et~al.} (\bibinfo {collaboration} {Particle Data Group}),\ }\href {\doibase 10.1093/ptep/ptac097} {\bibfield  {journal} {\bibinfo  {journal} {PTEP}\ }\textbf {\bibinfo {volume} {2022}},\ \bibinfo {pages} {083C01} (\bibinfo {year} {2022})}\BibitemShut {NoStop}%
\bibitem [{\citenamefont {Aaij}\ \emph {et~al.}(2020{\natexlab{a}})\citenamefont {Aaij} \emph {et~al.}}]{LHCb:2020xds}%
  \BibitemOpen
  \bibfield  {author} {\bibinfo {author} {\bibfnamefont {R.}~\bibnamefont {Aaij}} \emph {et~al.} (\bibinfo {collaboration} {LHCb}),\ }\href {\doibase 10.1103/PhysRevD.102.092005} {\bibfield  {journal} {\bibinfo  {journal} {Phys. Rev. D}\ }\textbf {\bibinfo {volume} {102}},\ \bibinfo {pages} {092005} (\bibinfo {year} {2020}{\natexlab{a}})},\ \Eprint {http://arxiv.org/abs/2005.13419} {arXiv:2005.13419 [hep-ex]} \BibitemShut {NoStop}%
\bibitem [{\citenamefont {Aaij}\ \emph {et~al.}(2020{\natexlab{b}})\citenamefont {Aaij} \emph {et~al.}}]{LHCb:2020fvo}%
  \BibitemOpen
  \bibfield  {author} {\bibinfo {author} {\bibfnamefont {R.}~\bibnamefont {Aaij}} \emph {et~al.} (\bibinfo {collaboration} {LHCb}),\ }\href {\doibase 10.1007/JHEP08(2020)123} {\bibfield  {journal} {\bibinfo  {journal} {JHEP}\ }\textbf {\bibinfo {volume} {08}},\ \bibinfo {pages} {123} (\bibinfo {year} {2020}{\natexlab{b}})},\ \Eprint {http://arxiv.org/abs/2005.13422} {arXiv:2005.13422 [hep-ex]} \BibitemShut {NoStop}%
\bibitem [{\citenamefont {Voloshin}(2004)}]{Voloshin:2003nt}%
  \BibitemOpen
  \bibfield  {author} {\bibinfo {author} {\bibfnamefont {M.~B.}\ \bibnamefont {Voloshin}},\ }\href {\doibase 10.1016/j.physletb.2003.11.014} {\bibfield  {journal} {\bibinfo  {journal} {Phys. Lett. B}\ }\textbf {\bibinfo {volume} {579}},\ \bibinfo {pages} {316} (\bibinfo {year} {2004})},\ \Eprint {http://arxiv.org/abs/hep-ph/0309307} {arXiv:hep-ph/0309307} \BibitemShut {NoStop}%
\bibitem [{\citenamefont {Fleming}\ \emph {et~al.}(2021)\citenamefont {Fleming}, \citenamefont {Hodges},\ and\ \citenamefont {Mehen}}]{Fleming:2021wmk}%
  \BibitemOpen
  \bibfield  {author} {\bibinfo {author} {\bibfnamefont {S.}~\bibnamefont {Fleming}}, \bibinfo {author} {\bibfnamefont {R.}~\bibnamefont {Hodges}}, \ and\ \bibinfo {author} {\bibfnamefont {T.}~\bibnamefont {Mehen}},\ }\href {\doibase 10.1103/PhysRevD.104.116010} {\bibfield  {journal} {\bibinfo  {journal} {Phys. Rev. D}\ }\textbf {\bibinfo {volume} {104}},\ \bibinfo {pages} {116010} (\bibinfo {year} {2021})},\ \Eprint {http://arxiv.org/abs/2109.02188} {arXiv:2109.02188 [hep-ph]} \BibitemShut {NoStop}%
\bibitem [{\citenamefont {Dai}\ \emph {et~al.}(2023)\citenamefont {Dai}, \citenamefont {Fleming}, \citenamefont {Hodges},\ and\ \citenamefont {Mehen}}]{Dai:2023mxm}%
  \BibitemOpen
  \bibfield  {author} {\bibinfo {author} {\bibfnamefont {L.}~\bibnamefont {Dai}}, \bibinfo {author} {\bibfnamefont {S.}~\bibnamefont {Fleming}}, \bibinfo {author} {\bibfnamefont {R.}~\bibnamefont {Hodges}}, \ and\ \bibinfo {author} {\bibfnamefont {T.}~\bibnamefont {Mehen}},\ }\href@noop {} {\  (\bibinfo {year} {2023})},\ \Eprint {http://arxiv.org/abs/2301.11950} {arXiv:2301.11950 [hep-ph]} \BibitemShut {NoStop}%
\bibitem [{\citenamefont {Guo}\ \emph {et~al.}(2018)\citenamefont {Guo}, \citenamefont {Hanhart}, \citenamefont {Mei\ss{}ner}, \citenamefont {Wang}, \citenamefont {Zhao},\ and\ \citenamefont {Zou}}]{Guo:2017jvc}%
  \BibitemOpen
  \bibfield  {author} {\bibinfo {author} {\bibfnamefont {F.-K.}\ \bibnamefont {Guo}}, \bibinfo {author} {\bibfnamefont {C.}~\bibnamefont {Hanhart}}, \bibinfo {author} {\bibfnamefont {U.-G.}\ \bibnamefont {Mei\ss{}ner}}, \bibinfo {author} {\bibfnamefont {Q.}~\bibnamefont {Wang}}, \bibinfo {author} {\bibfnamefont {Q.}~\bibnamefont {Zhao}}, \ and\ \bibinfo {author} {\bibfnamefont {B.-S.}\ \bibnamefont {Zou}},\ }\href {\doibase 10.1103/RevModPhys.90.015004} {\bibfield  {journal} {\bibinfo  {journal} {Rev. Mod. Phys.}\ }\textbf {\bibinfo {volume} {90}},\ \bibinfo {pages} {015004} (\bibinfo {year} {2018})},\ \Eprint {http://arxiv.org/abs/1705.00141} {arXiv:1705.00141 [hep-ph]} \BibitemShut {NoStop}%
\bibitem [{\citenamefont {Kaplan}\ \emph {et~al.}(1998{\natexlab{a}})\citenamefont {Kaplan}, \citenamefont {Savage},\ and\ \citenamefont {Wise}}]{Kaplan:1998we}%
  \BibitemOpen
  \bibfield  {author} {\bibinfo {author} {\bibfnamefont {D.~B.}\ \bibnamefont {Kaplan}}, \bibinfo {author} {\bibfnamefont {M.~J.}\ \bibnamefont {Savage}}, \ and\ \bibinfo {author} {\bibfnamefont {M.~B.}\ \bibnamefont {Wise}},\ }\href {\doibase 10.1016/S0550-3213(98)00440-4} {\bibfield  {journal} {\bibinfo  {journal} {Nucl. Phys. B}\ }\textbf {\bibinfo {volume} {534}},\ \bibinfo {pages} {329} (\bibinfo {year} {1998}{\natexlab{a}})},\ \Eprint {http://arxiv.org/abs/nucl-th/9802075} {arXiv:nucl-th/9802075} \BibitemShut {NoStop}%
\bibitem [{\citenamefont {Guo}\ \emph {et~al.}(2019)\citenamefont {Guo}, \citenamefont {Jing}, \citenamefont {Mei\ss{}ner},\ and\ \citenamefont {Sakai}}]{Guo:2019fdo}%
  \BibitemOpen
  \bibfield  {author} {\bibinfo {author} {\bibfnamefont {F.-K.}\ \bibnamefont {Guo}}, \bibinfo {author} {\bibfnamefont {H.-J.}\ \bibnamefont {Jing}}, \bibinfo {author} {\bibfnamefont {U.-G.}\ \bibnamefont {Mei\ss{}ner}}, \ and\ \bibinfo {author} {\bibfnamefont {S.}~\bibnamefont {Sakai}},\ }\href {\doibase 10.1103/PhysRevD.99.091501} {\bibfield  {journal} {\bibinfo  {journal} {Phys. Rev. D}\ }\textbf {\bibinfo {volume} {99}},\ \bibinfo {pages} {091501} (\bibinfo {year} {2019})},\ \Eprint {http://arxiv.org/abs/1903.11503} {arXiv:1903.11503 [hep-ph]} \BibitemShut {NoStop}%
\bibitem [{\citenamefont {Liu}\ \emph {et~al.}(2013)\citenamefont {Liu}, \citenamefont {Orginos}, \citenamefont {Guo}, \citenamefont {Hanhart},\ and\ \citenamefont {Mei{\ss}ner}}]{Liu:2012zya}%
  \BibitemOpen
  \bibfield  {author} {\bibinfo {author} {\bibfnamefont {L.}~\bibnamefont {Liu}}, \bibinfo {author} {\bibfnamefont {K.}~\bibnamefont {Orginos}}, \bibinfo {author} {\bibfnamefont {F.-K.}\ \bibnamefont {Guo}}, \bibinfo {author} {\bibfnamefont {C.}~\bibnamefont {Hanhart}}, \ and\ \bibinfo {author} {\bibfnamefont {U.-G.}\ \bibnamefont {Mei{\ss}ner}},\ }\href {\doibase 10.1103/PhysRevD.87.014508} {\bibfield  {journal} {\bibinfo  {journal} {Phys. Rev. D}\ }\textbf {\bibinfo {volume} {87}},\ \bibinfo {pages} {014508} (\bibinfo {year} {2013})},\ \Eprint {http://arxiv.org/abs/1208.4535} {arXiv:1208.4535 [hep-lat]} \BibitemShut {NoStop}%
\bibitem [{\citenamefont {Weinberg}(1965)}]{Weinberg:1965zz}%
  \BibitemOpen
  \bibfield  {author} {\bibinfo {author} {\bibfnamefont {S.}~\bibnamefont {Weinberg}},\ }\href {\doibase 10.1103/PhysRev.137.B672} {\bibfield  {journal} {\bibinfo  {journal} {Phys. Rev.}\ }\textbf {\bibinfo {volume} {137}},\ \bibinfo {pages} {B672} (\bibinfo {year} {1965})}\BibitemShut {NoStop}%
\bibitem [{\citenamefont {Baru}\ \emph {et~al.}(2004)\citenamefont {Baru}, \citenamefont {Haidenbauer}, \citenamefont {Hanhart}, \citenamefont {Kalashnikova},\ and\ \citenamefont {Kudryavtsev}}]{Baru:2003qq}%
  \BibitemOpen
  \bibfield  {author} {\bibinfo {author} {\bibfnamefont {V.}~\bibnamefont {Baru}}, \bibinfo {author} {\bibfnamefont {J.}~\bibnamefont {Haidenbauer}}, \bibinfo {author} {\bibfnamefont {C.}~\bibnamefont {Hanhart}}, \bibinfo {author} {\bibfnamefont {Y.}~\bibnamefont {Kalashnikova}}, \ and\ \bibinfo {author} {\bibfnamefont {A.~E.}\ \bibnamefont {Kudryavtsev}},\ }\href {\doibase 10.1016/j.physletb.2004.01.088} {\bibfield  {journal} {\bibinfo  {journal} {Phys. Lett. B}\ }\textbf {\bibinfo {volume} {586}},\ \bibinfo {pages} {53} (\bibinfo {year} {2004})},\ \Eprint {http://arxiv.org/abs/hep-ph/0308129} {arXiv:hep-ph/0308129} \BibitemShut {NoStop}%
\bibitem [{\citenamefont {Yan}\ and\ \citenamefont {Valderrama}(2022)}]{Yan:2021wdl}%
  \BibitemOpen
  \bibfield  {author} {\bibinfo {author} {\bibfnamefont {M.-J.}\ \bibnamefont {Yan}}\ and\ \bibinfo {author} {\bibfnamefont {M.~P.}\ \bibnamefont {Valderrama}},\ }\href {\doibase 10.1103/PhysRevD.105.014007} {\bibfield  {journal} {\bibinfo  {journal} {Phys. Rev. D}\ }\textbf {\bibinfo {volume} {105}},\ \bibinfo {pages} {014007} (\bibinfo {year} {2022})},\ \Eprint {http://arxiv.org/abs/2108.04785} {arXiv:2108.04785 [hep-ph]} \BibitemShut {NoStop}%
\bibitem [{\citenamefont {Guo}(2019)}]{Guo:2019qcn}%
  \BibitemOpen
  \bibfield  {author} {\bibinfo {author} {\bibfnamefont {F.-K.}\ \bibnamefont {Guo}},\ }\href {\doibase 10.1103/PhysRevLett.122.202002} {\bibfield  {journal} {\bibinfo  {journal} {Phys. Rev. Lett.}\ }\textbf {\bibinfo {volume} {122}},\ \bibinfo {pages} {202002} (\bibinfo {year} {2019})},\ \Eprint {http://arxiv.org/abs/1902.11221} {arXiv:1902.11221 [hep-ph]} \BibitemShut {NoStop}%
\bibitem [{\citenamefont {Guo}\ \emph {et~al.}(2011)\citenamefont {Guo}, \citenamefont {Hanhart}, \citenamefont {Li}, \citenamefont {Mei{\ss}ner},\ and\ \citenamefont {Zhao}}]{Guo:2010ak}%
  \BibitemOpen
  \bibfield  {author} {\bibinfo {author} {\bibfnamefont {F.-K.}\ \bibnamefont {Guo}}, \bibinfo {author} {\bibfnamefont {C.}~\bibnamefont {Hanhart}}, \bibinfo {author} {\bibfnamefont {G.}~\bibnamefont {Li}}, \bibinfo {author} {\bibfnamefont {U.-G.}\ \bibnamefont {Mei{\ss}ner}}, \ and\ \bibinfo {author} {\bibfnamefont {Q.}~\bibnamefont {Zhao}},\ }\href {\doibase 10.1103/PhysRevD.83.034013} {\bibfield  {journal} {\bibinfo  {journal} {Phys. Rev. D}\ }\textbf {\bibinfo {volume} {83}},\ \bibinfo {pages} {034013} (\bibinfo {year} {2011})},\ \Eprint {http://arxiv.org/abs/1008.3632} {arXiv:1008.3632 [hep-ph]} \BibitemShut {NoStop}%
\bibitem [{\citenamefont {Kaplan}\ \emph {et~al.}(1998{\natexlab{b}})\citenamefont {Kaplan}, \citenamefont {Savage},\ and\ \citenamefont {Wise}}]{Kaplan:1998tg}%
  \BibitemOpen
  \bibfield  {author} {\bibinfo {author} {\bibfnamefont {D.~B.}\ \bibnamefont {Kaplan}}, \bibinfo {author} {\bibfnamefont {M.~J.}\ \bibnamefont {Savage}}, \ and\ \bibinfo {author} {\bibfnamefont {M.~B.}\ \bibnamefont {Wise}},\ }\href {\doibase 10.1016/S0370-2693(98)00210-X} {\bibfield  {journal} {\bibinfo  {journal} {Phys. Lett. B}\ }\textbf {\bibinfo {volume} {424}},\ \bibinfo {pages} {390} (\bibinfo {year} {1998}{\natexlab{b}})},\ \Eprint {http://arxiv.org/abs/nucl-th/9801034} {arXiv:nucl-th/9801034} \BibitemShut {NoStop}%
\bibitem [{\citenamefont {Jing}\ \emph {et~al.}(2021)\citenamefont {Jing}, \citenamefont {Shen},\ and\ \citenamefont {Guo}}]{Jing:2020tth}%
  \BibitemOpen
  \bibfield  {author} {\bibinfo {author} {\bibfnamefont {H.-J.}\ \bibnamefont {Jing}}, \bibinfo {author} {\bibfnamefont {C.-W.}\ \bibnamefont {Shen}}, \ and\ \bibinfo {author} {\bibfnamefont {F.-K.}\ \bibnamefont {Guo}},\ }\href {\doibase 10.1016/j.scib.2020.10.009} {\bibfield  {journal} {\bibinfo  {journal} {Science Bulletin}\ }\textbf {\bibinfo {volume} {66}},\ \bibinfo {pages} {653} (\bibinfo {year} {2021})},\ \Eprint {http://arxiv.org/abs/2005.01942} {arXiv:2005.01942 [hep-ph]} \BibitemShut {NoStop}%
\end{thebibliography}%


%

\end{document}